\documentclass[fleqn,usenatbib]{mnras}

\usepackage{newtxtext,newtxmath}

\usepackage[T1]{fontenc}

\DeclareRobustCommand{\VAN}[3]{#2}
\let\VANthebibliography\thebibliography
\def\thebibliography{\DeclareRobustCommand{\VAN}[3]{##3}\VANthebibliography}

\usepackage{graphicx}	
\usepackage{amsmath}	

\usepackage{longtable, booktabs}
\usepackage{threeparttable}
\usepackage{threeparttablex}
\usepackage{subcaption}
\usepackage{endnotes}

\title[Hydrogen-rich Type II Supernovae]{The optical properties of three type II supernovae: 2014cx, 2014cy and 2015cz}

\author[R Dastidar et al.]{R. Dastidar$^{1,2}$\thanks{E-mail: rayadastidar@aries.res.in, rdastidr@gmail.com},
K. Misra$^{1}$, M. Singh$^{3,1}$, A. Pastorello$^{4}$, D. K. Sahu$^{5}$, X. Wang$^{6,7}$, \newauthor A. Gangopadhyay$^{1,8}$, L. Tomasella$^{4}$, J. Zhang$^{9,10}$, S. Bose$^{11,12}$, J. Mo$^{6}$, N. Elias-Rosa$^{4,13}$, \newauthor L. Tartaglia$^{14}$, S. Yan$^{6}$, B. Kumar$^{1}$, G. C. Anupama$^{5}$, S. B. Pandey$^{1}$, L. Rui$^{6}$, T. Zhang$^{15}$, \newauthor G. Terreran$^{16}$, P. Ochner$^{4}$ and F. Huang$^{17}$
\\
$^{1}$Aryabhatta Research Institute of observational sciencES, Manora Peak, Nainital 263 001 India\\
$^{2}$Department of Physics \& Astrophysics, University of Delhi, Delhi-110 007\\
$^{3}$Korea Astronomy and Space Science Institute, 776 Daedeokdae-ro, Yuseong-gu, Daejeon 34055, Republic of Korea\\
$^{4}$INAF-Osservatorio Astronomico di Padova, Vicolo dell’Osservatorio 5, 35122 Padova, Italy\\
$^{5}$Indian Institute of Astrophysics, Koramangala, Bangalore 560 034, India\\
$^{6}$Physics Department and Tsinghua Center for Astrophysics, Tsinghua University, Beijing, 100084, China\\
$^{7}$Beijing Planetarium, Beijing Academy of Sciences and Technology, Beijing, 100044, China\\
$^{8}$Pt.Ravi Shankar Shukla University, Raipur 492 010,  India\\
$^{9}${Yunnan Observatories, Chinese Academy of Sciences, Kunming 650216, China}\\
$^{10}${Key Laboratory for the Structure and Evolution of Celestial Objects, Chinese Academy of Sciences, Kunming 650216, China} \\
$^{11}$Department of Astronomy, The Ohio State University, 140 W. 18th Avenue, Columbus, OH 43210, USA.\\
$^{12}$Center for Cosmology and AstroParticle Physics (CCAPP), The Ohio State University, 191 W. Woodruff Avenue, Columbus, OH 43210, USA.\\
$^{13}$Institute of Space Sciences (ICE, CSIC), Campus UAB, Carrer de Can Magrans s/n, 08193 Barcelona, Spain\\
$^{14}$Department of Astronomy and The Oskar Klein Centre, AlbaNova University Centre, Stockholm University, SE-106 91 Stockholm, Sweden\\
$^{15}$Key Laboratory of Optical Astronomy, National Astronomical Observatories, Chinese Academy of Sciences, 100012, Beijing, China\\
$^{16}$Center for Interdisciplinary Exploration and Research in Astrophysics (CIERA) and Department of Physics and Astronomy, Northwestern University, \\~~~~Evanston, IL 60208, USA \\
$^{17}$Department of Astronomy, Shanghai Jiaotong University, Shanghai, 200240, China\\
}

\date{Accepted XXX. Received YYY; in original form ZZZ}

\pubyear{2020}

\begin{document}
\label{firstpage}
\pagerange{\pageref{firstpage}--\pageref{lastpage}}
\maketitle

\begin{abstract}
We present the photometric and spectroscopic analysis of three Type II SNe: 2014cx, 2014cy and 2015cz. SN\,2014cx is a conventional Type IIP with shallow slope (0.2\,mag/50d) and an atypical short plateau ($\sim$86\,d). SNe\,2014cy and 2015cz show relatively large decline rates (0.88 and 1.64\,mag/50d, respectively) at early times before settling to the plateau phase, unlike the canonical Type IIP/L SN light curves. All of them are normal luminosity SN II with an absolute magnitude at mid-plateau of M$_{V,14cx}^{50}$\,=\,$-$16.6$\,\pm\,$0.4$\,\rm{mag}$, M$_{V,14cy}^{50}$\,=\,$-$16.5$\,\pm\,$0.2$\,\rm{mag}$ and M$_{V,15cz}^{50}$\,=\,$-$17.4$\,\pm\,$0.3$\,\rm{mag}$. A relatively broad range of $^{56}$Ni masses is ejected in these explosions (0.027-0.070\,M$_\odot$). The spectra shows the classical evolution of Type\,II SNe, dominated by a blue continuum with broad H lines at early phases and narrower metal lines with P\,Cygni profiles during the plateau. High velocity H\,{\sc i} features are identified in the plateau spectra of SN\,2014cx at 11600\,km\,s$^{-1}$, possibly a sign of ejecta-circumstellar interaction. The spectra of SN\,2014cy exhibit strong absorption profile of H\,{\sc i} similar to normal luminosity events whereas strong metal lines akin to sub-luminous SNe. The analytical modelling of the bolometric light curve of the three events yields similar radii for the three objects within errors (478, 507 and 608\,R$_\odot$ for SNe\,2014cx, 2014cy and 2015cz, respectively) and a range of ejecta masses (15.0, 22.2 and 18.7\,M$_\odot$ for SNe\,2014cx, 2014cy and 2015cz), and a modest range of explosion energies (3.3 - 6.0 foe where 1 foe = 10$^{51}$ erg).

\end{abstract}




\section{Introduction}

Core-collapse supernovae (SNe) are the terminal explosion of massive stars ($\gtrsim$ 8\,M$_\odot$) giving rise to a spectacular display in various domains of the electromagnetic spectrum. The initial classification of SNe based on the presence or absence of hydrogen features in their spectra dates back to 1941 \citep{1941PASP...53..224M}. The hydrogen-rich sub-class of core collapse SNe designated as SNe\,II, is a heterogeneous group exhibiting diverse observed properties in terms of their rise time, peak luminosity, decline rate, ejected $^{56}$Ni mass and expansion velocity (e.g. \citealt{2014ApJ...786...67A, 2016MNRAS.459.3939V}). In the broader picture, these events are characterized by strong hydrogen features in their spectra, however, the peculiarity associated with the event becomes apparent when looked at closely. The hydrogen-rich SNe II can be broadly classified into SNe\,IIP, which shows a $\sim$100$\,\rm{d}$ constant luminosity (plateau) phase and SNe\,IIL, which exhibit a linearly declining phase lasting for around 80$\,\rm{d}$ \citep{1979A&A....72..287B,2014MNRAS.442..844F,2014MNRAS.445..554F}. The other three identified classes of SNe\,II are Type IIn, displaying narrow emission lines in the spectra, Type IIb, with diminishing hydrogen and 1987A-likes, also known as SNe\,II-pec, with long rise times in their light curve. The progenitors of classical SNe IIP are well-constrained through direct detection of Red Supergiants (RSGs) in pre-explosion archival images \citep{2009ARA&A..47...63S,2015PASA...32...16S}. The proposed progenitors of 1987A-like transients are blue supergiants \citep{1989ARA&A..27..629A}, while the progenitors of SNe IIL are still uncertain. For example, the likely progenitor of Type IIL SN\,2009kr was proposed to be yellow supergiant in a binary system \citep{2010ApJ...714L.254E,2010ApJ...714L.280F}. However, with late time imaging it was discovered that the progenitor of SN\,2009kr is rather a small compact cluster \citep{2015MNRAS.447.3207M}. The impetus for observing SNe\,II events is fuelled by their higher recurrence rate among core-collapse SN \citep{2011MNRAS.412.1441L}, triggering the SN community to investigate their utility as standard candles \citep{1974ApJ...193...27K,2001ApJ...558..615H,2005A&A...439..671D} and metallicity indicators \citep{2014MNRAS.440.1856D,2016A&A...589A.110A}. 

In this paper, we present the optical photometric and spectroscopic analysis of three SNe\,II: SN\,2014cx, a normal luminosity SN\,II with a relatively short plateau length than the conventional SNe\,IIP, SN\,2014cy, a normal luminosity SN\,II with relatively low velocity and SN\,2015cz, a bright SN\,II with an early declining light curve. The SN and host galaxy parameters are provided in Table\,\ref{sn_detail}. The paper is organized as follows. In Section\,\ref{sec2}, the data reduction procedure is outlined and the SN and the host galaxy parameters are discussed in Section\,\ref{sec3}. The light curve properties, colour and temperature evolution and absolute magnitudes of the three SNe are discussed in Section\,\ref{sec:4}. The analysis of spectroscopic data, ejecta velocity and comparison with other SNe II are carried out in Section\,\ref{sec:5}. The distance, modelling of the bolometric light curve and the derived parameters are discussed in Section\,\ref{sec:6}. Finally, we discuss the implications of the observed properties in Section\,\ref{sec:7} and summarise the paper in Section\,\ref{sec:8}. 

\section{Data Reduction}
\label{sec2}
The photometric and spectroscopic observations of SNe\,2014cx, 2014cy and 2015cz were carried out using the 1.04\,m Sampurnanand Telescope, 1.3\,m Devasthal Fast Optical Telescope, ARIES Observatory and 2.0\,m Himalayan Chandra Telescope, Indian Astronomical Observatory, India; 80\,cm Tsinghua-NAOC (National Astronomical Observatories of China) Telescope and 2.16\,m Telescope, Xinglong Observatory and 2.4\,m Lijiang Telescope, Yunnan Astronomical Observatory, China; Schmidt 67/92\,cm and 1.82\,m Copernico Telescope, Asiago, Mt. Ekar and 1.22\,m Galileo Telescope, Pennar, Italy; 2.56\,m Nordic Optical Telescope and 3.58\,m Telescopio Nazionale Galileo, La Palma, Spain.

\begin{table}
    \caption{Basic information on SNe 2014cx, 2014cy and 2015cz and their host galaxies. The host galaxy parameters are taken from NED.\label{sn_detail}}
    \scalebox{0.75}{
     \begin{tabular}{llll}
\hline
 & SN 2014cx & SN 2014cy & SN 2015cz\\
 \hline
Host galaxy & NGC\,337 & NGC\,7742 & NGC\,582 \\
Galaxy type &  SBd & SAb & SBb\\  
Redshift & 0.00549  & 0.005547  & 0.014517 \\ 
Major Diameter (arcmin) &  2.9 & 1.7 & 2.2\\
Minor Diameter (arcmin) &  1.8 & 1.7 & 0.6\\
Helio. Radial Velocity (km s$^{-1}$) & 1646$\,\pm\,$2  & 1663$\,\pm\,$1 & 4352$\,\pm\,$4\\
\hline
SN type & II & II & II\\
RA (J2000.0) & 00$^h$59$^m$47$^s$.83 & 23$^h$44$^m$16$^s$.03 & 01$^h$31$^m$59$^s$.45\\
Dec (J2000.0) & $-$07$^\circ$34$^{\prime}$18$^{\prime\prime}$.6 & $+$10$^\circ$46$^{\prime}$12$^{\prime\prime}$.5 & $+$33$^\circ$28$^{\prime}$45$^{\prime\prime}$.8\\
Offset from nucleus & 33$^{\prime\prime}$.7 W, 21$^{\prime\prime}$.7 N & 3$^{\prime\prime}$.8 E, 10$^{\prime\prime}$.4 N & 15$^{\prime\prime}$.8 E, 10$^{\prime\prime}$.3 N\\
Date of Discovery (JD) & 2456903.1 & 2456901.2 & 2457298.4\\
Estimated date of explosion (JD) & 2456902.4 $\,\pm\,$ 0.5 & 2456900.0 $\,\pm\,$ 1.0 & 2457291.6 $\,\pm\,$ 4.3 \\
Distance (Mpc) & 18.5 $\,\pm\,$ 1.1 & 23.4 $\,\pm\,$ 1.6 & 63.7 $\,\pm\,$ 5.7 \\
Total Extinction E(B-V) (mag) & 0.096 $\,\pm\,$ 0.002 & 0.36 $\,\pm\,$ 0.05 & 0.48 $\,\pm\,$ 0.06 \\
\hline
\end{tabular}}
\end{table}

\begin{figure*}
  \centerline{\includegraphics[width=0.332\textwidth, trim={0.0cm -2.0cm 0.0cm 0cm}]{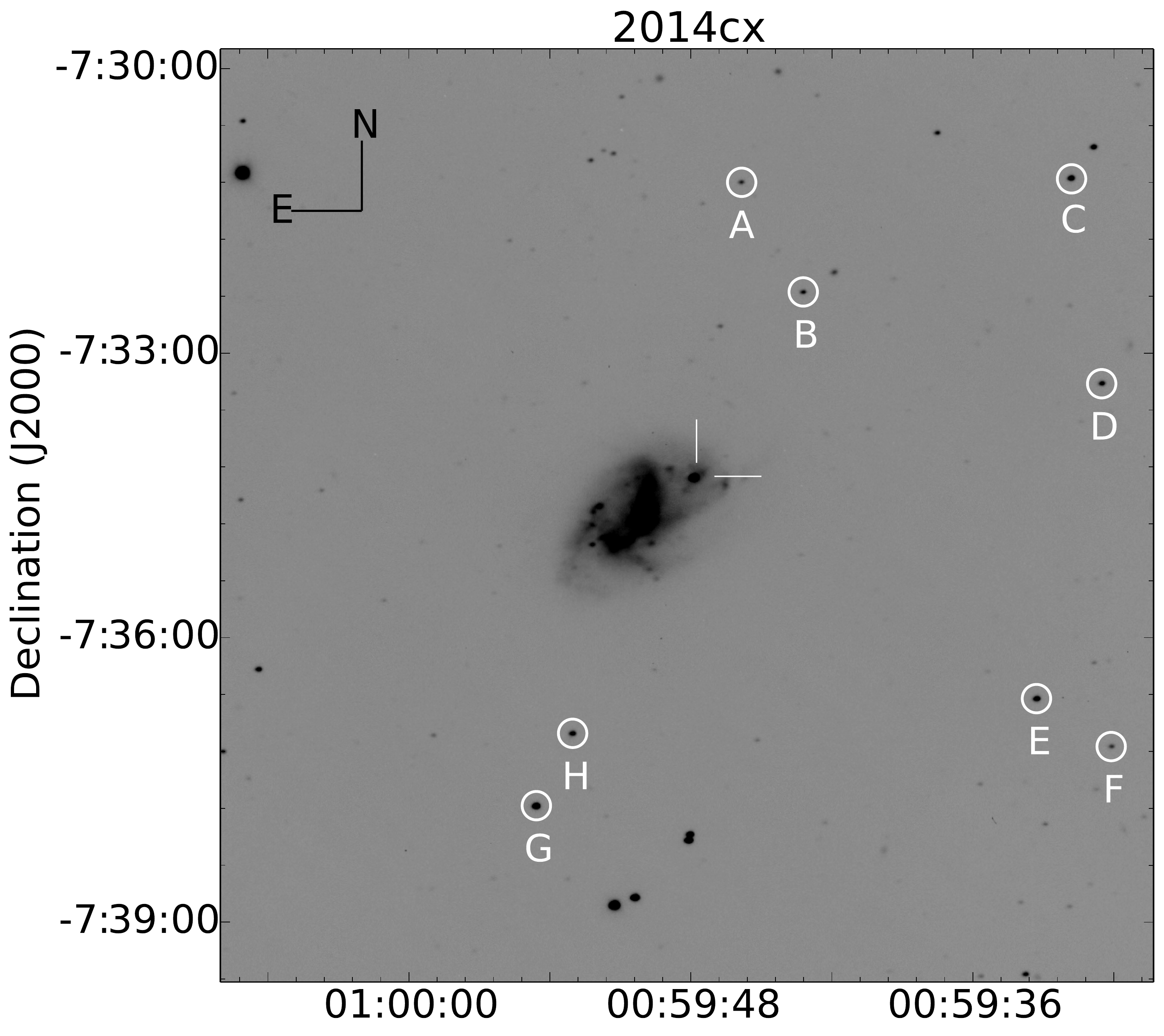}
              \includegraphics[width=0.332\textwidth,trim={0.0cm -0.7cm 0.0cm 0cm}]{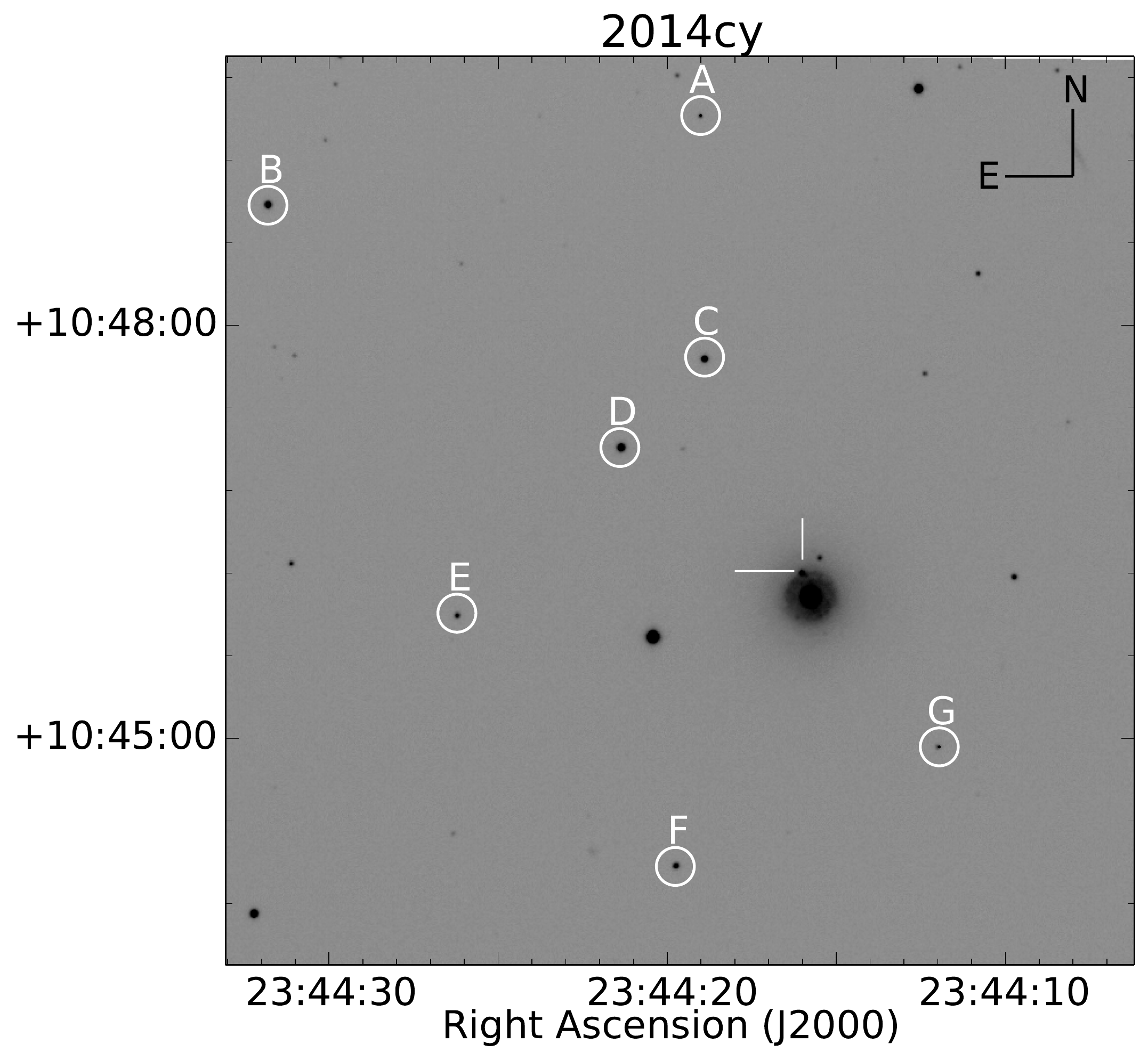}
              \includegraphics[width=0.332\textwidth, trim={0.0cm -2.0cm 0.0cm 0cm}]{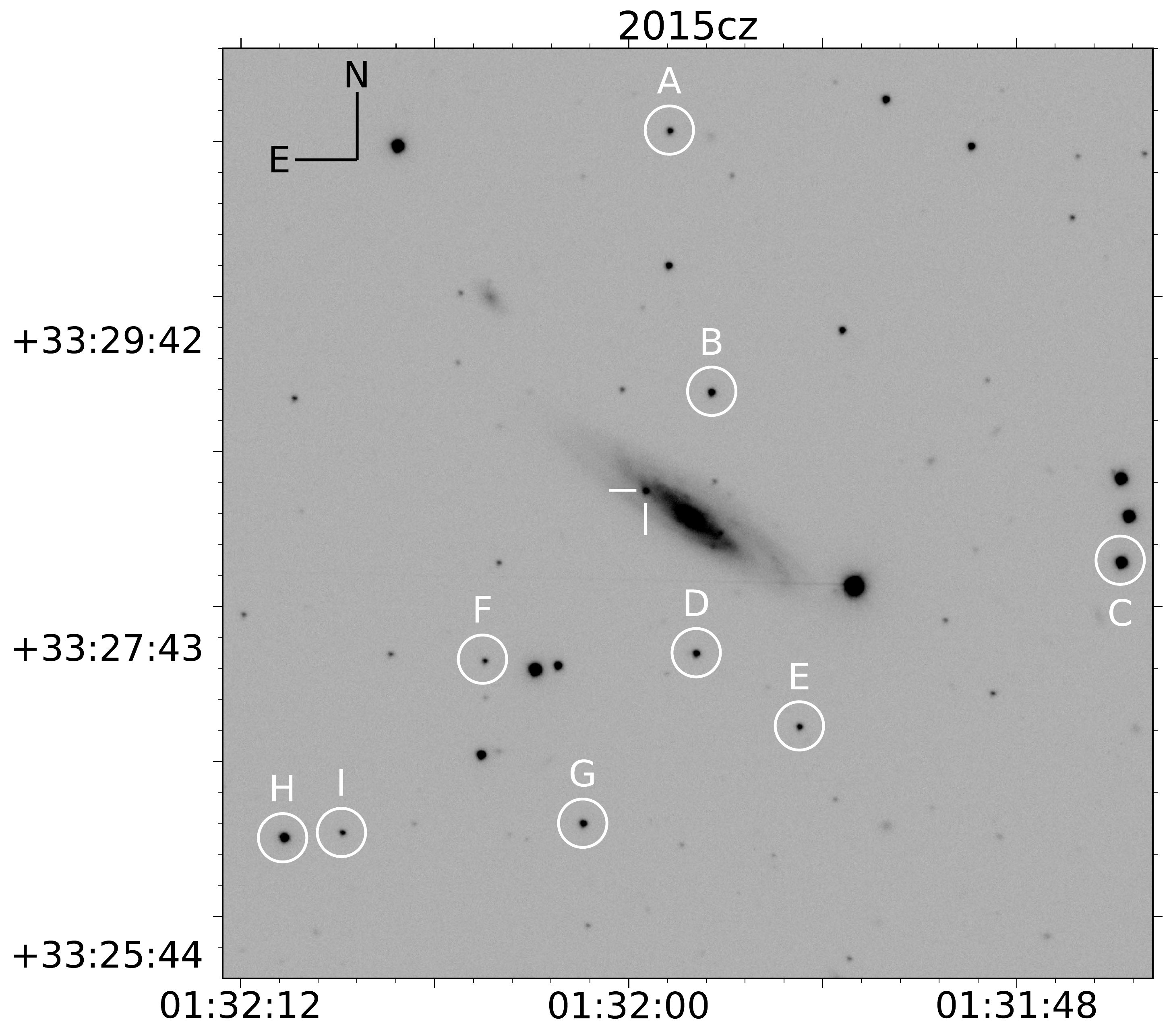}}
  \caption{The local standards along with the SNe 2014cx, 2014cy and 2015cz in their respective fields are marked.}
  \label{ch6:Fig1}
\end{figure*}

The follow up of SNe\,2014cx and 2014cy started nearly a month after discovery and that of SN\,2015cz one day after its discovery. SNe\,2014cx, 2014cy and 2015cz were observed up to $\sim$163, 165 and 150$\,\rm{d}$, respectively, in the broadband {\it BVRI} and Sloan {\it griz} filters (in the AB system). The raw data were pre-processed and reduced using standard tasks in {\sc iraf}\footnote{IRAF stands for Image Reduction and Analysis Facility distributed by the National Optical Astronomy Observatories which is operated by the Association of Universities for research in Astronomy, Inc., under cooperative agreement with the National Science Foundation.}. SNe\,2014cy and 2015cz exploded close to their respective host galaxy nucleus. Hence, template subtraction was performed to eliminate the host galaxy contamination from the SN magnitudes. To estimate the instrumental magnitudes, we used the standalone {\sc daophot ii} program \citep{1987PASP...99..191S} to perform point spread function photometry. Local standard stars in the field of the SN (shown in Figure\,\ref{ch6:Fig1} and tabulated for each field in Table\,\ref{local_std_cx}) were generated using the colour terms and zeropoints, that are evaluated using a set of Landolt \citep{2009AJ....137.4186L} standards observed on photometric nights. The Sloan magnitudes of the local standards were obtained from the AAVSO Photometric All-Sky Survey (APASS\footnote{https://www.aavso.org/apass}) catalog. The nightly zero-points were computed using the local standards and the SN magnitudes were differentially calibrated. The calibrated magnitudes of the SNe are provided in Table\,\ref{snmag}. 

\begin{table*}
\centering
\caption{Star ID, coordinates and calibrated magnitudes of secondary standard stars in the field of SNe 2014cx, 2014cy and 2015cz.\label{local_std_cx}}
\begin{tabular}{c c c c c c c c }
\hline
 ID & $\alpha_{J2000.0}$ & $\delta_{J2000.0}$ & $B$     &  $V$  &  $R$  & $I$    \\
    & (hh:mm:ss)         & (dd:mm:ss)         & (mag)   & (mag) & (mag) & (mag)  \\
\hline
\multicolumn{7}{c}{SN 2014cx}\\
\cline{4-5}
A            & 00:59:45.8  &  -07:31:12.0   & 19.484 $\,\pm\,$ 0.029 &   18.598 $\,\pm\,$ 0.026 & 17.987 $\,\pm\,$ 0.023    &   17.367 $\,\pm\,$ 0.019  \\
B            & 00:59:43.2  &  -07:32:21.3   & 20.203 $\,\pm\,$ 0.014 &   18.632 $\,\pm\,$ 0.006 & 17.633 $\,\pm\,$ 0.008    &   16.716 $\,\pm\,$ 0.010 \\
C            & 00:59:31.8 &  -07:31:09.8    & 18.832 $\,\pm\,$ 0.026 &   17.506 $\,\pm\,$ 0.005 & 16.598 $\,\pm\,$ 0.005    &   15.568 $\,\pm\,$ 0.008  \\
D            & 00:59:30.5  &  -07:33:19.4   & 16.999 $\,\pm\,$ 0.008 &   17.151 $\,\pm\,$ 0.007 & 17.173  $\,\pm\,$ 0.008   &   --                                         \\
E            & 00:59:33.3  &  -07:36:38.6    & 17.853 $\,\pm\,$ 0.006 &   17.081 $\,\pm\,$ 0.003  &  16.614 $\,\pm\,$ 0.005   &   16.206 $\,\pm\,$ 0.010  \\
F           & 00:59:30.1  &  -07:37:08.9    & 18.710 $\,\pm\,$ 0.007 &   18.363 $\,\pm\,$ 0.005 &  18.109 $\,\pm\,$ 0.008 &   17.822 $\,\pm\,$ 0.013 \\
G           & 00:59:54.6  &  -07:37:46.3    & 17.378 $\,\pm\,$ 0.006 &   16.409 $\,\pm\,$ 0.004  &  15.756 $\,\pm\,$ 0.003   &   15.309 $\,\pm\,$ 0.008  \\
H          & 00:59:53.0   &  -07:37:00.6    &  18.132 $\,\pm\,$ 0.008 &   17.405 $\,\pm\,$ 0.004  &  16.970 $\,\pm\,$ 0.004   &   16.625 $\,\pm\,$ 0.009  \\
\hline
\multicolumn{7}{c}{SN 2014cy}\\
\cline{4-5}
A           &  23:44:19  &  +10:49:31        & 18.928 $\,\pm\,$ 0.010 &   18.135 $\,\pm\,$ 0.006  &  17.709 $\,\pm\,$ 0.005   &   17.264 $\,\pm\,$ 0.008 \\
B           &   23:44:32  &   +10:48:53        & 16.447 $\,\pm\,$ 0.008 &   15.722 $\,\pm\,$ 0.005  &  15.334 $\,\pm\,$ 0.003   &    14.952 $\,\pm\,$ 0.007  \\  
C            & 23:44:19  &  +10:47:45     & 16.454 $\,\pm\,$ 0.008 &   15.893 $\,\pm\,$ 0.004  & 15.576 $\,\pm\,$ 0.004    &   15.232 $\,\pm\,$ 0.007  \\
D            &  23:44:21 &  +10:47:07     & 15.834 $\,\pm\,$ 0.008 &   15.225 $\,\pm\,$ 0.004  & 14.880 $\,\pm\,$ 0.004    &   14.506 $\,\pm\,$ 0.008 \\
E            & 23:44:26  &  +10:45:53     & 17.885 $\,\pm\,$ 0.008 &   17.194 $\,\pm\,$ 0.004  & 16.813  $\,\pm\,$ 0.004   &   16.433 $\,\pm\,$ 0.008  \\
F            & 23:44:20  &  +10:44:04      & 17.907 $\,\pm\,$ 0.009 &   16.777 $\,\pm\,$ 0.005  &  16.138 $\,\pm\,$ 0.004   &   15.553 $\,\pm\,$ 0.007  \\
G            & 23:44:12 &   +10:44:56      & 18.983 $\,\pm\,$ 0.010 &  18.118 $\,\pm\,$ 0.006  & 17.662 $\,\pm\,$ 0.005   &   17.193 $\,\pm\,$ 0.008  \\     
\hline
\multicolumn{7}{c}{SN 2015cz}\\
\cline{4-5}
A     & 01:31:58.7 &  +33:31:04.5 & 18.290 $\,\pm\,$ 0.011 &   17.772 $\,\pm\,$ 0.009  &  17.443 $\,\pm\,$ 0.010   &   17.090 $\,\pm\,$ 0.011 \\
B      & 01:31:57.4 &  +33:29:23.4 & 17.739 $\,\pm\,$ 0.007 &   17.141 $\,\pm\,$ 0.007  &  16.774 $\,\pm\,$ 0.007   &   16.386 $\,\pm\,$ 0.008  \\
C      & 01:31:58.8 &  +33:30:13.0 & 15.823 $\,\pm\,$ 0.004 &   15.203 $\,\pm\,$ 0.004  &  14.825 $\,\pm\,$ 0.004   &   14.466 $\,\pm\,$ 0.004  \\  
D    & 01:31:57.9 &  +33:27:42.2 & 18.558 $\,\pm\,$ 0.013 &   17.587 $\,\pm\,$ 0.007 & 17.030  $\,\pm\,$ 0.008   &   16.483 $\,\pm\,$ 0.009  \\
E      & 01:31:54.7 &  +33:27:13.9 & 18.380 $\,\pm\,$ 0.010 &   17.839 $\,\pm\,$ 0.009  &  17.171 $\,\pm\,$ 0.011   &   17.498 $\,\pm\,$ 0.009  \\
F    & 01:32:04.5 &  +33:27:39.7 & 19.302 $\,\pm\,$ 0.018 &  18.428 $\,\pm\,$ 0.010  & 17.459 $\,\pm\,$ 0.011    &   17.931 $\,\pm\,$ 0.012  \\
G      & 01:32:01.4 &  +33:26:36.2  & 18.165 $\,\pm\,$ 0.009 &   17.375 $\,\pm\,$ 0.007  & 16.445  $\,\pm\,$ 0.008   &   16.902 $\,\pm\,$ 0.007   \\
H      & 01:32:10.7 &  +33:26:30.6  & 17.260 $\,\pm\,$ 0.006 &   16.335 $\,\pm\,$ 0.006  &  15.345 $\,\pm\,$ 0.006   &   15.807 $\,\pm\,$ 0.005  \\
I    & 01:32:08.9 &  +33:26:32.6  & 18.843 $\,\pm\,$ 0.012 &  18.186 $\,\pm\,$ 0.009  & 17.402 $\,\pm\,$ 0.013    &   17.798 $\,\pm\,$ 0.010  \\
\hline
\end{tabular}
\end{table*}

\begin{table*}
\centering
\caption{Optical photometry of SNe\,2014cx, 2014cy and 2015cz. (Complete table available as supplementary material)\label{snmag}}
\begin{tabular}{c c c c c c c c }
\hline
\multicolumn{1}{c}{UT Date} & \multicolumn{1}{c}{JD} & \multicolumn{1}{l}{Phase$^a$} & \multicolumn{1}{c}{B} & \multicolumn{1}{c}{V} & \multicolumn{1}{c}{R} & \multicolumn{1}{c}{I}  & \multicolumn{1}{c}{Tel$^b$}\\
\multicolumn{1}{c}{(yyyy-mm-dd)} & \multicolumn{1}{l}{2456900+} & \multicolumn{1}{c}{(days)} & \multicolumn{1}{c}{(mag)} & \multicolumn{1}{c}{(mag)} & \multicolumn{1}{c}{(mag)} & \multicolumn{1}{c}{(mag)} & \multicolumn{1}{c}{}\\ 
\hline
\multicolumn{7}{c}{SN 2014cx}\\
\cline{4-5}
2014-10-04.7 & 35.5 & 33.1   & 15.50 $\pm$ 0.03 & 15.00 $\pm$ 0.02 & 14.62 $\pm$ 0.03 & 14.44 $\pm$ 0.02 & 1 \\
2014-10-07.9 & 38.4 & 36.0   & 15.63 $\pm$ 0.04 & 14.76 $\pm$ 0.03 & --               &  --             & 1 \\
2014-10-10.7 & 41.5 & 39.1   & 15.69 $\pm$ 0.02 & 14.99 $\pm$ 0.01 & 14.58 $\pm$ 0.02 & 14.39 $\pm$ 0.02 & 1 \\
2014-10-11.7 & 42.5 & 40.1   & 15.73 $\pm$ 0.02 & 14.97 $\pm$ 0.02 & 14.60 $\pm$ 0.03 & 14.35 $\pm$	0.03 & 1 \\
-- & -- & --   & -- & -- & -- & -- & -- \\
\hline
\end{tabular}
\end{table*}

Spectroscopic observations were acquired for SNe\,2014cx, 2014cy and 2015cz at 6, 7 and 14 epochs respectively and the log of spectroscopic observations is provided in Table\,\ref{sn_spectra_log}. The spectroscopic data were cleaned by debiasing the 2D spectral frames, flat-fielding and eliminating cosmic hits. One dimensional spectra extracted using the {\sc apall} task in IRAF were calibrated in wavelength using arc-lamp spectra. Spectrophotometric standard stars observed on the same night were used for flux-calibration and removal of telluric features using standard IRAF tasks. To account for the slit losses, the spectroscopic continuum was scaled to match the photometric continuum at coeval epochs. Finally the spectra were Doppler-corrected for the redshift of the host galaxy.

\begin{table}
\centering
\caption{Log of spectroscopic observations of SNe 2014cx, 2014cy and 2015cz.\label{sn_spectra_log}}
\begin{threeparttable}
\setlength{\tabcolsep}{3pt}
\footnotesize{
\begin{tabular}{c c c c c c}
\hline 
\multicolumn{1}{c}{UT Date} & \multicolumn{1}{c}{JD} & \multicolumn{1}{l}{Phase\tnote{a}} & \multicolumn{1}{l}{Instrument} & \multicolumn{1}{l}{Grism} & \multicolumn{1}{c}{Tel\tnote{b}}\\
\multicolumn{1}{c}{(yyyy-mm-dd)} & \multicolumn{1}{l}{2456900+} & \multicolumn{1}{l}{(days)} & \multicolumn{1}{c}{} & \multicolumn{1}{c}{} & \multicolumn{1}{c}{}\\
\hline
\multicolumn{6}{c}{\underline {SN 2014cx}}\\
2014-09-06.0 & 6.5  & 4.1  & ALFOSC & Gr4 & 6 \\
2014-10-22.9 & 53.4 & 51.0 & AFOSC  & Gr4,VPH6 & 2 \\
2014-10-26.9 & 57.4 & 55.0 & B\&C    & 300tr/mm & 1 \\
2014-10-28.9 & 59.4 & 57.0 & AFOSC  & Gr4,VPH6 & 2 \\
2014-12-18.8 & 110.3 & 107.9 & AFOSC  & Gr4,VPH6 & 2 \\
2015-01-12.8 & 135.3 & 132.9 & AFOSC  & Gr4,VPH6 & 2 \\
\hline
\multicolumn{6}{c}{\underline {SN 2014cy}}\\
2014-10-13.9 & 44.4 & 44.4 & ALFOSC & Gr4 & 6 \\
2014-10-26.7 & 57.2 & 57.2 & B\&C    & 300tr/mm & 1\\
2014-10-28.9 & 59.4 & 59.4 & AFOSC  & Gr4,VPH6 & 2 \\
2014-11-07.5  & 69.0 & 69.0 & HFOSC & Gr7,Gr8 & 3 \\
2014-11-09.5  & 71.0 & 71.0 & HFOSC & Gr7,Gr8 & 3 \\
2014-11-23.7  & 85.2 & 85.2 & HFOSC & Gr7,Gr8 & 3 \\
2014-12-18.8  & 110.2 & 110.2 & AFOSC  & Gr4 & 2 \\
\hline
\multicolumn{6}{c}{\underline {SN 2015cz}}\\
2015-10-03.7 & 299.2 & 7.6 & YFOSC & Gr3 & 5 \\
2015-10-06.7 & 302.2 & 10.6 & YFOSC & Gr3 & 5 \\
2015-10-08.9  & 304.4 & 12.8 & HFOSC & Gr7,Gr8 & 3 \\
2015-10-15.9  & 311.4 & 19.8 & HFOSC & Gr7,Gr8 & 3 \\
2015-10-19.8  & 315.3 & 23.7 & HFOSC & Gr7,Gr8 & 3 \\
2015-10-28.8  & 324.3 & 32.7 & HFOSC & Gr7,Gr8 & 3 \\
2015-11-04.0  & 330.1 & 38.5 & BFOSC & Gr4 & 4\\
2015-11-09.8  & 336.3 & 44.7 & HFOSC & Gr7,Gr8 & 3 \\
2015-11-27.6  & 354.1 & 62.5 & HFOSC & Gr7,Gr8 & 3 \\
2015-12-17.9  & 374.4 & 82.8 & AFOSC  & Gr4 & 2 \\
2015-12-21.6 & 377.0 & 85.4 & HFOSC & Gr7,Gr8 & 3 \\
2015-12-23.9 & 380.1 & 88.5 & HFOSC & Gr7,Gr8 & 3 \\
2016-01-11.7 & 399.2 & 107.6 & HFOSC & Gr7,Gr8 & 3 \\
2016-01-16.9 & 404.0 & 112.4 & HFOSC & Gr7,Gr8 & 3 \\
\hline
\end{tabular}
}
\begin{tablenotes}
\setlength\labelsep{0pt}
\scriptsize{\item[a]since the explosion epoch t$_0$\,=\,JD\,2456902.4 (SN\,2014cx), JD\,2456900.0 (SN\,2014cy), JD\,2457291.6 (SN\,2015cz)
\item[b](1) 1.22$\,\rm{m}$ Galileo Telescope, Pennar, Asiago Astrophysical Observatory, Italy, (2) 1.82$\,\rm{m}$ Copernico Telescope, Mt. Ekar, Italy, (3) 2.0$\,\rm{m}$ Himalayan Chandra Telescope, Indian Astronomical Observatory, India, (4) 2.16$\,\rm{m}$ telescope, Xinglong Observatory, China, (5) Lijiang 2.4$\,\rm{m}$ telescope (LJT) Yunnan Observatories, China, (6) 2.56$\,\rm{m}$ Nordic Optical Telescope, La Palma, Canaris, Spain.}
\end{tablenotes}
\end{threeparttable}
\end{table}

\section{Supernova and host galaxy parameters}
\label{sec3}
\subsection{Discovery, explosion epoch and distance}
SN\,2014cx was discovered in the galaxy NGC\,337 on 2014\,September\,2\,UT (JD\,2456902.9) independently by \cite{2014CBET.3963....1N} and \cite{2014ATel.6436....1H}. The classification spectra obtained at both optical and near-infrared wavelengths \citep{2014ATel.6440....1E,2014ATel.6442....1M} are those of a young SN II. The photometric observations by \cite{2015ATel.7084....1A} confirmed this event to be a Type IIP SN, with the $R$-band plateau lasting $\sim$86$\,\rm{d}$. Optical photometric and spectroscopic data of SN~2014cx have been previously reported by \cite{2016ApJ...832..139H} and \cite{2016MNRAS.459.3939V}. The explosion epoch is constrained to JD\,2456902.4$\,\pm\,$0.5, from the last non-detection of the event on 2014\,September\,1 by Katzman Automatic Imaging Telescope at a limiting magnitude of 19.1$\,\rm{mag}$. The distance to the SN is constrained to 18.5$\,\pm\,$1.1$\,\rm{Mpc}$ using the expanding photosphere method (see Section \ref{sec:6}), which is consistent (within errors) with that given by the Tully-Fisher method (19.6$\,\pm\,$1.8$\,\rm{Mpc}$, \citealt{2016AJ....152...50T}). We will refer to JD\,2456902.4$\,\pm\,$0.5 as the explosion epoch of SN\,2014cx and 18.5$\,\pm\,$1.1$\,\rm{Mpc}$ as the distance to SN\,2014cx in the rest of the paper.

SN\,2014cy was discovered in the galaxy NGC\,7742 on 2014\,August\,31.7\,UT (2456901.2\,JD; \citealt{2014CBET.3964....1N}). Classification spectra of the SN were obtained at optical wavelengths on 2014\,September\,2.46\,UT and at near infrared wavelengths on 2014\,September\,3\,UT \citep{2014ATel.6437....1V,2014ATel.6442....1M}, both matching the spectra of SNe II. The explosion epoch is constrained from the last non-detection (2014\,August\,29.3\,UT) and a prediscovery LOSS detection (2014\,August\,31.3\,UT) to August\,30.3 (2456900$\,\pm\,$1\,JD), which is hereafter adopted as the explosion epoch of SN\,2014cy. Photometric and spectroscopic data of SN~2014cy have been presented in \cite{2016MNRAS.459.3939V} and \cite{2019MNRAS.490.2799D}. Using the expanding photosphere method, we estimate the explosion epoch and the distance to the host galaxy of SN\,2014cy to be 2456898.7$\,\pm\,$1.0\,JD and 23.4$\,\pm\,$1.6$\,\rm{Mpc}$, respectively (see Section \ref{sec:6}). This corresponds to a distance modulus of 31.85$\,\pm\,$0.34$\,\rm{mag}$ which we adopt in this paper. This is in agreement with the distance modulus (31.87$\,\pm\,$0.15$\,\rm{mag}$) adopted in \cite{2016MNRAS.459.3939V}.

SN\,2015cz (PNV J01315945+3328458) was discovered in the galaxy NGC\,582 by Jiaming Liao, Guoyou Sun and Xing Gao on 2015\,October\,02.9\,UT (2457298.4\,JD)\footnote{www.cbat.eps.harvard.edu/unconf/followups/J01315945+3328458.html}. A classification spectrum of the event, obtained on 2015\,October\,3.7\,UT with the 2.4$\,\rm{m}$ telescope at LiJiang Gaomeigu Station of Yunnan Observatories, confirmed SN\,2015cz to be a Type II SN \citep{2015ATel.8120....1Z}. We used {\sc snid} \citep{2007ApJ...666.1024B} to estimate the explosion epoch of SN\,2015cz using the spectrum obtained on 2015\,October\,08. The best four matches assessed with quality of fit parameter, r$_{lap}$ \textgreater{} 6.5, were found with the spectra of SNe\,2003bn, 2004et, 2006bp and 2006iw at 12.6$\,\pm\,$1.0, 15.1$\,\pm\,$3.0, 11.0$\,\pm\,$1.0 and 12.0$\,\pm\,$3.0$\,\rm{d}$, respectively from explosion. We used the mean of these values to constrain the explosion epoch of SN\,2015cz to JD\,2457291.6$\,\pm\,$4.3, which will be referred to as 0\,d of SN\,2015cz in this paper. The distance to NGC\,582 is 63.7$\,\pm\,$5.7$\,\rm{Mpc}$ \citep{2016AJ....152...50T} which corresponds to a distance modulus of 34.0$\,\pm\,$0.4$\,\rm{mag}$ and will be used in this paper. 

\subsection{Total line-of-sight extinction}

\begin{figure}
\includegraphics[scale=0.59, clip, trim={1.4cm 0.1cm 0.6cm 1.2cm}]{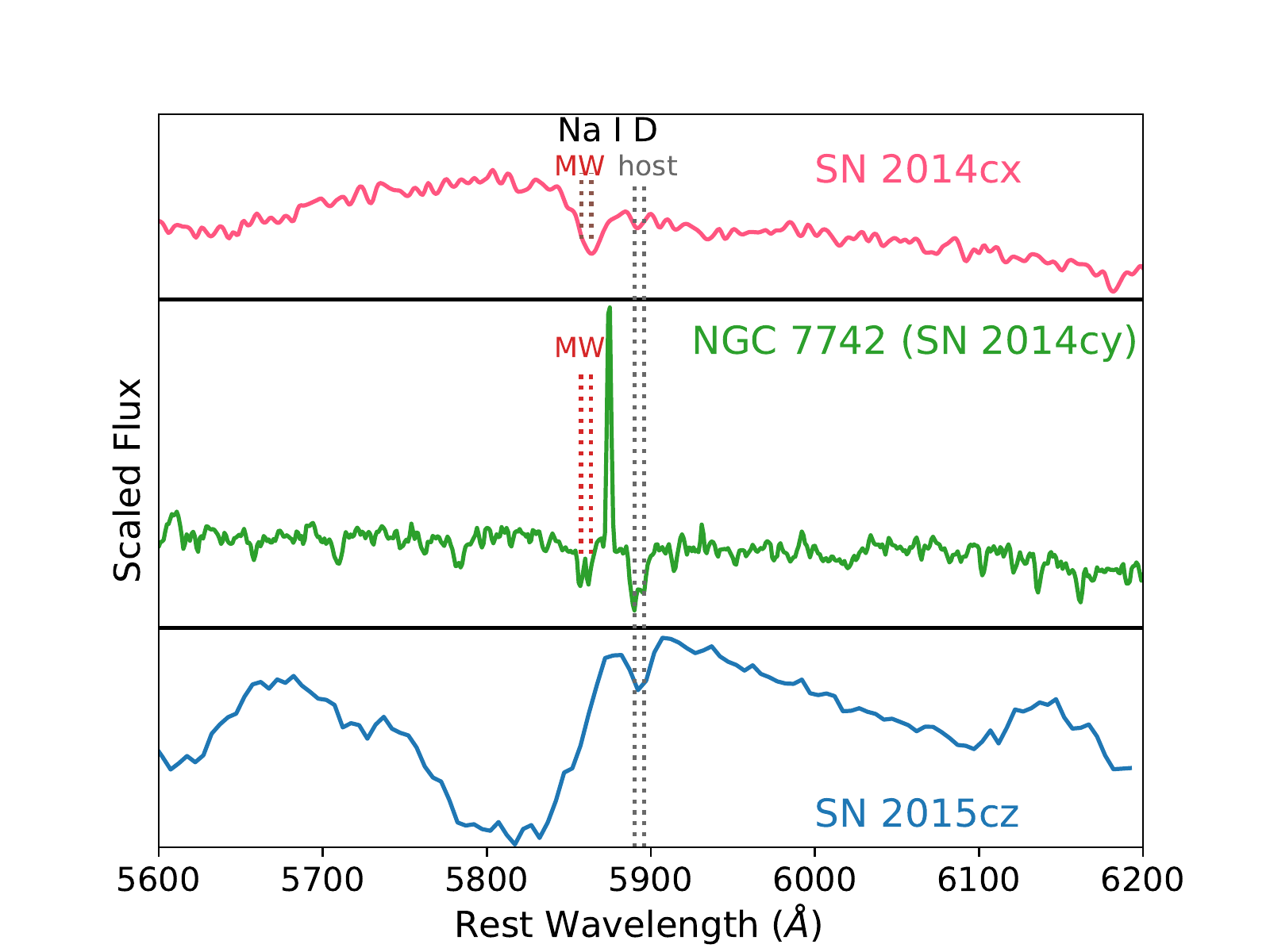}
\caption{The 4.1\,d spectrum of SN\,2014cx, the spectrum of H\,II region in NGC\,7742 nearest to SN\,2014cy \citep{2016MNRAS.455.4087G} and the composite spectrum of SN\,2015cz. Both Galactic and host galaxy Na\,{\sc i}\,D narrow absorption feature are marked with dashed lines.}
\label{ch6:spectra_NaID}
\end{figure}

The Milky Way reddening towards SNe\,2014cx, 2014cy and 2015cz are E($B$$-$$V$)$_{MW,14cx}$\,=\,0.096$\,\pm\,$0.002$\,\rm{mag}$, E($B$$-$$V$)$_{MW,14cy}$\,=\,0.048$\,\pm\,$0.002$\,\rm{mag}$ and E($B$$-$$V$)$_{MW,15cz}$ = 0.046$\,\pm\,$0.001$\,\rm{mag}$ \citep{2011ApJ...737..103S}. The Na\,{\sc i}\,D feature from the Milky Way is conspicuous in the early spectra of SN\,2014cx (Figure\,\ref{ch6:spectra_NaID}). The equivalent width (EW) of this feature is 0.738\,\AA, which corresponds to Galactic reddening $E(B-V)_{MW}$\,=\,0.10$\,\pm\,$0.02$\,\rm{mag}$ following \cite{2012MNRAS.426.1465P}. Narrow Na\,{\sc i}\,D absorption feature at the host galaxy redshift is insignificant in the spectra of SN\,2014cx. From the peripheral location of the SN in the host galaxy, it is unlikely that the object is significantly extinguished by additional dust reddening from the host.

 In the spectra of both SNe\,2014cy and 2015cz, narrow Na\,{\sc i}\,D absorption lines arising from their respective host galaxies can be discerned. We used the spectrum of the H\,II region in NGC\,7742 nearest to SN\,2014cy location from \cite{2016MNRAS.455.4087G} to determine the dust reddening using the Na\,{\sc i}\,D diagnostic and Balmer decrement \citep{2004A&A...420..475S}. The Galactic as well as the host galaxy component of Na\,{\sc i}\,D is visible in this spectrum (Figure\,\ref{ch6:spectra_NaID}). The equivalent width of the Na\,{\sc i}\,D line from our Galaxy is 0.47\,\AA{} and that from host galaxy is 1.14\,\AA. Using equation 9 of \cite{2012MNRAS.426.1465P}, we estimated the Galactic ($E(B-V)_{MW}$\,=\,0.05$\,\pm\,$0.01$\,\rm{mag}$) and host galaxy ($E(B-V)_{host}$\,=\,0.31$\,\pm\,$0.06$\,\rm{mag}$) extinction components. A colour excess $E(B-V)_{tot}$\,=\,0.36$^{+0.09}_{-0.08}\,\rm{mag}$ is estimated through the Balmer decrement, assuming an intrinsic Case B recombination value of (H$\alpha$/H$\beta$)$_{int}$\,=\,2.86$^{+0.18}_{-0.11}$ \citep{2006agna.book.....O} and Fitzpatrick's extinction curve \citep{1999PASP..111...63F}. This is in excellent agreement with that obtained with Na\,{\sc i}\,D diagnostic. The weighted mean of the above two methods yield a total colour excess $E(B-V)_{tot}$\,=\,0.36$\,\pm\,$0.05$\,\rm{mag}$.
 
We used the Na\,{\sc i}\,D diagnostic to estimate the line-of-sight reddening of SN\,2015cz contributed by the host galaxy dust. A composite spectrum (as shown in Figure\,\ref{ch6:spectra_NaID}) is constructed by stacking a set of best signal-to-noise ratio spectra, each being resampled to the same binset and weighted by the signal-to-noise ratio before combining. The EW of the Na\,{\sc i}\,D absorption line in the composite spectrum is 1.6\,\AA. Using the empirical relations from \cite{1990AA...237...79B,2003fthp.conf..200T,2011MNRAS.415L..81P,2012MNRAS.426.1465P}, the estimated mean value of colour excess $E(B-V)_{host}$\,=\,0.42$\,\pm\,$0.18$\,\rm{mag}$ for SN\,2015cz, where the errors correspond to the standard deviation of the values obtained from the empirical relations. The extinction can also be estimated from the scatter in the colour during the recombination phase. We estimated the host galaxy colour excess contribution $E(B-V)_{host}$\,=\,0.44$\,\pm\,$0.07$\,\rm{mag}$ for SN\,2015cz from the Galactic reddening corrected ($V-I$) colour at 89.5 and 95.5$\,\rm{d}$ using the \textquoteleft colour method\textquoteright\, \citep{2010ApJ...715..833O}. The weighted mean value of colour excess from the above two methods is $E(B-V)_{host}$\,=\,0.44$\,\pm\,$0.06$\,\rm{mag}$ and the total colour excess is $E(B-V)_{tot}$\,=\,0.48$\,\pm\,$0.06$\,\rm{mag}$.

\begin{figure}
\begin{subfigure}{0.55\textwidth}
   \includegraphics[scale=0.45, clip, trim={0.5cm 8.8cm 1.0cm 1.5cm}]{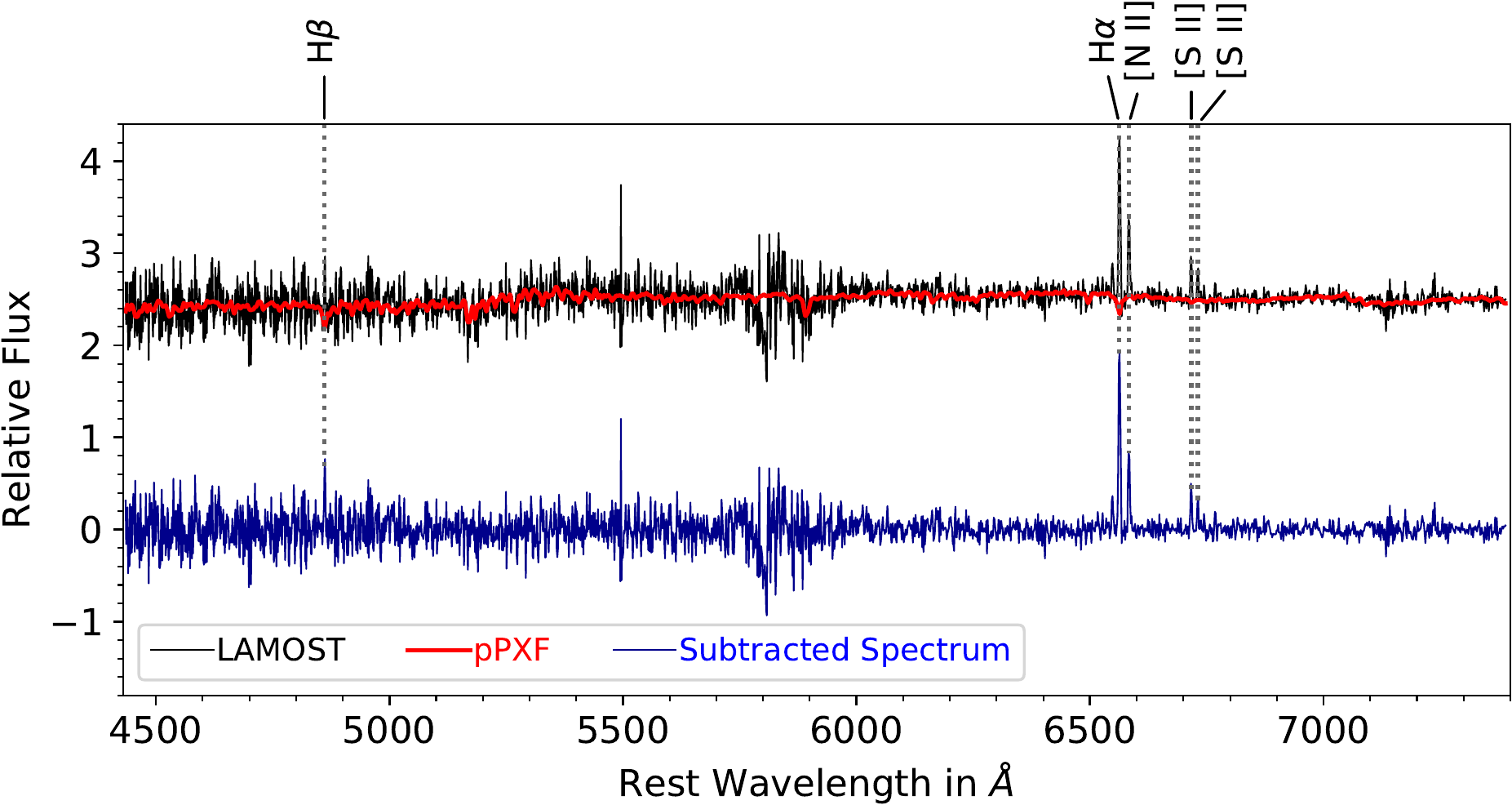}
   \caption{}
   \label{fig:Ng1} 
\end{subfigure}

\begin{subfigure}{0.55\textwidth}
   \includegraphics[scale=0.55, clip, trim={0.4cm 0.0cm 1.4cm 1.4cm}]{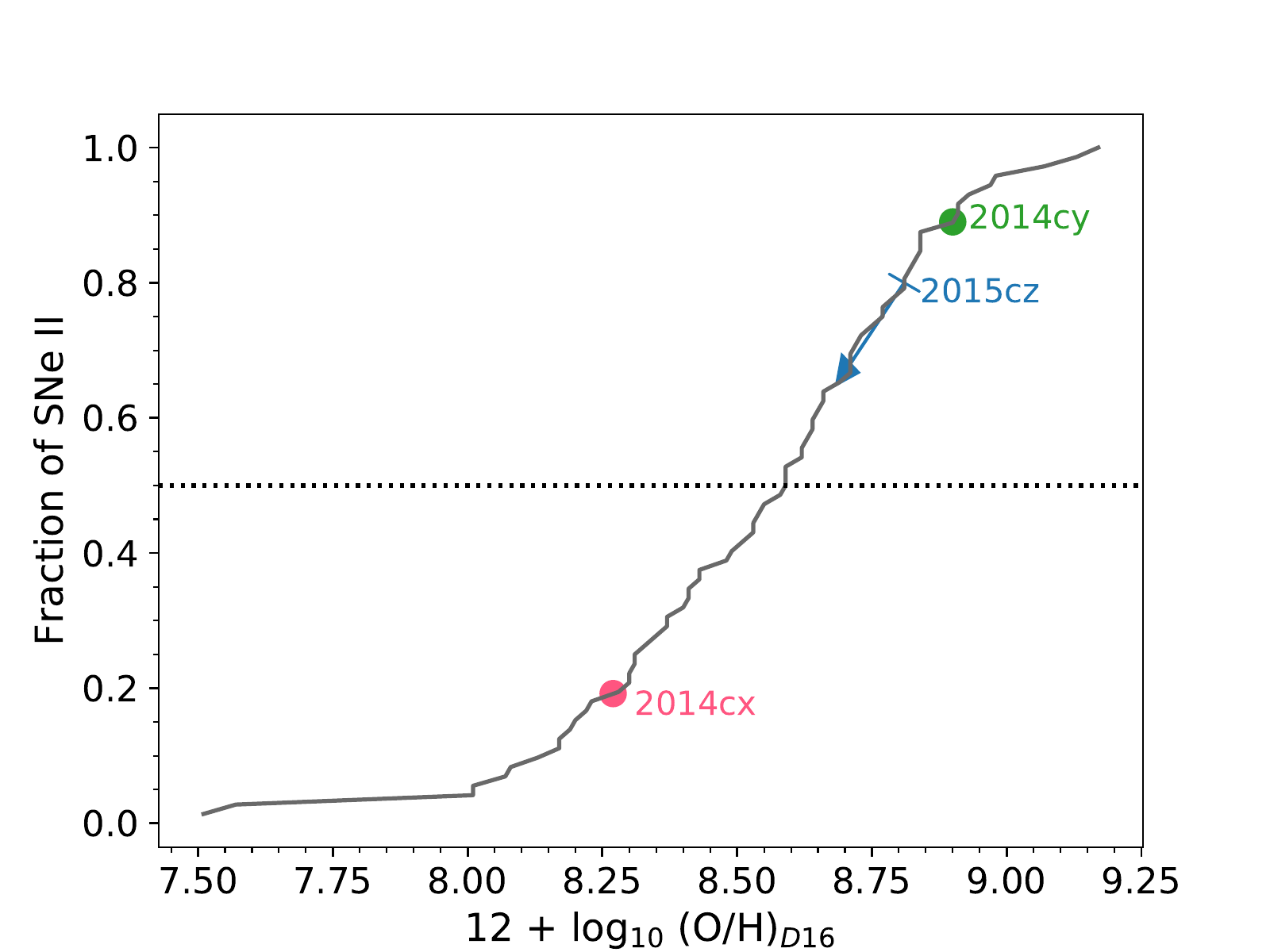}
   \caption{}
   \label{fig:Ng2}
\end{subfigure}

\caption[]{(a) The spectrum of NGC\,582 (in black) along with the pPXF fit (in red). The pure nebular emission line spectrum is shown in blue. (b) The oxygen abundance of the host galaxy of the SNe\,2014cx, 2014cy and 2015cz overplotted on the normalized cumulative distribution of the oxygen abundance of a sample of 72 SNe II from \citet{2018ApJ...855..107G}. The metallicity for SN 2014cx is an average metallicity at the radius of the SN, that of SN 2014cy is a local metallicity estimate and that of SN 2015cz is for the host nucleus that is shown as an arrow to indicate an upper limit. The median value of the distribution is shown with the horizontal dashed line at 0.5 fraction.}
\end{figure}

\begin{figure*}
\includegraphics[scale=0.46, clip, trim={4.0cm 0.5cm 4.5cm 1.8cm}]{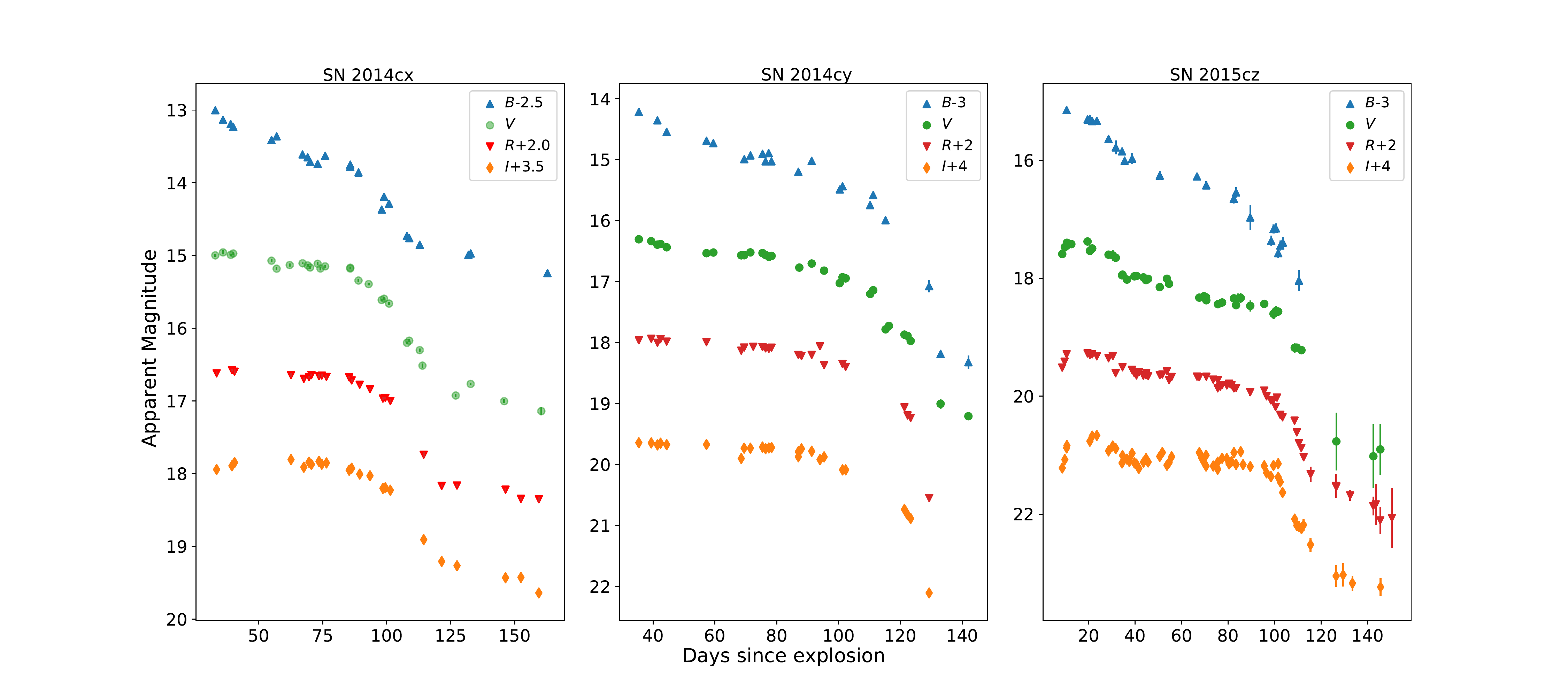}
\caption{$BVRI$ light curves of SNe\,2014cx, 2014cy and 2015cz offset by values as shown in the legend.}
\label{ch6:LC}
\end{figure*}

 Thus, the total colour excess due to Galactic plus host galaxy extinction for SN\,2014cx is $E(B-V)_{tot}$\,=\,0.096$\,\pm\,$0.002$\,\rm{mag}$, for SN\,2014cy is $E(B-V)_{tot}$\,=\,0.36$\,\pm\,$0.05$\,\rm{mag}$ and for SN\,2015cz is $E(B-V)_{tot}$\,=\,0.48$\,\pm\,$0.06$\,\rm{mag}$, which will be adopted hereafter in the paper.
 
 \subsection{Host galaxy metallicity}

The gas-phase metallicity of the host galaxy can be estimated from the strong emission line fluxes, such as H$\alpha$, [N\,{\sc ii}]\,$\lambda$6583 and [S\,{\sc ii}]\,$\lambda$6717,6731 doublet. To estimate the true fluxes, pure nebular emission line spectrum of host galaxy free from any contribution from the different single stellar populations is required. The stellar continuum normalized host galaxy spectra of SNe\,2014cx and 2014cy are available in the literature \citep{2010ApJS..190..233M, 2016MNRAS.455.4087G}. The spectrum of the host galaxy nucleus of SN\,2015cz (NGC\,582) was published in the second data release of LAMOST (Large Sky Area Multi-Object Fiber Spectroscopic Telescope, also known as Guo Shou Jing
telescope). We used Penalized Pixel-Fitting method (pPXF, \citealt{2017MNRAS.466..798C}) and the MILES spectral library \citep{2006MNRAS.371..703S} for stellar continuum fitting in NGC\,582. The pPXF fits (as shown in Figure\,\ref{fig:Ng1}) were subtracted from the observed spectrum to accurately measure the emission line fluxes. The \cite{2016Ap&SS.361...61D} calibrator based on photoionization models was used to determine the oxygen abundance of the host galaxies from the dereddened emission line fluxes and compared to a sample of 72 SNe II hosts in the PMAS/PPAK Integral-field Supernova hosts
COmpilation (PISCO) \citep{2018ApJ...855..107G} as shown in Figure\,\ref{fig:Ng2}.
The average oxygen abundance of SNe\,II in the PISCO sample is 8.54$\,\pm\,$0.04$\,\rm{dex}$. To estimate the local oxygen abundance of SN\,2014cx, we have used the dereddened emission line fluxes given in \cite{2010ApJS..190..233M} which is computed from the integrated optical spectrum of a radial strip covering the SN site of NGC\,337. The dereddened fluxes of the emission lines from NGC\,7742 are measured from the spectrum of H\,II region in NGC\,7742 closest to SN site from \cite{2016MNRAS.455.4087G} and that from NGC\,582 using the spectrum of the host galaxy centre. The oxygen abundance of NGC\,337 (SN\,2014cx) is 8.27$\,\pm\,$0.02$\,\rm{dex}$, NGC\,7742 (SN\,2014cy) is 8.90$\,\pm\,$0.02$\,\rm{dex}$ and NGC\,582 (SN\,2015cz) is 8.81$\,\pm\,$0.02$\,\rm{dex}$. The host oxygen abundance of SN\,2014cx is 0.22$\,\rm{dex}$ lower while that of SNe\,2014cy and 2015cz are 0.36$\,\rm{dex}$ and 0.27$\,\rm{dex}$ higher than the average oxygen abundance of the PISCO SN\,II sample. The host oxygen abundance of SN\,2015cz can only constrain the upper bound of the local environment metallicity of the SN considering the metallicity gradient in spiral galaxies \citep{2000MNRAS.312..497B, 2005PASP..117..227K}. As compared to the solar metallicity (8.69$\,\pm\,$0.05\,dex; \citealt{2009ARA&A..47..481A}), SN\,2014cx exploded in sub-solar and SNe\,2014cy and 2015cx exploded in marginally super-solar metallicity environment. 

\section{Light Curve Analysis} \label{sec:4}
\begin{figure}
\begin{center}
\includegraphics[scale=0.55, clip, trim={0.4cm 0.2cm 1.6cm 1.2cm}]{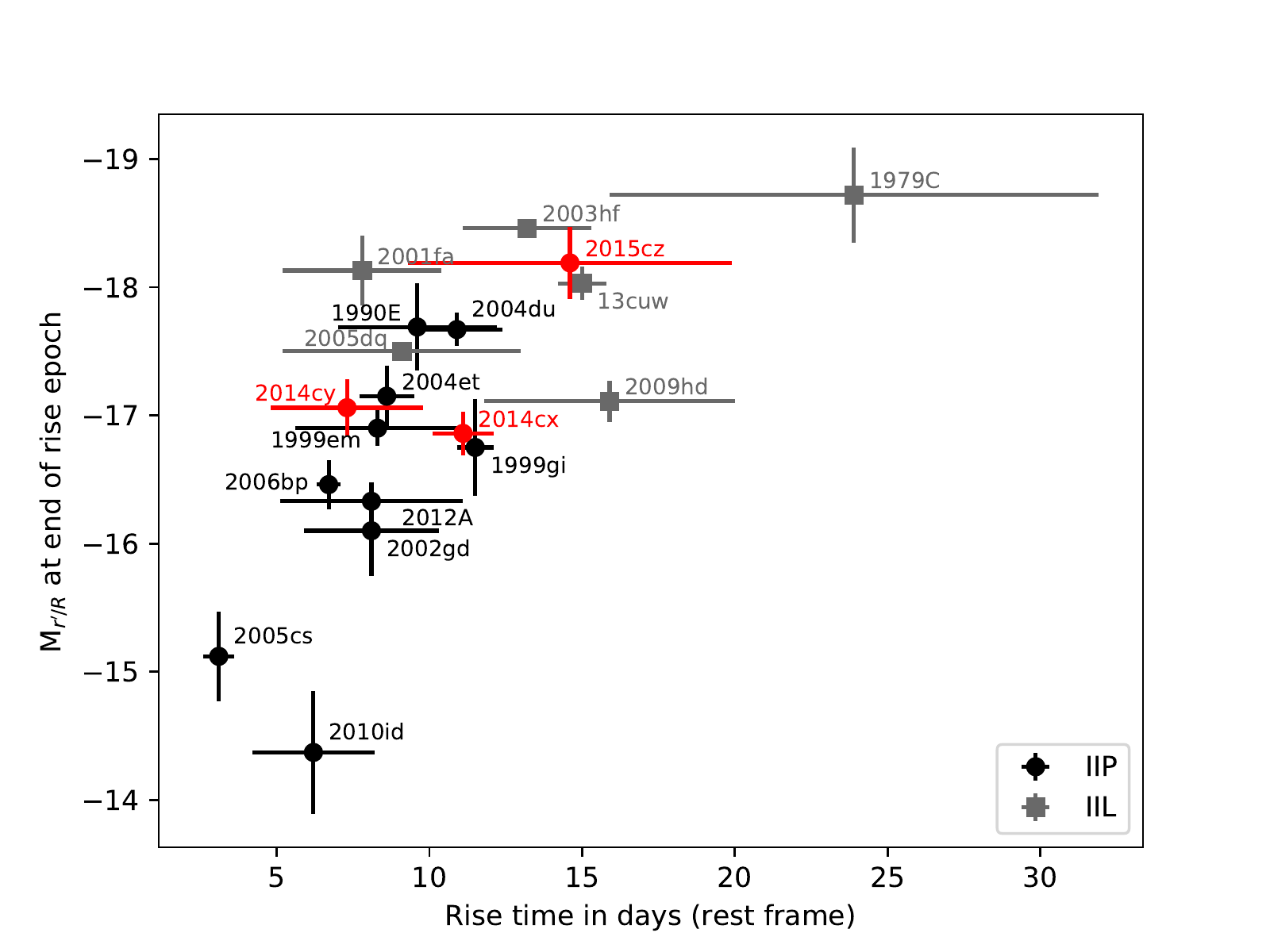}
\end{center}
\caption{The absolute magnitude at end-of-rise epoch versus the rise time in $r^\prime/R$ band of SNe~2014cx, 2014cy and 2015cz is compared to the sample of SNe from \citet{2015A&A...582A...3G}}
\label{ch6:rise}
\end{figure}

The light curves of SNe 2014cx, 2014cy and 2015cz are shown in Figure~\ref{ch6:LC}. We estimated the end-of-rise epoch in $r^\prime$/$R$ band, where the light curve rises by less than 0.01$\,\rm{mag/d}$, by fitting a low-order polynomial to the data in the same way as done in \cite{2015A&A...582A...3G}. We used the early time $r^\prime$-band data of SNe\,2014cx and 2014cy from \cite{2016MNRAS.459.3939V} and $R$-band data of SN\,2015cz from our observations. The rise time and the absolute magnitude at peak, thus obtained, for the three SNe are: t$_{r^\prime}$\,=\,11.1$\,\pm\,$1.0$\,\rm{d}$, M$_{r^\prime}$\,=\,$-$16.86$\,\pm\,$0.17$\,\rm{mag}$ for SN\,2014cx; t$_{r^\prime}$\,=\,7.3$\,\pm\,$2.5$\,\rm{d}$, M$_{r^\prime}$\,=\,$-$17.06$\,\pm\,$0.22$\,\rm{mag}$ for SN\,2014cy; t$_R$\,=\,14.6$\,\pm\,$5.3$\,\rm{d}$, M$_{R}$\,=\,$-$18.19$\,\pm\,$0.28$\,\rm{mag}$ for SN\,2015cz. In Figure~\ref{ch6:rise}, we compare the rise times and the absolute magnitudes of SNe\,2014cx, 2014cy and 2015cz with a sample of SNe IIP and IIL from \cite{2015A&A...582A...3G}. SN\,2014cx has similar rise time and peak magnitude as SN\,1999gi (t$_R$\,=\,11.5$\,\pm\,$0.6$\,\rm{d}$, M$_{R}$\,=\,$-$16.75$\,\pm\,$0.38$\,\rm{mag}$), SN\,2014cy lies between SNe\,2004et and 1999em, while SN\,2015cz shows a longer rise time and a brighter peak magnitude similar to the IIL SN LSQ13cuw (t$_{r^\prime}$\,=\,15.0$\,\pm\,$0.8$\,\rm{d}$, M$_{r^\prime}$\,=\,$-$18.03$\,\pm\,$0.13$\,\rm{mag}$; \citealt{2015A&A...582A...3G}). The rise time gives an idea about the initial radius for a spherically symmetric explosion powered by a central source, while the ejecta mass and the explosion energy are expected to only marginally affect the rise time \citep{2015A&A...582A...3G, 2015MNRAS.451.2212G}. SNe 2014cx, 2014cy and 2015cz have comparable rise times (within error) and hence more likely to have similar pre-supernova radii.

\begin{figure*}
\centering
      \includegraphics[scale=0.5]{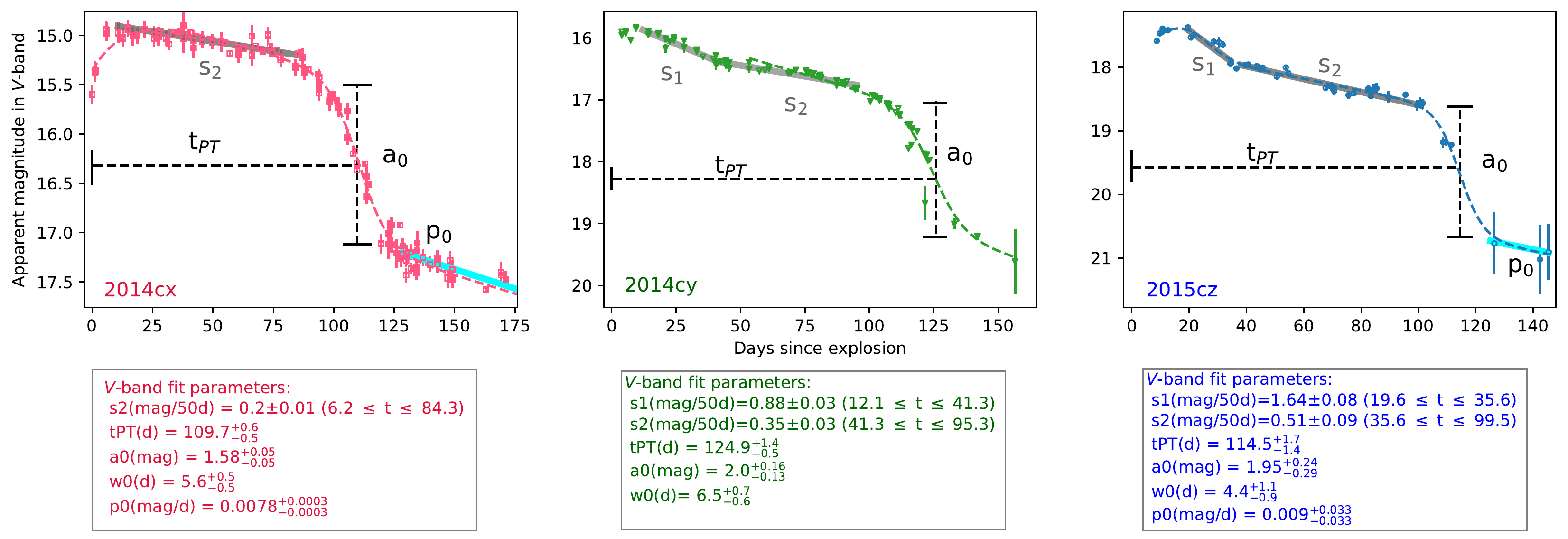}
  \caption{Fits to the $V$-band light curves of SNe 2014cx, 2014cy and 2015cz with best-fit parameters being tabulated below.}
  \label{ch6:fit_val}
\end{figure*}

\begin{figure}
    \subfloat{{\includegraphics[width=4.45cm, height=3.8cm, clip, trim={0.0cm 0.5cm 0cm 0cm}]{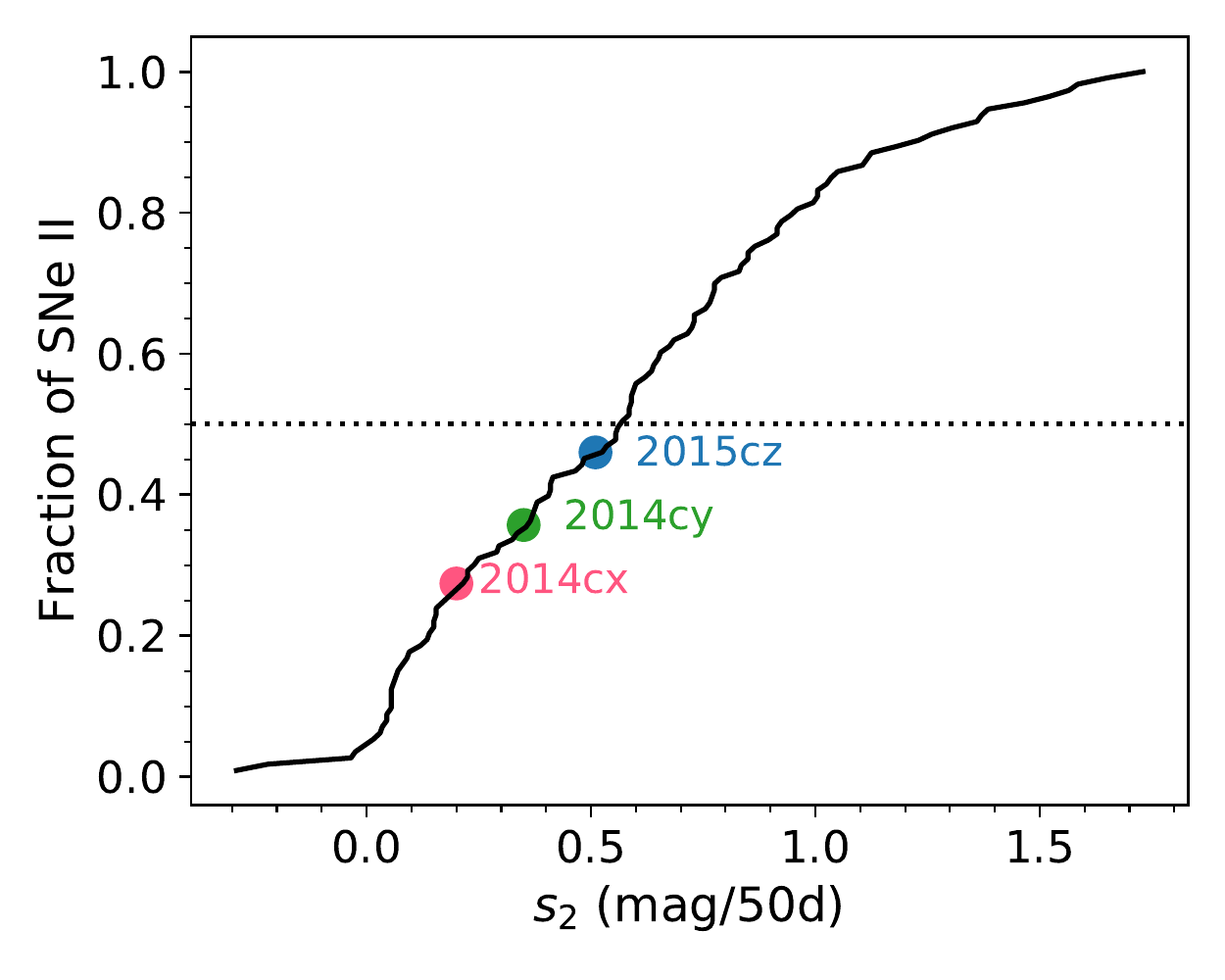} }}%
    \subfloat{{\includegraphics[width=4.05cm, height=3.8cm, clip, trim={0.4cm 0.5cm 0cm 0cm}]{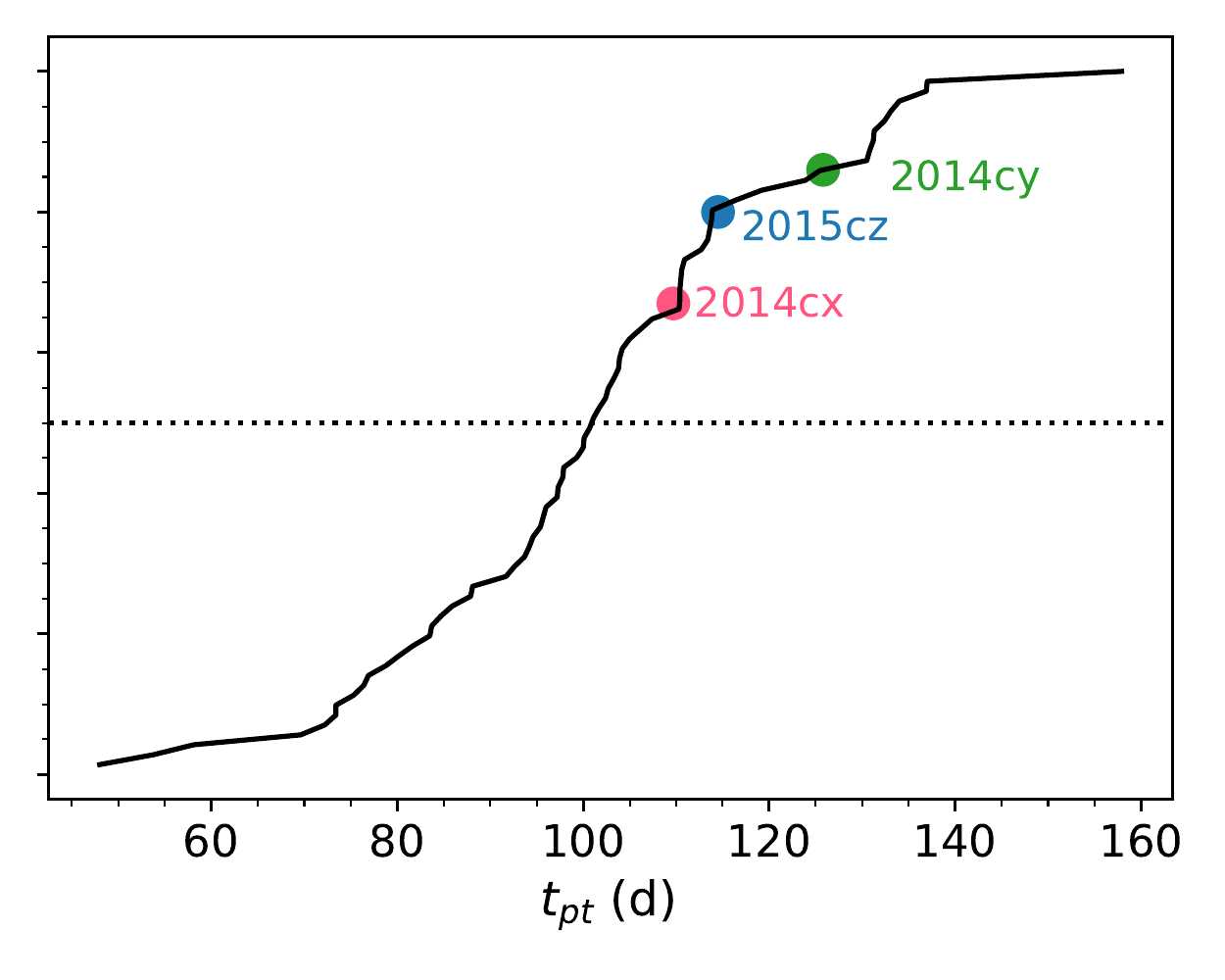} }}%
    \caption{The slope $s_2$ and $t_{PT}$ of SNe\,2014cx, 2014cy and 2015cz overplotted on the normalized cumulative distribution of a sample of SNe\,II from \citet{2014ApJ...786...67A}. The dashed line at 0.5 fraction shows the median value of these quantities in the distribution.}%
    \label{fig:example}
\end{figure}

To estimate the light curve parameters of SNe\,2014cx and 2014cy, we have included the data of \cite{2016MNRAS.459.3939V} to our $V$-band dataset. The $V$-band data of SNe\,2014cx and 2014cy from \cite{2016MNRAS.459.3939V} are linearly interpolated to our observation dates and an average shift (difference between the interpolated data and our data) of \textless$\delta$V\textgreater = $-$0.10\,$\,\pm\,$\,0.19 in SN\,2014cx and 0.02\,$\,\pm\,$\,0.17 in SN\,2014cy are estimated. To account for the systematic differences in $V$-band magnitudes between the two datasets, likely arising from different calibrations, the average shift was applied to the literature dataset before merging them.

While the post-maximum $V$-band light curve of SN\,2014cx exhibits a continuous, slow decline (plateau) at a rate 0.20$\,\pm\,$0.01$\,\rm{mag/50d}$ lasting for $\sim$84$\,\rm{d}$, the $V$-band light curve of SNe\,2014cy and 2015cz shows an early declining part ($s_1$) prior to the plateau phase ($s_2$). The two slopes are simultaneously fitted with a smoothed piecewise linear function:
\begin{equation}
\label{equ2}
m_v = a + a_1t + a_2\Bigg(\frac{t-T_{tran}}{2} + \sqrt{\frac{(t-T_{tran})^2}{4} + b}\Bigg)
\end{equation}
 The transition time from the early declining phase to the plateau phase is indicated by the break time ($T_{tran}$), $a_1$ is the slope of the early declining phase, ($a_1$+$a_2$) is the slope of the plateau phase and $b$ is the smoothing factor. The post maximum cooling phase ($s_1$) declines at a rate of 0.88$\,\pm\,$0.03$\,\rm{mag/50d}$ in SN\,2014cy up to 41.3$\,\pm\,$1.0$\,\rm{d}$ and at a rate of 1.64$\,\pm\,$0.08$\,\rm{mag/50d}$ in SN\,2015cz up to 35.6$\,\pm\,$4.3$\,\rm{d}$. The two slopes ($s_1$ and $s_2$) may be attributed to density discontinuity in the SN ejecta. Other relevant light curve parameters are estimated by fitting the $V$-band light curves of the three SNe to equation 4 of \cite{2010ApJ...715..833O}. The fit parameters are shown in Figure \ref{ch6:fit_val}, where $a_0$ indicates the magnitude drop from the recombination phase to the linear tail, $w_0$ indicates the width of this transition phase ($\sim$6w$_0$), $t_{PT}$ is a proxy for plateau length and $p_0$ is the slope of the radioactive tail. In SN\,2014cx, the magnitude drops by 1.58$\,\pm\,$0.05$\,\rm{mag}$ in $\sim$34$\,\rm{days}$ from the plateau to the tail phase. The light curves of SNe\,2014cy and 2015cz show larger declines (2.00$\,\rm{mag}$ in 39$\,\rm{d}$ and 1.95$\,\rm{mag}$ in 26$\,\rm{d}$, respectively), although these values are uncertain due to the modest number of data points in this phase. During the plateau phase, the SN is powered by the recombination energy of ionised hydrogen and in the tail phase, the dominant source of energy is the radioactive decay energy. Thus, it is expected that the later the fall from the plateau, the larger will be a$_0$, since the radioactive decay energy is decaying with time. Here, we find that SN 2014cx which exhibit a shorter plateau indeed shows a lower magnitude drop in the transition phase than SNe 2014cy and 2015cz. However, no such correlation between a$_0$ and t$_{PT}$ was found by \citep{2016MNRAS.459.3939V}, where they found that a$_0$ is rather constrained between 1 - 2.6$\,\rm{mag}$ for the normal luminosity events. The a$_0$ values of the three SNe 2014cx, 2014cy and 2015cz are well within this range.
 
 We plotted the normalized cumulative distribution of two parameters ($s_2$ and $t_{PT}$) in a sample of SNe\,II from \cite{2014ApJ...786...67A} and overplotted the respective estimated values of SNe\,2014cx, 2014cy and 2015cz. All the three SNe have plateau slopes ($s_2$) lower than the median value of the distribution and $t_{PT}$ higher than the median value of the distribution. SN\,2014cy has the longest plateau, followed by SNe\,2015cz and 2014cx, while SN\,2015cz has the largest plateau slope followed by SNe\,2014cy and 2014cx. Overall, this indicates that the three SNe conforms to the characteristic of the population, i.e., slowly declining SNe II tends to have longer plateau phase \textgreater 100\,d \citep{2014ApJ...786...67A}.

As SNe\,2014cy and 2015cz have large decline rates in their early light curves (\textgreater 0.5$\,\rm{mag/50d}$), according to the criteria of \cite{2014MNRAS.445..554F}, they would be classified as SNe\,IIL. However, \cite{2014ApJ...786...67A} supported the idea of a physical continuity between the Type IIP and IIL sub-class and endorsed identifying SNe\,II based on their plateau slope. They estimated an average decline rate of s$_2$\,=\,0.64$\,\pm\,$0.47$\,\rm{mag/50d}$ in the plateau phase for a sample of SNe II. This suggests that SNe\,2014cx (s$_2$\,=\,0.20$\,\pm\,$0.01$\,\rm{mag/50d}$), 2014cy (s$_2$\,=\,0.35$\,\pm\,$0.03$\,\rm{mag/50d}$) and 2015cz (s$_2$\,=\,0.51$\,\pm\,$0.09$\,\rm{mag/50d}$) are consistent with the slowly declining SN II. After the drop, the linear declining phase in the light curve is powered by the radioactive decay of $^{56}$Co $\rightarrow$ $^{56}$Fe with an expected decay rate of 0.98$\,\rm{mag/100d}$ in case of full trapping of $\gamma$-rays. The $V$-band decline rates in the radioactive tail phase of SNe\,2014cx and 2015cz are 0.78 and 0.91$\,\rm{mag/100d}$. The decline rate of SN 2014cx is marginally slower than the $^{56}$Co $\rightarrow$ $^{56}$Fe decay rate. For SN 2015cz, the large errors on the tail data introduces a large error of 3.3$\,\rm{mag/100d}$ on the quoted value of the slope. So, we refrain from making any comments on the tail decline rate of SN 2015cz.

\begin{figure*}
             \centerline{\includegraphics[scale=0.6, height=0.3\textheight,width=0.5\textwidth,clip, trim={0.4cm -0.0cm 0.3cm 0.35cm}]{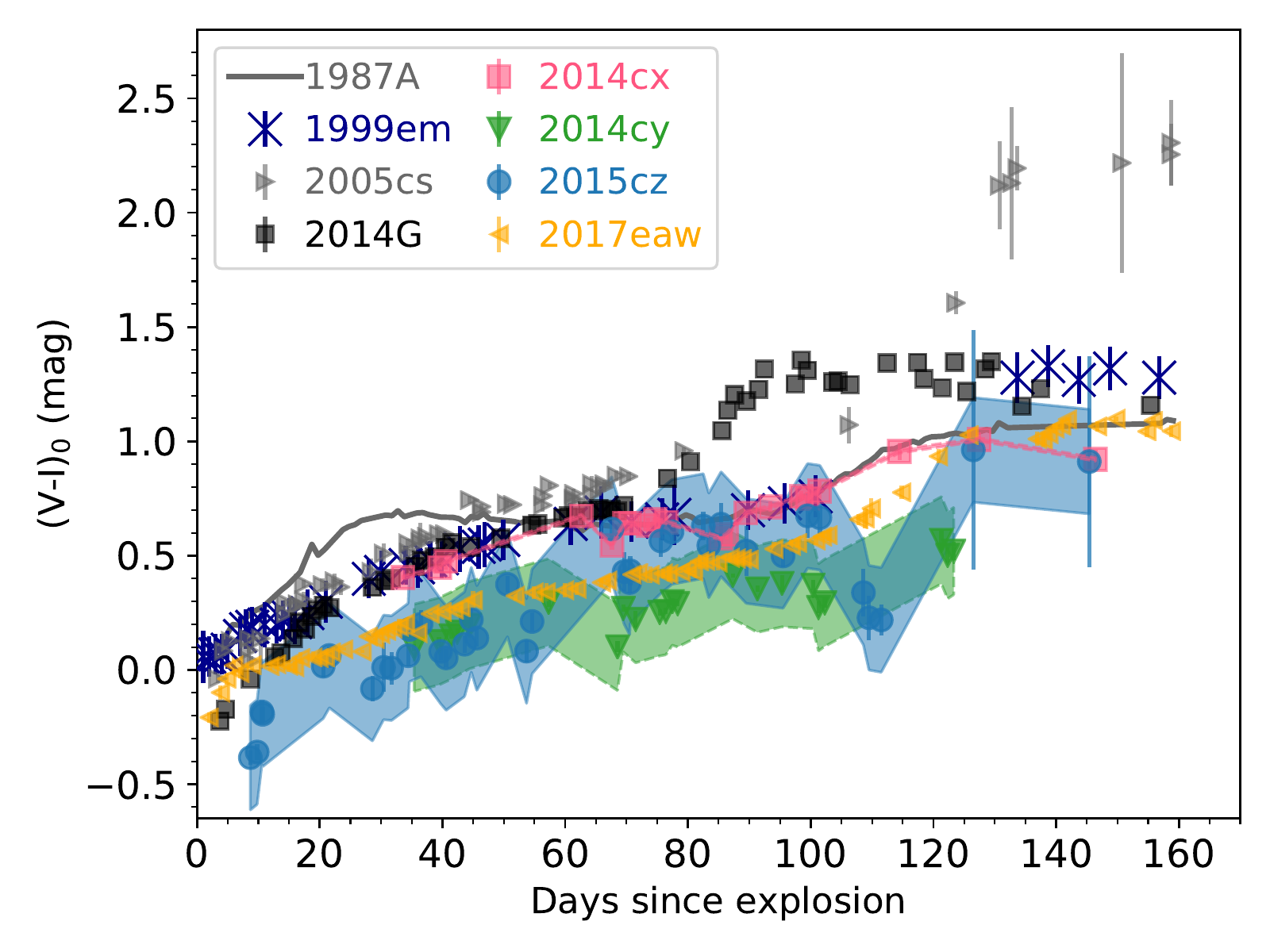}
\hspace{0.5pt}
               \includegraphics[scale=0.5, width=0.5\textwidth, height=0.3\textheight,clip, trim={0.4cm 0.0cm 0.0cm 0.35cm}]{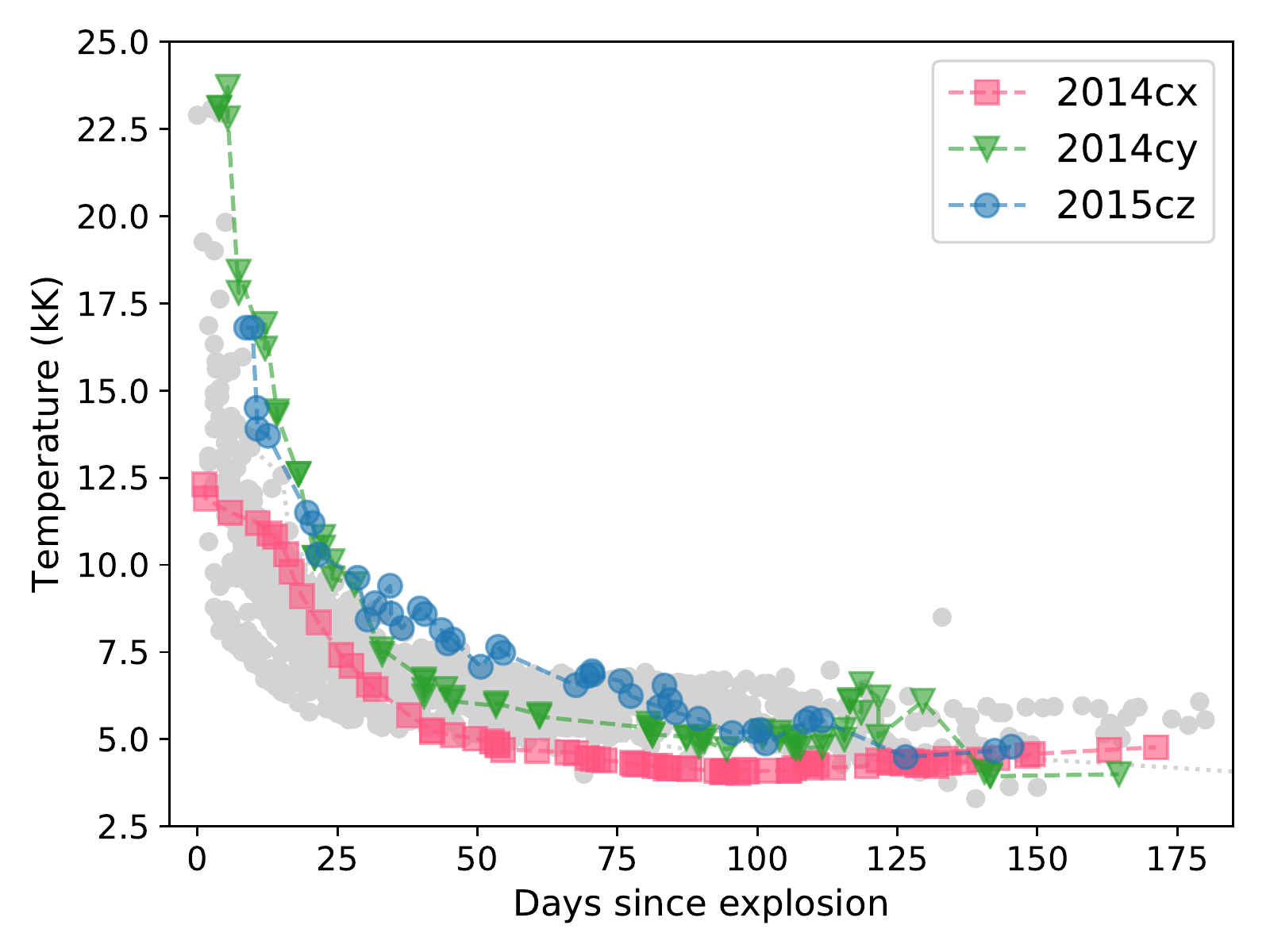}}
  \caption{{\bf Left:} The ($V-I$)$_0$ colour evolution of SNe~2014cx, 2014cy and 2015cz is compared with a sample of SNe II. All the ($V-I$) colours are corrected for Galactic plus host galaxy reddening. Here the reddening uncertainty for SNe 2014cx, 2014cy and 2015cz are shown with the shaded region and the error bars on the individual points are the random photometric error. Due to small reddening error in SN 2014cx, the shaded region appears as a line. {\bf Right:} Comparison of the temperature evolution of SNe~2014cx, 2014cy and 2015cz obtained by fitting blackbody to the SED constructed from the observed photometric fluxes with a sample of SNe II from \citet{2018MNRAS.473..513F}.}
  \label{ch6:colour_mag}
\end{figure*}

\subsection{Comparison with other SNe II}
For comparing the photometric properties of SNe\,2014cx, 2014cy and 2015cz, we assembled a sample of six SNe that captures the overall light curve morphology and the luminosity range of SNe II family, that is, 1987A: a Type II-pec event (d\,=\,0.05$\,\rm{Mpc}$, $E(B-V)$\,=\,0.19$\,\rm{mag}$; \citealt{1990AJ.....99.1146H}); 1999em: prototypical Type IIP SN (d\,=\,11.7$\,\pm\,$0.1$\,\rm{Mpc}$, $E(B-V)$\,=\,0.10$\,\rm{mag}$; \citealt{2002PASP..114...35L}); 2005cs: a low luminosity, low velocity Type IIP SN (d\,=\,7.1$\,\pm\,$1.2$\,\rm{Mpc}$, E($B-V$)=0.11$\,\rm{mag}$; \citealt{2006MNRAS.370.1752P,2009MNRAS.394.2266P}), the fast declining Type IIL SN 2014G (d\,=\,24.5$\,\pm\,$3.9$\,\rm{Mpc}$, E($B-V$)\,=\,0.21$\,\rm{mag}$; \citealt{2016MNRAS.455.2712B,2016MNRAS.462..137T}); 2016ija: one of the brightest SN II (d\,=\,20.3$\,\pm\,$4.1$\,\rm{Mpc}$, E($B-V$)\,=\,1.95$\,\rm{mag}$; \citealt{2018ApJ...853...62T}); a recent Type IIP SN\,2017eaw (d\,=\,5.6$\,\pm\,$0.1$\,\rm{Mpc}$, E($B-V$)\,=\,0.41$\,\rm{mag}$; \citealt{2019ApJ...876...19S}). In addition, we have used 29 SNe~II from \cite{2018MNRAS.473..513F}, sample of SNe II from \cite{2003ApJ...582..905H,2014MNRAS.439.2873S,2016MNRAS.459.3939V, 2018MNRAS.480.2475S} and luminous with low expansion velocity (LLEV) events (SNe 1983K, 2008bm, 2009aj, 2009au, LSQ13fn) from \cite{2020MNRAS.494.5882R}.

\subsubsection{Colour and Temperature Evolution}
 The ($V-I$)$_0$ colour evolution of SNe\,2014cx, 2014cy and 2015cz is compared with five SNe from the reference sample in the left panel of Figure \ref{ch6:colour_mag}. The colours are corrected for Galactic plus host galaxy reddening. SNe\,2005cs and 2017eaw determine the red and blue extremes of the colour distribution of this sample. The ($V-I$)$_0$ colour of SNe~II rapidly becomes redder at the initial phases, followed by a slower change towards redder colours as they enter the plateau phase. Then, the colour changes rapidly to redder colours during the transitional phase from plateau to the tail and subsequently evolves slowly over the tail phase. 
 The colour evolution of SN\,2014cx traces SN\,1999em up to $\sim$100$\,\rm{d}$. However, in the tail phase, SN\,2014cx is bluer than SN\,1999em by $\sim$0.3$\,\rm{mag}$. SN\,2014cy show bluer colours similar to SN\,2017eaw (within error) during the plateau phase and is bluer than SN\,2017eaw during the transition from plateau to the tail. The colour at the early phases of SN\,2015cz is bluer than SN\,2017eaw by $\sim$0.4$\,\rm{mag}$. Nonetheless, SN\,2015cz moves rapidly to redder colours with the expansion and cooling of the ejecta.

The temperature evolution of SNe\,2014cx, 2014cy and 2015cz is compared with a sample of 29 SNe\,II from \cite{2018MNRAS.473..513F} in the right panel of Figure~\ref{ch6:colour_mag}. The temperatures are estimated by fitting a blackbody function to the SED constructed using photometric fluxes corrected for total reddening. The temperature of the sample at $\sim$10\,d spans from 7200 to 14500\,K and has a median value of 9900\,K. At similar phases, SN\,2014cy has the highest temperature ($\sim$17400\,K), followed by SN\,2015cz ($\sim$14500\,K) and SN\,2014cx ($\sim$11300\,K). This is in agreement with the bluer ($V-I$)$_0$ colour of SN\,2015cz at early times. The decline in the photospheric temperature of SN\,2015cz is slower than SN\,2014cy and exceeds the temperture of SN\,2014cy after $\sim$30\,d. The temperature drops to 6000\,K at $\sim$35\,d in SN\,2014cx, at 50\,d in SN\,2014cy and at 80\,d in SN\,2015cz, and declines slowly thereafter. Post plateau phase ($\sim$86\,d), the temperature of SN\,2014cx remains nearly constant at $\sim$4000\,K, while that of SNe\,2014cy and 2015cz drops to $\sim$5000\,K after $\sim$100\,d. 

\begin{figure}
\begin{center}
\includegraphics[scale=0.54, clip, trim={0.2cm 0.5cm 1cm 2.1cm}]{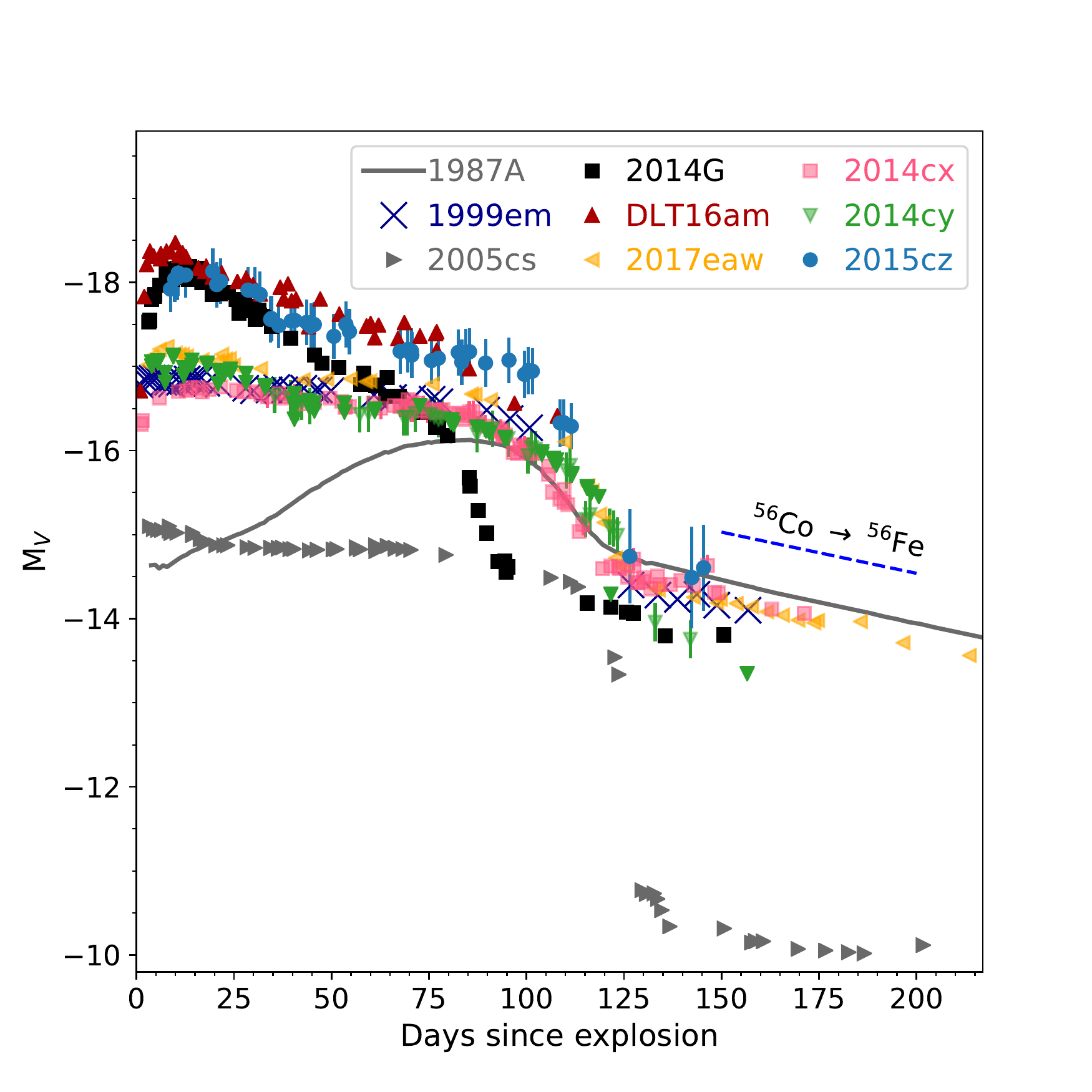}
\end{center}
\caption{The absolute $V$-band magnitudes of SNe\,2014cx, 2014cy and 2015cz are shown with those of a sample of SNe\,II.}
\label{ch6:abs_mag}
\end{figure}

\subsubsection{Absolute Magnitude and $^{56}$Ni mass}
We compare the absolute $V$-band light curves of SNe\,2014cx, 2014cy and 2015cz with six SNe from the reference sample as shown in Figure~\ref{ch6:abs_mag}. We estimated the $^{56}$Ni mass from the tail luminosity of the $V$-band light-curve following \cite{2003ApJ...582..905H}. The weighted average of the tail luminosity of SN\,2014cx is (1.6$\,\pm\,$0.1)$\times$10$^{41}$\,erg\,s$^{-1}$, SN\,2014cy is (0.9$\,\pm\,$0.2)$\times$10$^{41}$\,erg\,s$^{-1}$ and SN\,2015cz is (2.9$\,\pm\,$0.4)$\times$10$^{41}$\,erg\,s$^{-1}$, which gives $^{56}$Ni masses of 0.050$\,\pm\,$0.003\,M$_\odot$, 0.027$\,\pm\,$0.006~M$_\odot$ and 0.07$\,\pm\,$0.01\,M$_\odot$, respectively. The light curve evolution of SN\,2014cx is similar to the prototypical Type IIP SNe\,1999em and 2017eaw, with an absolute $V$-band magnitude M$_V^{50}$\,=\,$-$16.6$\,\pm\,$0.4$\,\rm{mag}$. The $^{56}$Ni yield of SN\,2014cx is also similar to SNe\,1999em and 2017eaw within error. The peak magnitude of SN\,2014cy is M$_V^{peak}$\,=\,$-$17.1$\,\pm\,$0.2$\,\rm{mag}$ that declines rapidly by 0.5$\,\rm{mag}$ in the first $\sim$50$\,\rm{d}$ before settling in the shallower declining phase. SN\,2014cy has an absolute $V$-band magnitude at 50$\,\rm{d}$: M$_V^{50}$\,=\,$-$16.5$\,\pm\,$0.2$\,\rm{mag}$, which is also similar to SNe\,1999em and 2017eaw, but with a lower $^{56}$Ni mass yield. SN\,2015cz falls on the brighter end of the sample of SNe\,II with similar magnitudes (within errors) as SN\,2014G and DLT16am, and has a shallower slope post 50\,days of its explosion and a longer plateau. The absolute $V$-band magnitude at 50$\,\rm{d}$ of SN\,2015cz is M$_V^{50}$\,=\,$-$17.4$\,\pm\,$0.3$\,\rm{mag}$. In essence, from the comparison plot it turns out that SNe\,2014cx, 2014cy and 2015cz are within the range of normal luminosity SNe\,II.

\begin{figure}
    \includegraphics[scale=0.46, clip, trim={0.2cm 0.9cm 0.45cm 2.4cm}]{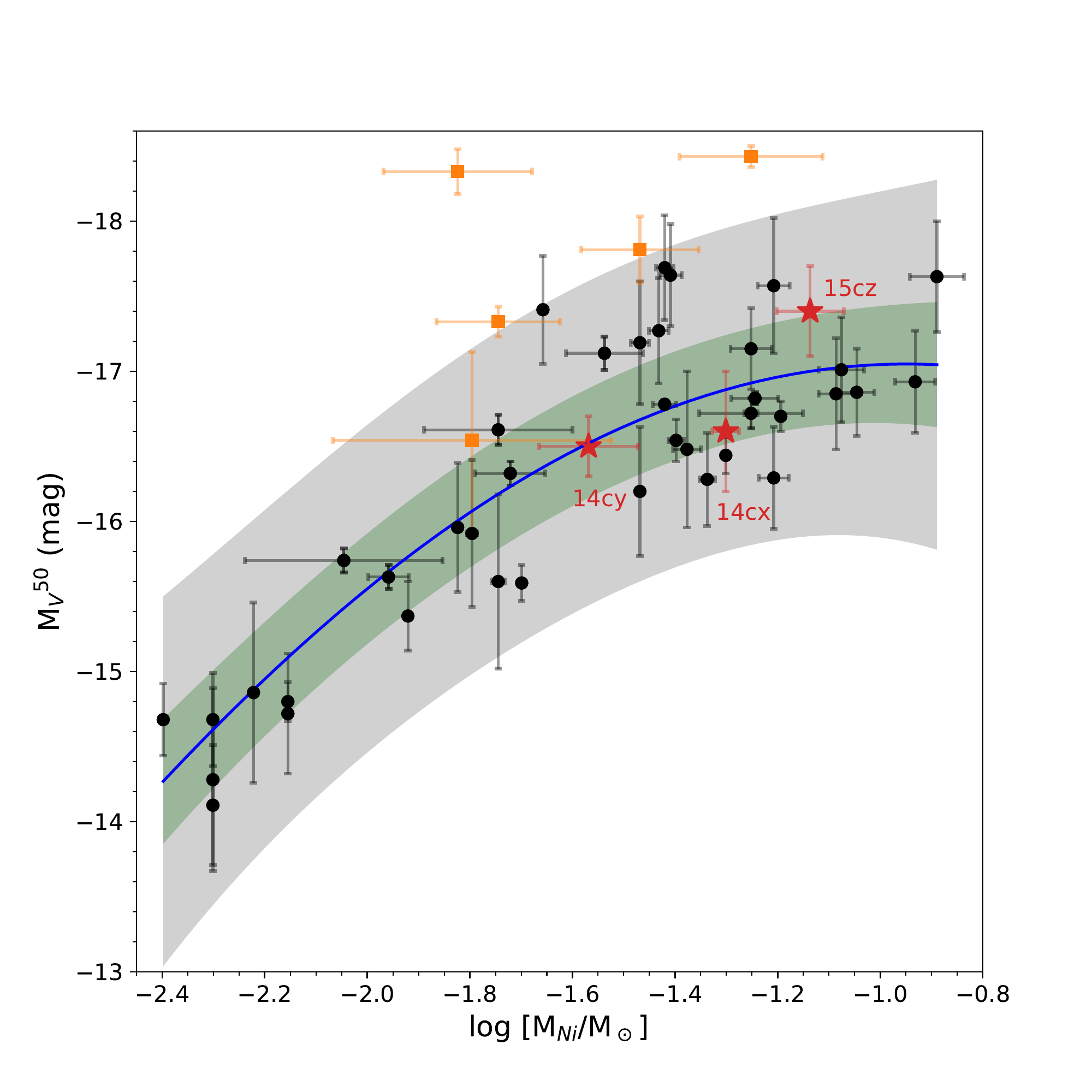}
  \caption{Position of SNe\,2014cx, 2014cy and 2015cz (labelled with red stars) in the $V$-band absolute magnitude at 50$\,\rm{d}$ (M$_V^{50}$) and $^{56}$Ni mass diagram. The yellow squares are the luminous with low expansion velocity SNe. The gaussian fit is shown by the solid blue line and the 1$\sigma$ and 3$\sigma$ fits are shown with shaded green and grey regions respectively.}
  \label{ch6:Mni_Mv}
\end{figure}

In Figure\,\ref{ch6:Mni_Mv}, the absolute magnitude in $V$-band at 50$\,\rm{d}$ is plotted against the logarithm of $^{56}$Ni mass for a sample of SNe II from \cite{2003ApJ...582..905H,2014MNRAS.439.2873S,2016MNRAS.459.3939V, 2018MNRAS.480.2475S,2020MNRAS.494.5882R} and SNe\,2014cx, 2014cy and 2015cz. The non-parametric regression line (gaussian process fit) for the sample, is shown in blue and the green and grey shaded region gives 1$\sigma$ and 3$\sigma$ error limits for the regression model. SNe\,2014cx, 2014cy and 2015cz stay within $\pm$1$\sigma$ error limit. This correlation is expected for typical SNe since higher yield of $^{56}$Ni is the outcome of a more energetic explosion and subsequently would manifest higher optical luminosity (e.g. \citealt{2001ApJ...558..615H}). We note that only one of the five LLEV events (shown with yellow squares) lie within 1$\sigma$ error of the fit and two lie outside the 3$\sigma$ limit. The ejecta circumstellar interaction is thought to be responsible for the comparatively higher luminosity of LLEV events as compared to the $^{56}$Ni mass yield in these explosions \citep{2020MNRAS.494.5882R}. 

\section{Spectral Analysis} \label{sec:5}

\begin{figure}
\begin{center}
\includegraphics[scale=0.45, clip, trim={1.8cm 21.5cm 0cm 5.45cm}]{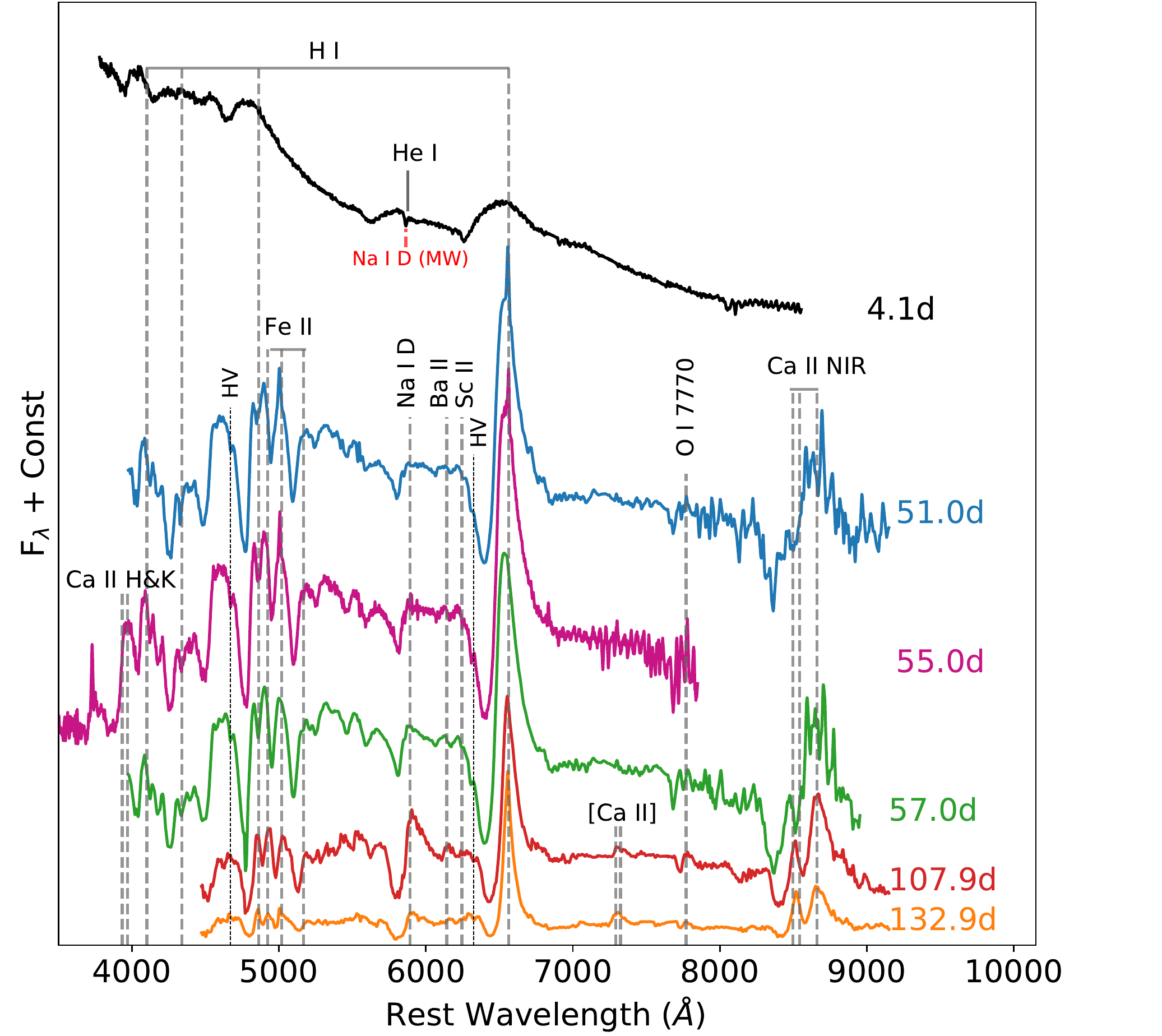}
\end{center}
\caption{The spectral evolution of SN\,2014cx is shown and the prominent spectral features are marked with dashed lines.}
\label{ch6:spectra_cx}
\end{figure}

\begin{figure}
\begin{center}
\includegraphics[scale=0.4,  width=0.45\textwidth, clip, trim={1.5cm 0.3cm 1.0cm 1.0cm}]{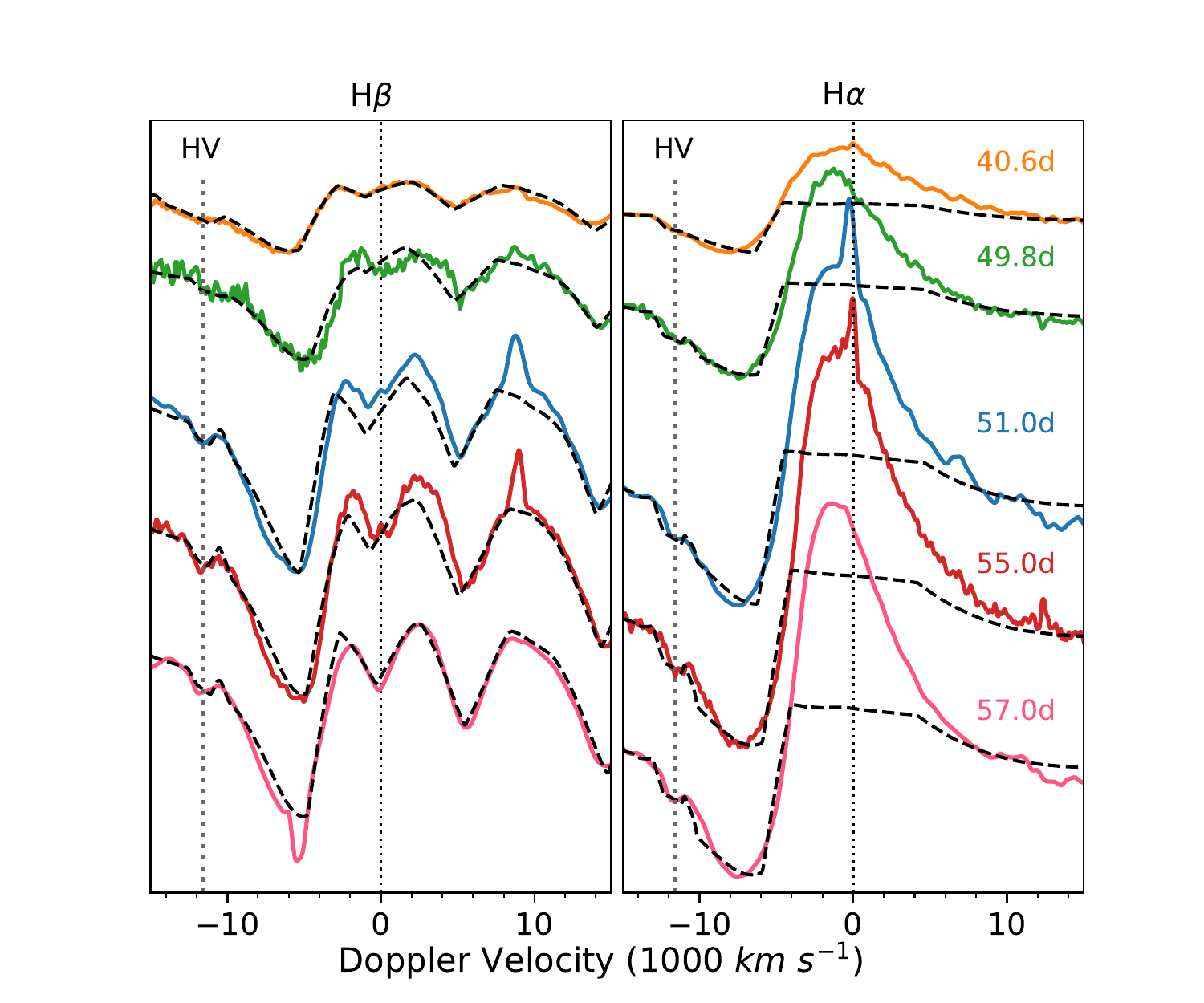}
\end{center}
\caption{The {\sc syn++} fits (dashed lines) to the plateau phase spectra of SN\,2014cx with a normal and high velocity (HV) component of H\,{\sc i} are shown. The first two spectra (40.6 and 49.8\,d) are taken from \citet{2016ApJ...832..139H}} 
\label{ch6:spectra_cx_syn}
\end{figure}

\begin{figure}
\begin{center}
\includegraphics[scale=0.45, clip, trim={1.8cm 3.5cm 0cm 23.5cm}]{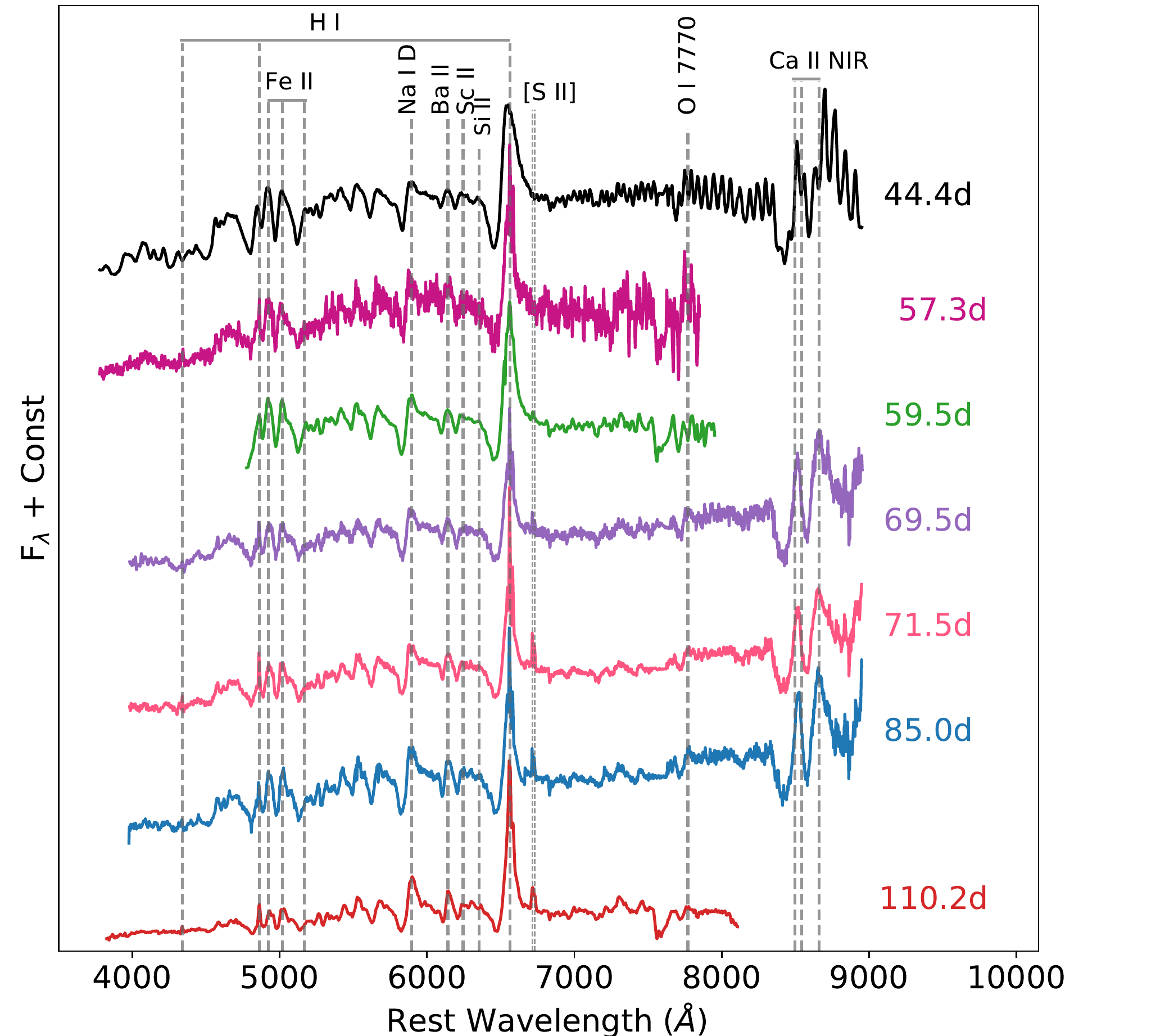}
\end{center}
\caption{The spectral evolution of SN\,2014cy is shown and the prominent spectral features are marked with dashed lines.}
\label{ch6:spectra_cy}
\end{figure}

\begin{figure}
\begin{center}
\includegraphics[scale=0.48, clip, trim={4.5cm 1.6cm 4cm 3.6cm}]{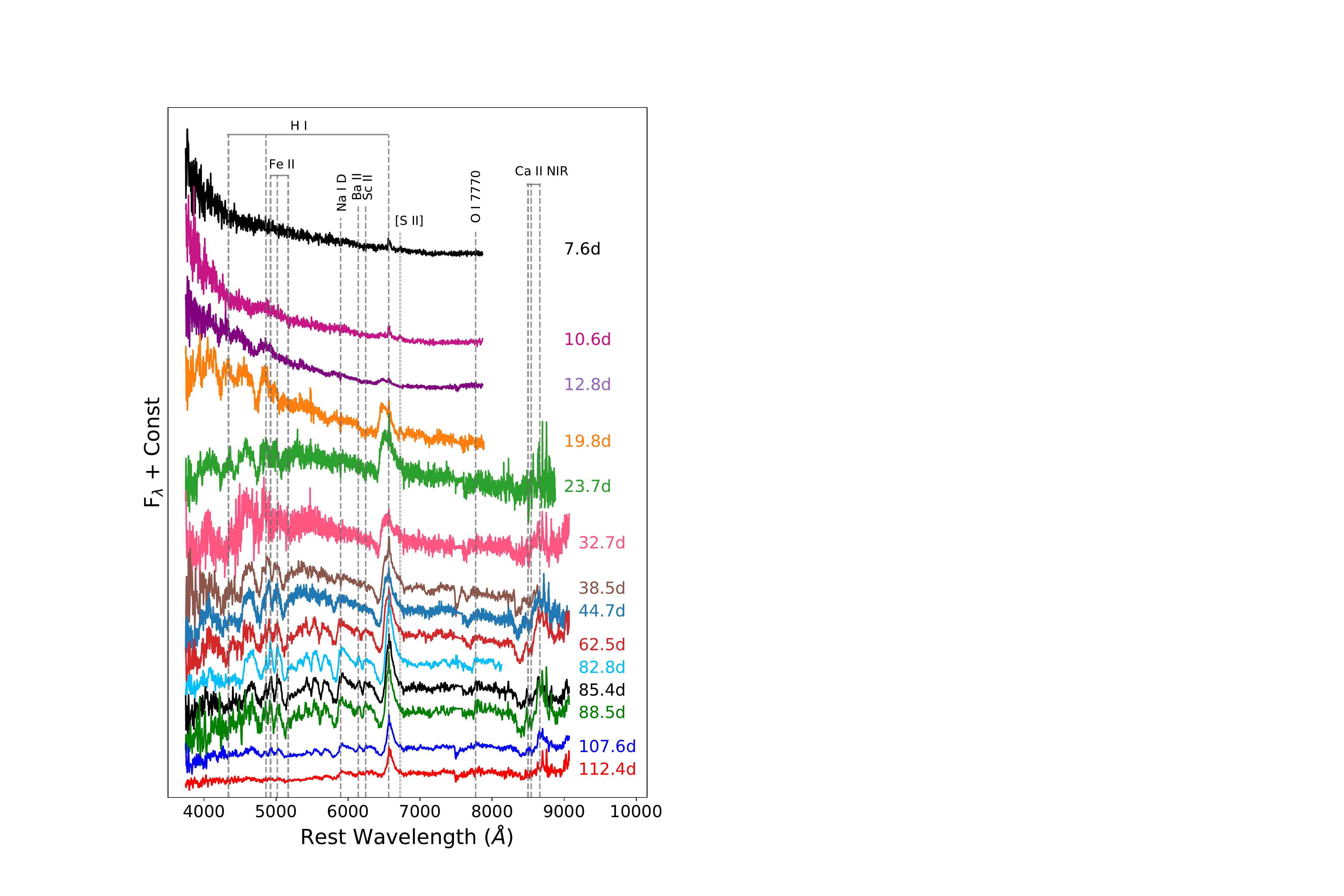}
\end{center}
\caption{Spectral evolution of SN\,2015cz with prominent spectral features marked with vertical lines.}
\label{ch6:spectra_cz}
\end{figure}

The spectral evolution of SNe\,2014cx, 2014cy and 2015cz is shown in Figures~\ref{ch6:spectra_cx}, \ref{ch6:spectra_cy} and \ref{ch6:spectra_cz} and prominent features were identified following \cite{2002PASP..114.1333L}. We have one early spectrum of SN~2014cx at 4.1$\,\rm{d}$ after explosion, three spectra during the plateau phase (53.4, 57.4 and 57.0$\,\rm{d}$), one corresponding to the transitioning phase (107.9$\,\rm{d}$) from plateau to the tail and and one in the nebular phase (132.9$\,\rm{d}$). The early spectrum of SN\,2014cx shows prominent H\,{\sc i} and He\,{\sc i} features superimposed on a blue continuum. The narrow Na\,{\sc i}\,D line from the Milky Way is conspicuous in this spectrum. The plateau phase spectra show broad H$\alpha$ and H$\beta$ absorption features with kinks at $\sim$6310 and 4670\,\AA, respectively. The {\sc syn++} model \citep{2011PASP..123..237T} fits with a normal component and a high velocity component (11600\,km\,s$^{-1}$) of H\,{\sc i} to the plateau phase spectra are shown in Figure\,\ref{ch6:spectra_cx_syn}. The features are at the same velocity ($\sim$11600\,km\,s$^{-1}$), where the velocities are computed by assuming that these are associated with H\,{\sc i}. Further, these do not show any appreciable evolution in velocity as expected from metal lines such as Si\,{\sc ii}. This suggests that it is possibly the high velocity (HV) component of H\,{\sc i} arising from the excitation of the outer layers of the H-rich ejecta by the high energy photons from the interaction of ejecta and circumstellar material \citep{2007ApJ...662.1136C}. The 132.9$\,\rm{d}$ spectrum of SN\,2014cx corresponds to the nebular phase in which [Ca\,{\sc ii}] 7291,7324 can be discerned. 

Spectroscopic follow-up of SN\,2014cy started $\sim$44$\,\rm{d}$ after explosion, when the SN was in the plateau phase. Si\,{\sc ii}\,$\lambda$6355 can be discerned at 44.4$\,\rm{d}$ and disappears at later phases. The plateau phase spectra of SN\,2014cy show well-developed metal lines, such as Fe\,{\sc ii}, Sc\,{\sc ii}, Ba\,{\sc ii}, O\,{\sc i}, Ca\,{\sc ii}\,NIR, as expected in SNe\,II. Due to the proximity of the SN to the host galaxy centre, narrow emission lines of H$\alpha$ and [S\,{\sc ii}]\,$\lambda\lambda$\,6717,6731 are clearly visible in the spectra of SN\,2014cy.

The early spectra (7.6 to 19.8$\,\rm{d}$) of SN\,2015cz show a blue continuum, with shallow H\,{\sc i} features. Metal features start developing from 12.8$\,\rm{d}$ onwards. The spectrum remains blue up to 19.8$\,\rm{d}$ after which the spectral slope changes and the SN becomes redder. Sc\,{\sc ii} and Ba\,{\sc ii} become prominent from 62.5$\,\rm{d}$ to 107.6$\,\rm{d}$. The narrow emission features (H$\alpha$, [S~{\sc ii}] 6717,6731) in the spectra of SN\,2015cz are likely from an H\,II region in the host galaxy. The final spectrum of SN\,2015cz (112.4$\,\rm{d}$) shows emission features with a weak absorption component and corresponds to the transition from the recombination to the radioactive tail. 

\begin{figure}
\begin{center}
\includegraphics[scale=0.7, clip, trim={0cm 0cm 0cm 0.3cm}]{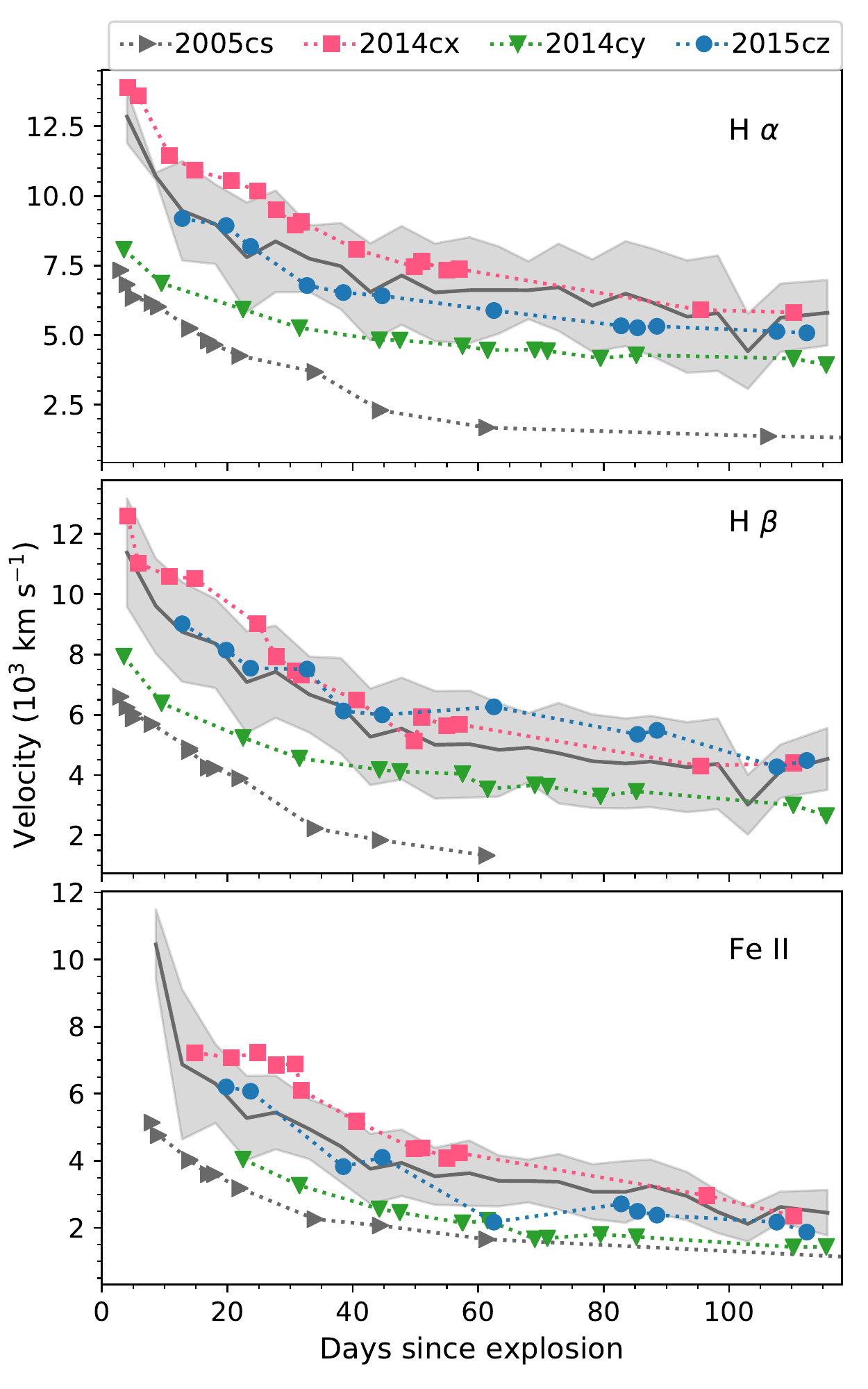}
\end{center}
\caption{Velocity evolution of H$\alpha$ (top panel), H$\beta$ (middle panel) and Fe\,{\sc ii}\,$\lambda$5169 (bottom panel) features of SNe\,2014cx, 2014cy and 2015cz compared to the mean velocity of 122 SNe\,II from \citet{2017ApJ...850...90G} and those of the sub-luminous SN\,2005cs \citep{2006MNRAS.370.1752P,2009MNRAS.394.2266P}. The mean velocities of the features in the sample are shown with grey solid line and the standard deviations of the mean velocities are shown with light grey regions.}
\label{ch6:vel}
\end{figure}

We estimated the velocity of H$\alpha$, H$\beta$ and Fe\,{\sc ii} lines from the position of the blue-shifted absorption minima and the evolution of velocity is shown in Figure\,\ref{ch6:vel}. The mean velocity evolution and the standard deviation for a sample of 122 SNe\,II from \cite{2017ApJ...850...90G} is shown with a grey solid line and grey shaded region. As an interpolation to the missing phases, we used the available spectra of SNe\,2014cx and 2014cy from \cite{2016ApJ...832..139H} and \cite{2019MNRAS.490.2799D}, respectively. H$\alpha$, H$\beta$ and Fe\,{\sc ii} lines of SN\,2014cx show higher velocity at the initial phases (13600\,km\,s$^{-1}$ at 5.8$\,\rm{d}$ for H$\alpha$) than the mean velocity of the sample (12000\,km\,s$^{-1}$ at 6$\,\rm{d}$ for H$\alpha$), but later the velocities approaches the mean velocity of the sample. SN\,2014cy exhibits much lower velocity than the sample (e.g. 8100\,km\,s$^{-1}$ at 6$\,\rm{d}$ as compared to the sample mean velocity of 12000\,km\,s$^{-1}$ at 6$\,\rm{d}$ for H$\alpha$) in the first 50\,d of the evolution. Afterwards, SN\,2014cy traces the lower end of the velocity distribution of the sample and lies between the normal luminosity SNe\,II and the low-luminosity SN\,2005cs. The H$\alpha$, H$\beta$ and Fe\,{\sc ii} velocity evolution of SN\,2015cz is consistent with the mean velocity of the sample.

\begin{figure}
\begin{center}
\includegraphics[scale=0.43, clip, trim={3.0cm 1.4cm 0cm 3cm}]{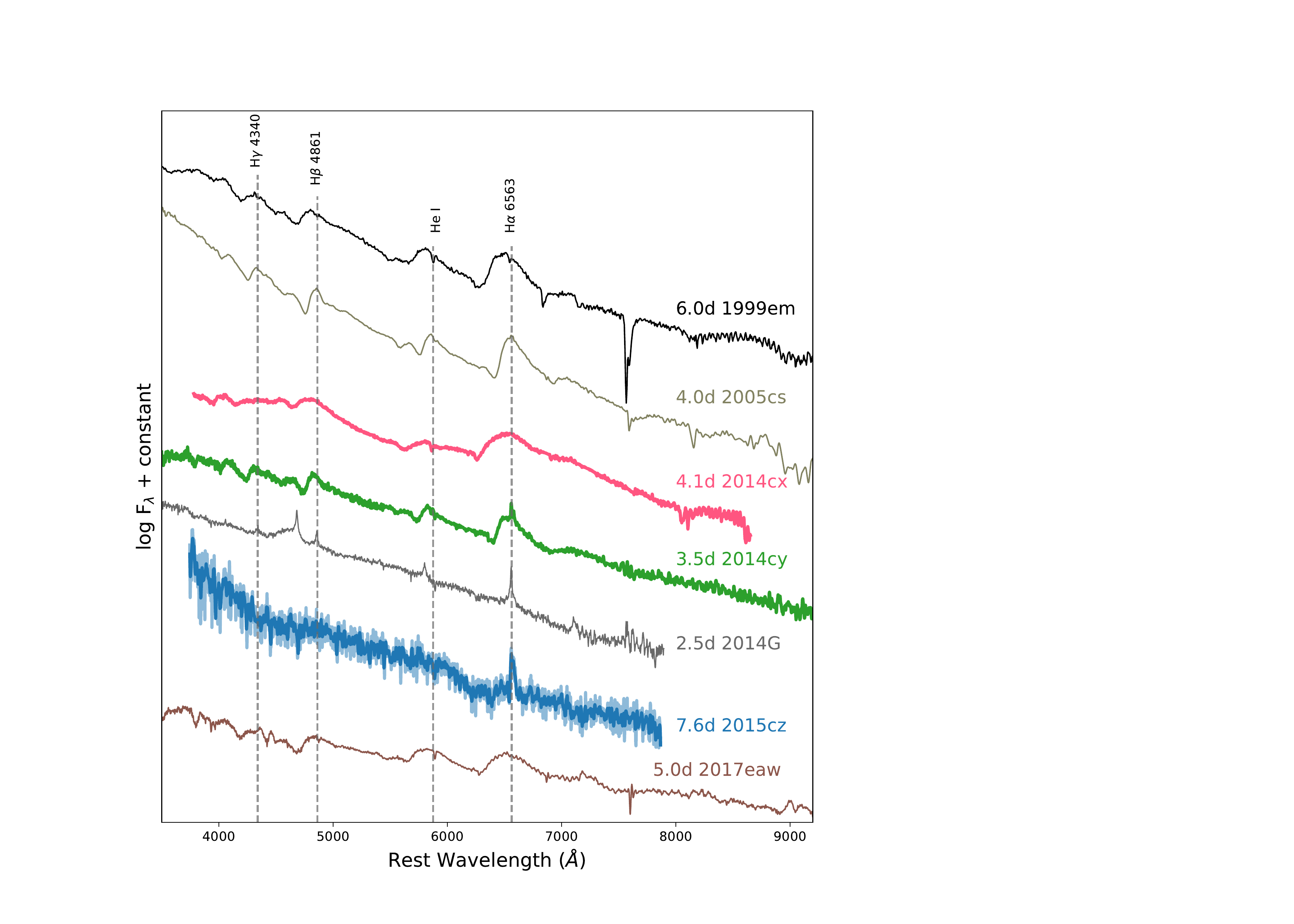}
\end{center}
\caption{The comparison of early spectra of SNe\,2014cx, 2014cy and 2015cz with a low-luminosity SN\,2005cs, a IIL SN\,2014G and a IIP SN\,2017eaw. The early spectrum of SN\,2014cy is taken from \citet{2019MNRAS.490.2799D}.}
\label{ch6:spectra_comp1}
\end{figure}

\begin{figure}
\begin{center}
\includegraphics[scale=0.43, clip, trim={3.0cm 1.4cm 0cm 3cm}]{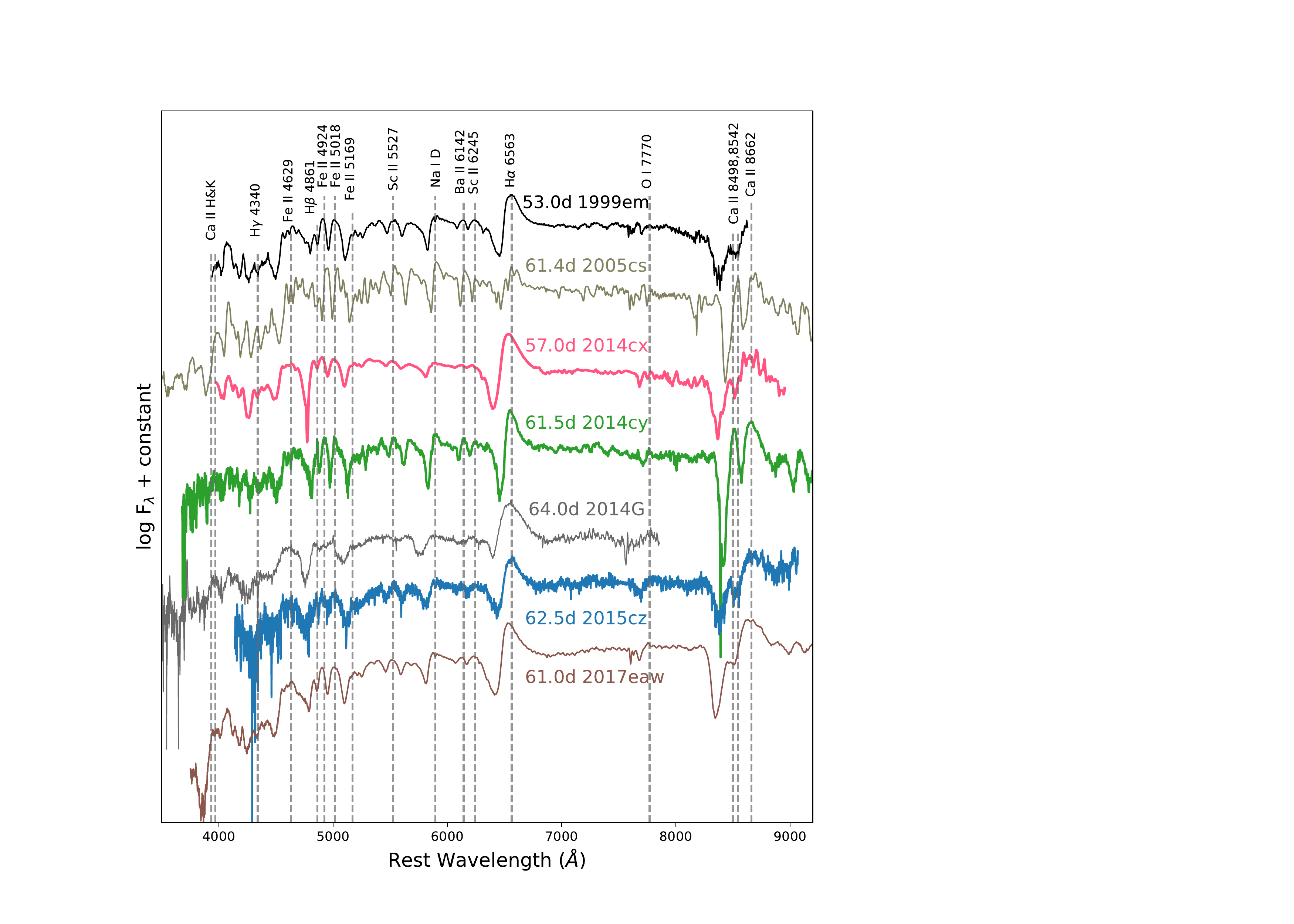}
\end{center}
\caption{The comparison of plateau spectra of SNe\,2014cx, 2014cy and 2015cz with a low-luminosity SN\,2005cs, a IIL SN\,2014G and a classical IIP SN\,2017eaw.}
\label{ch6:spectra_comp2}
\end{figure}

\begin{figure*}
\includegraphics[scale=0.48, clip, trim={4.1cm 0cm 0cm 0cm}]{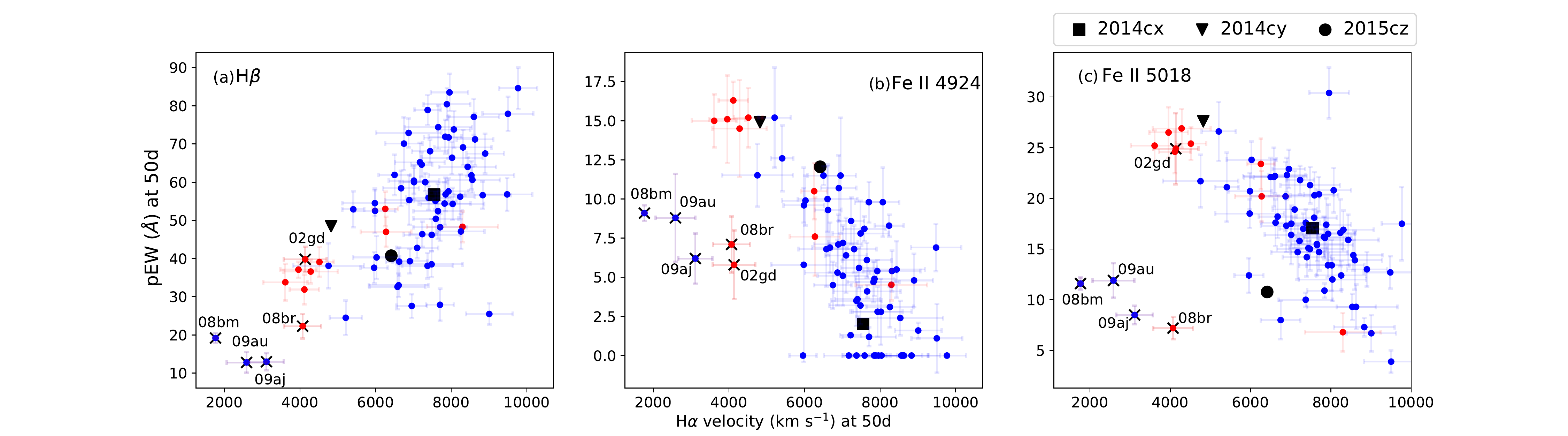}
\caption{The pseudo equivalent width (pEW) of H$\beta$, Fe\,{\sc ii}\,4924 and Fe\,{\sc ii}\,5018 absorption features of SNe\,2014cx, 2014cy and 2015cz at 50\,d are plotted against the H$\alpha$ velocity estimated from the shift of the absorption minima at 50\,d and compared with the sample of \citet{2017ApJ...850...90G}. The low-luminosity SNe ($-$14 \textgreater{} M$_V^{max}$ \textgreater $-$15.5) are shown in red and LLEV events (SNe\,2002gd, 2008bm, 2008br, 2009aj and 2009au) are marked with \textquoteleft$\times$\textquoteright.}
\label{ch6:pEW_comp}
\end{figure*}

We compare the early and plateau phase spectra of SNe\,2014cx, 2014cy and 2015cz, with the Type IIP SN\,1999em, low luminosity, low velocity SN\,2005cs \citep{2009MNRAS.394.2266P}, the Type IIL SN\,2014G \citep{2016MNRAS.462..137T} and the normal Type IIP SN\,2017eaw \citep{2019ApJ...876...19S} in Figures\,\ref{ch6:spectra_comp1} and \ref{ch6:spectra_comp2}. The spectra of SN\,2014cx closely matches SNe\, 1999em and 2017eaw during the early times and plateau phase. SNe\,2005cs and 2014cy show stronger H~{\sc i} and He~{\sc i} features at early phases as compared to the rest of the sample. The early spectrum of SNe\,2014G and 2015cz is dominated by a blue continuum with weak features. In the plateau phase, emission peak of H$\alpha$ is blue-shifted in the spectra of SNe\,2014cx and 2017eaw by $\sim$1200\,km\,s$^{-1}$ while in other SNe it is close to the rest wavelength of H$\alpha$. Although the origin of this offset is debated, \cite{2005A&A...437..667D} and \cite{2014MNRAS.441..671A} suggested that faster declining SNe\,II exhibit larger offset likely arising from their steeper density structure. The H$\alpha$ absorption component of SN\,2005cs shows sub-structures which are suggested to be due to the overlap of Ba\,{\sc ii}\,$\lambda$6496.9 and Sc\,{\sc ii} lines \citep{2017MNRAS.466...34L}. SN\,2014cy displays narrow absorption features, indicating low expansion velocity of the ejecta. The plateau spectrum of SN\,2015cz exhibits strong absorption component of H$\alpha$ unlike Type IIL SN\,2014G. The weak H$\alpha$ absorption component in SN\,2014G is the distinguishing feature commonly found in typical SN\,IIL, which has an otherwise similar spectrum as Type IIP.

\cite{2017ApJ...850...90G} analysed a sample of 122 SNe\,II exhibiting a range of H$\alpha$ velocities (1500\,\textless\,v$_{H\alpha}$\,\textless\,9600\,km\,s$^{-1}$) and pseudo equivalent widths (pEWs) and noted that H$\alpha$ velocity at 50\,d exhibits a positive correlation with the pEW of the absorption component of H$\beta$ and negative correlation with pEW of Fe\,{\sc ii}\,$\lambda$4924 and Fe\,{\sc ii}\,$\lambda$5018. We plotted the pEW of these features of SNe\,2014cx, 2014cy and 2015cz against the H$\alpha$ velocity both measured at 50$\,\rm{d}$ and compared with this sample in Figure\,\ref{ch6:pEW_comp}. SNe\,2014cx, 2014cy and 2015cz follow a similar trend as the sample in Figure\,\ref{ch6:pEW_comp}(a). In Figure\,\ref{ch6:pEW_comp}(a), we find that in general sub-luminous events show weaker absorption component of H$\beta$ while more canonical events display stronger H$\beta$ absorption. In Figures\,\ref{ch6:pEW_comp}(b) and \ref{ch6:pEW_comp}(c), we notice that low-luminosity events typically show stronger metal lines than the normal luminosity events. The strength of metal lines in SN\,2014cy is similar to the low-luminosity events while that of SNe\,2014cx and 2015cz are comparable to the normal luminosity events. All three of them show relatively stronger hydrogen feature akin to the normal luminosity events.

\section{Distance and Bolometric light-curve modelling} \label{sec:6}
To determine the distances to the host galaxies, we used the expanding photosphere method (EPM; \citealt{1974ApJ...193...27K}), a variant of the Baade-Wesselink method to estimate SN distances. At the early phases, when the ejecta is fully ionized and the opacity at the photosphere is dominated by electron scattering, the ejecta can be assumed to be homologously expanding and radiating as a diluted blackbody. By fitting a diluted blackbody function to the early photometric data, the colour temperature (T$_c$) and the angular radius ($\theta$) can be derived. The early spectroscopic observations are used to estimate the photospheric velocity ($v_{ph}$) from the minima of the absorption of Fe\,{\sc ii}\,$\lambda$5169. The distance $D$ and the time of explosion $t_0$ can be estimated from the slope and y-intercept, respectively, of the following expression:
\begin{equation}
t = D(\theta/v_{ph}) + t_0
\end{equation}

We apply EPM to determine the distances to SNe 2014cx and 2014cy using the same procedure as outlined in \cite{2018MNRAS.479.2421D}. For SN\,2014cx, we fixed $t_0$ to 2456902.4\,JD and estimated the distance to SN\,2014cx to be 18.5$\,\pm\,$1.1$\,\rm{Mpc}$. Both explosion epoch and distance are estimated for SN\,2014cy: 2456898.7$\,\pm\,$1.0\,JD and 23.4$\,\pm\,$1.6$\,\rm{Mpc}$. These values are consistent within error to the values adopted in \cite{2016ApJ...832..139H} and \cite{2016MNRAS.459.3939V}.

\begin{table*}
\centering
\caption{The range of priors on the parameters of SNe\,2014cx, 2014cy and 2015cz.}
\label{ch6:nagy_priors}
\begin{tabular}{ccccccccl}
\hline
 R$_{core}$ & M$_{core}$ & T$_{rec}$ & E$_{kin}$(core) & E$_{th}$(core) & R$_{shell}$ & M$_{shell}$ & E$_{kin}$(shell) & E$_{th}$(shell)\\
  (10$^{13}$~cm) & (M$_\odot$) & (K)  & (foe) & (foe) & (10$^{13}$~cm) & (M$_\odot$) & (foe) & (foe)\\
\hline
 1.5 - 6.0 & 8.0 - 30.0 & 5000 - 7000 & 0.1 - 9.0 & 0.1 - 2.0 & 2.0 - 12.0 & 0.01 - 0.90 & 0.001 - 0.50 & 0.001 - 0.50\\
\hline
\end{tabular}
\end{table*}

\begin{figure}
\begin{center}
\includegraphics[scale=0.5, trim={1.5cm 2.0cm 0cm 1.3cm}]{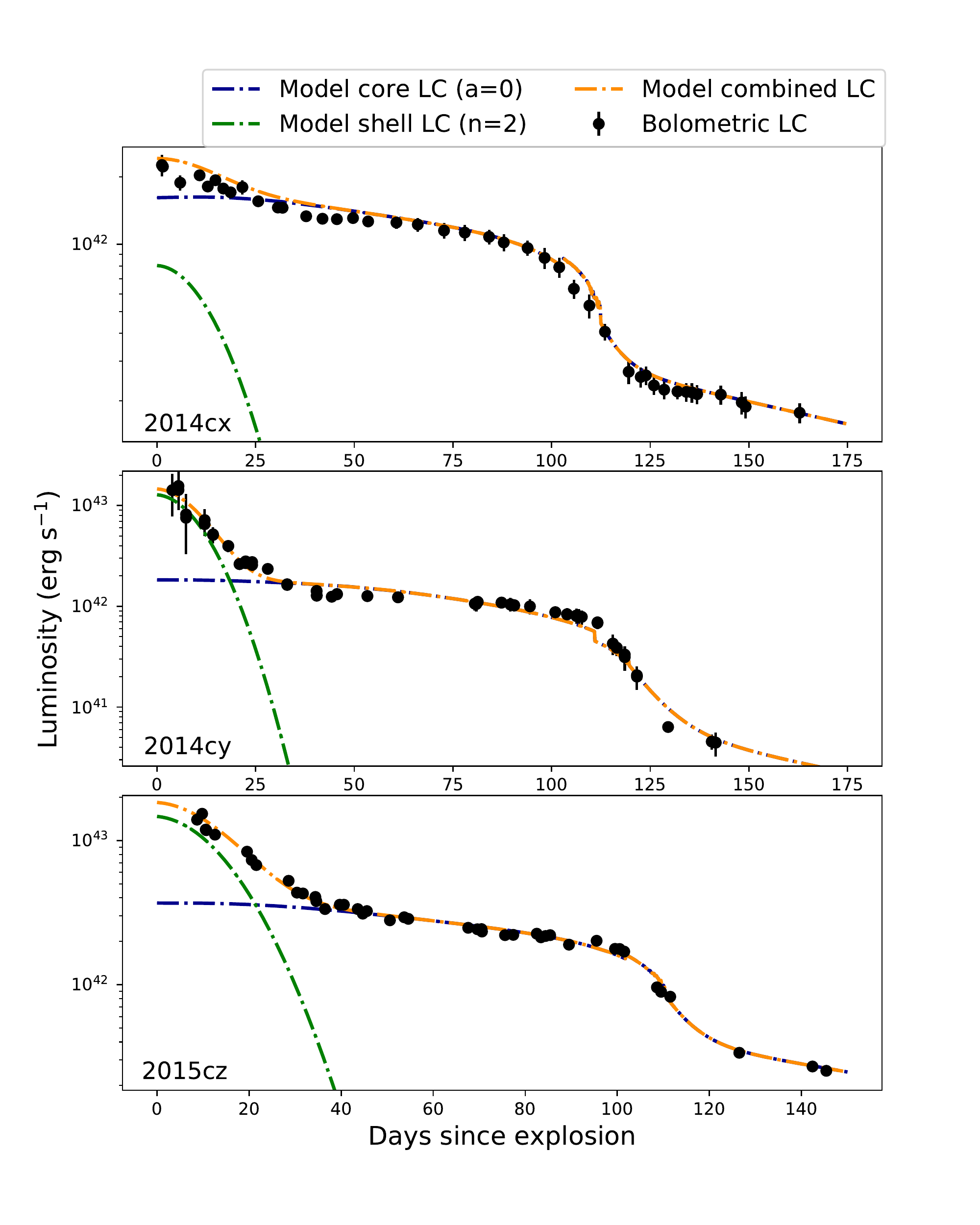}
\end{center}
\caption{The best-fit model light curve to the bolometric light curve of SNe\,2014cx, 2014cy and 2015cz using the semi-analytical modelling of \citet{2016A&A...589A..53N}.}
\label{ch6:nagy}
\end{figure}

\begin{table*}
\centering
\caption{The best fit core and shell parameters for true bolometric light curve of SNe\,2014cx, 2014cy and 2015cz using \citet{2016A&A...589A..53N}.\label{Nagy_ch6}}
\begin{tabular}{lccccccl}
\hline
                    & \multicolumn{2}{c}{2014cx} & \multicolumn{2}{c}{2014cy} & \multicolumn{2}{c}{2015cz} &\\
Parameter & Core & Shell & Core & Shell & Core & Shell & Remarks\\
\hline
R$_0$ (10$^{13}$~cm)       & 3.3$\,\pm\,$1.0   & 4.9$\,\pm\,$1.9     & 3.5$\,\pm\,$1.2   & 8.1$\,\pm\,$2.0     & 4.2$\,\pm\,$0.8  & 10.2$\,\pm\,$2.4     & Initial radius of ejecta\\
T$_{rec} (K)$              & 6430$\,\pm\,$275   & -               & 6904$\,\pm\,$82  & -               & 6902$\,\pm\,$89 & -               & Recombination temperature\\
M$_{ej}$ (M$_\odot$)       & 15.0$\,\pm\,$0.7   & 0.32$\,\pm\,$0.07  & 22.2$\,\pm\,$2.0   & 0.17$\,\pm\,$0.03 & 18.7$\,\pm\,$1.1 & 0.24$\,\pm\,$0.05 & Ejecta mass\\
E$_{th} (foe)$             & 0.73$\,\pm\,$0.24 & 0.03$\,\pm\,$0.01 & 1.12$\,\pm\,$0.38 & 0.14$\,\pm\,$0.04  & 1.58$\,\pm\,$0.26  & 0.18$\,\pm\,$0.05 & Initial thermal energy \\
E$_{kin} (foe)$            & 2.56$\,\pm\,$0.30 & 0.19$\,\pm\,$0.07 & 3.55$\,\pm\,$0.64 & 0.12$\,\pm\,$0.05  & 4.42$\,\pm\,$0.50  & 0.09$\,\pm\,$0.05 &  Initial kinetic energy \\
M$_{Ni}$ (M$_\odot$)       & 0.05          & -               & 0.023         & -               & 0.07         & -               & Initial $^{56}$Ni mass\\
$\kappa$ (cm$^2$ g$^{-1}$) & 0.2           & 0.38            & 0.2           & 0.38            & 0.2          & 0.38            & Opacity\\
A$_g$ (day$^2$)            & 3e5           & 1e10            & 1.1e4         & 1e10            & 5e4          & 1e10            & Gamma-ray leakage \\
\hline
\end{tabular}
\end{table*}

\begin{figure}
\begin{center}
\includegraphics[scale=0.6, trim={1.5cm 0.8cm 2.5cm 1.0cm}]{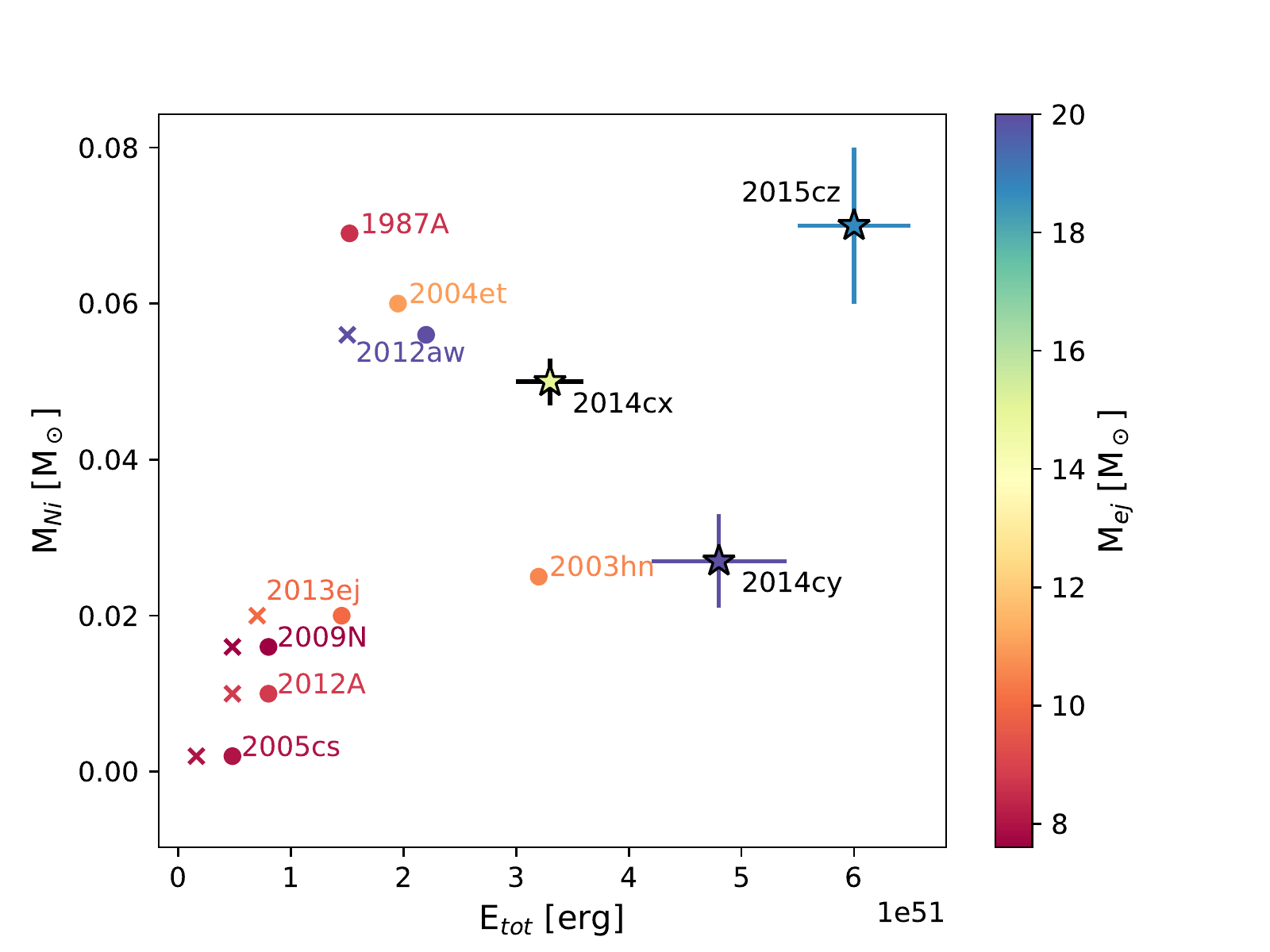}
\end{center}
\caption{The total explosion energies (E$_{tot}$\,=\,E$_{kin}$+E$_{th}$) of SNe\,2014cx, 2014cy and 2015cz are plotted against ejected $^{56}$Ni masses of a sample of SNe from \citet{2017MNRAS.464.3013P}. The sample is colour-coded with the ejecta masses (M$_{ej}$). The \textquoteleft x' and the \textquoteleft o' represents the total energy estimate from hydrodynamical modelling \citep{2017MNRAS.464.3013P} and semi-analytical modelling \citep{2014A&A...571A..77N,2016A&A...589A..53N}, respectively.}
\label{ch6:Efin_Mni_ej}
\end{figure}

We constructed the bolometric light curve of SNe\,2014cx, 2014cy and 2015cz using {\sc superbol} \citep{2018RNAAS...2..230N} as shown in Figure\,\ref{ch6:nagy}. The de-reddened magnitudes supplied to the routine are interpolated to a common set of epochs and converted to fluxes. The fluxes are used to construct the SED at all the epochs. The routine then fits a blackbody function to the SED, extrapolating to the UV and IR regimes to estimate the true bolometric luminosity. We used the semi-analytical model of \cite{2016A&A...589A..53N} to estimate the progenitor parameters of these SNe from the bolometric light curve. For the core, we used the constant density profile and for the shell a power-law density profile of index 2. We used the average opacity for core (0.2\,cm$^2$\,g$^{-1}$) and shell (0.38\,cm$^2$\,g$^{-1}$) recommended for SNe\,IIP in \cite{2018ApJ...862..143N}. The best fit light curves are shown in Figure\,\ref{ch6:nagy} and the corresponding parameters for the core and the shell for each SNe are tabulated in Table\,\ref{Nagy_ch6}. We have used the {\sc pymcmcstat} package \citep{Miles2019}, a Python program for running Markov Chain Monte Carlo simulations, for uncertainty quantification in nine free parameters of the model, namely, the initial radii of the core and the shell (R$_{core}$ and R$_{shell}$), the ejecta masses of the core and the shell (M$_{core}$ and M$_{shell}$), the kinetic and thermal energy of the explosion (E$_{kin}$(core), E$_{th}$(core), E$_{kin}$(shell) and E$_{th}$(shell)) and the temperature (T$_{rec}$). We started off this procedure with some initial parameter values guessed by tweaking the parameters to generate the best-fit-by-eye model light curve. We have used uniform probability distribution for the prior on the parameters. For the radii and masses of core and shell of the three SNe, we have chosen the priors based on the range of RSG radii (200 - 1800 R$_\odot$) and RSG masses (8 - 30 M$_\odot$) \citep{2010ASPC..425..103L}. The prior values for temperature lies in the range of the recombination temperature of hydrogen (5000 - 7000\,K). The priors for the energies (kinetic and thermal) span the range of energies estimated in the literature for the sub-luminous and the most luminous SNe II \citep{2005coex.conf..275Z}. The range of prior values used for each parameter is tabulated in Table \ref{ch6:nagy_priors}. 
We go through 10000 MCMC simulations. The trace plots of the parameters are shown in Figure\,\ref{ch6:trace}. In {\sc pymcmcstat}, Geweke\textquoteright s convergence diagnostic \citep{Geweke92evaluatingthe} is used to get an idea of the fraction of the chain to be rejected as burn-in in order to obtain the optimal parameter values and their standard deviations. The covariances between the different parameters are shown in a corner plot (using kalepy\footnote{https://github.com/lzkelley/kalepy}) in Figures\,\ref{ch6:corner_cx}, \ref{ch6:corner_cy} and \ref{ch6:corner_cz}. We find strong correlations between radius and thermal energy of the core and mass and kinetic energy of the core.

\begin{figure*}
               \includegraphics[scale=0.55,clip, trim={3.0cm 0.0cm 3.0cm 0.40cm}]{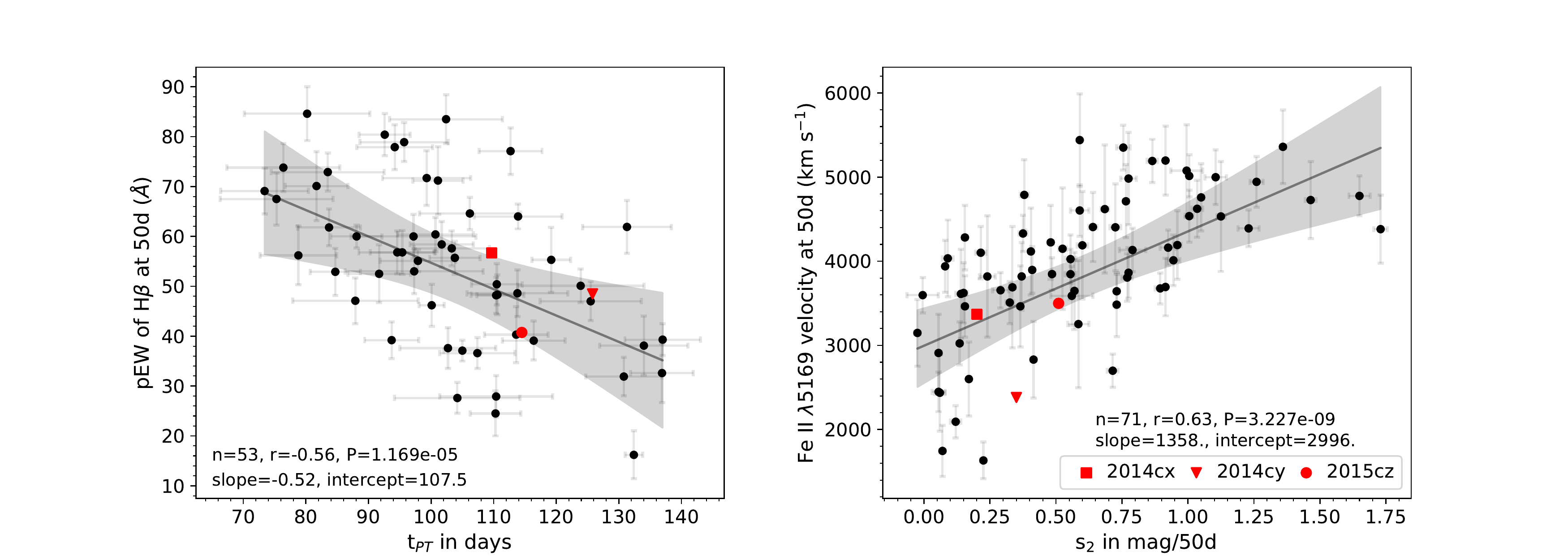}
  \caption{Position of SNe\,2014cx, 2014cy and 2015cz in the pEW of H$\beta$ and t$_{PT}$ diagram (left panel) and Fe\,{\sc ii} velocity and plateau slope (s$_2$) diagram (right panel) using the samples of SNe\,II from \citet{2014ApJ...786...67A} and \citet{2017ApJ...850...90G}. For each panel, $n$ is the number of events, $r$ is the Pearson correlation coefficient, $P$ is the probability of detecting a correlation by chance and slope and intercept are the parameters of the best fit linear regression model. The shaded grey region denotes the 3$\sigma$ confidence interval.\label{ch6:disc3}}
\end{figure*}

Also, we can get an idea about the expansion velocity of the ejecta from the model values of kinetic energy and ejecta mass. We obtain similar expansion velocity (within error) from the modelled core paramters for SNe 2014cx (5325\,km\,s$^{-1}$) and 2014cy (5177\,km\,s$^{-1}$), while 2015cz (6294\,km\,s$^{-1}$) has the highest velocity. From spectroscopy, we find that the expansion velocity is in the order vel$_{14cx}$ \textgreater vel$_{15cz}$ \textgreater vel$_{14cy}$. This kind of discrepancy probably arises because the correlation between the ejected mass and the kinetic energy introduce uncertainty in the explosion velocities derived from LC modelling \citep{2014A&A...571A..77N}, which may result in significant systematic offsets in the model velocities.

The $^{56}$Ni masses of SNe\,2014cx, 2014cy and 2015cz are plotted against the total explosion energies (E$_{tot}$\,=\,E$_{kin}$+E$_{th}$), along with that derived for other SNe using hydrodynamic modelling \citep{2017MNRAS.464.3013P} 
and semi-analytical modelling \citep{2014A&A...571A..77N,2016A&A...589A..53N} 
in Figure\,\ref{ch6:Efin_Mni_ej}. The plot is colour-coded with ejecta masses. It becomes apparent that the semi-analytical model of \cite{2014A&A...571A..77N} tend to over-estimate the total energy than those obtained from more sophisticated hydrodynamical models (also noted in \cite{2014A&A...571A..77N}).

\section{Discussion}
\label{sec:7}
The photometric and spectroscopic parameters of SNe\,2014cx, 2014cy and 2015cz are largely analogous to the Type II SN family, except for a few peculiarities. SN\,2014cx is a typical Type IIP SN with a rather atypical short plateau lasting $\sim$86\,d as compared to 100-120\,d in traditional Type IIP SNe. The absolute magnitudes, colour and spectral properties of SN\,2014cx is similar to archetypal Type IIP SNe\,1999em and 2017eaw. SN\,2014cy is a normal luminosity SN\,II with low expansion velocities. SN\,2014cy exhibits intermediate properties between normal and subluminous events with respect to its spectral features, viz. strong H lines similar to normal luminosity events whereas strong Fe\,{\sc ii} lines similar to sub-luminous events. The high metallicity environment of SN\,2014cy might play a role in the formation of strong metal lines \citep{2019ApJ...870L..16S}. SNe\,2014cy and 2015cz display an early steep declining light curve with slopes consistent with the decline rate of canonical SN\,IIL light curves. However, both SNe shows well-developed H$\alpha$ absorption component during the plateau phase unlike Type IIL SNe. Below we discuss the possible implications of the observed properties of these events and explore the analogies and differences of SNe\,2014cx, 2014cy and 2015cz with conventional SNe\,II by comparing the photometric and spectroscopic parameters.

\subsection{Ejecta - Circumstellar material interaction}
The interaction between ejecta and circumstellar material (CSM) is not very apparent in Type-IIP SNe, while Type IIL SNe are known to show some interaction. However, bolometric light curve modelling of SNe IIP \citep{2016A&A...589A..53N, 2017ApJ...838...28M} indicates that these events also display modest interaction. In the semi-analytical modelling of \cite{2016A&A...589A..53N}, a wind-like density profile is added above a constant density core to reproduce the early declining phase in Type II light curves. Similarly, in SNEC (SuperNova Explosion Code), the hydrodynamical modelling code of \cite{2017ApJ...838...28M, 2018ApJ...858...15M}, a wind-like density profile has been used to model the CSM and appended to the RSG model to simulate the intial decline after maximum light. Thus, the early declining phase in the bolometric light curve of SNe\,2014cx, 2014cy and 2015cz indicates that ejecta-CSM interaction exists in these events. 

Another possible interaction signature is the appearance of a notch bluewards of H Balmer absorption component due to the excitation of the outer recombined layer of hydrogen by high energy photons from the reverse shock. These features have been identified at both early and late photospheric epoch in a number of SNe II, and has been termed \textquoteleft cachito\textquoteright{} \citep{2017ApJ...850...89G}. The plateau phase spectra of SN\,2014cx shows such a non-evolving feature at 11600\,km\,s$^{-1}$ from the rest wavelength of H$\alpha$ and H$\beta$. The non-evolving nature as well as the {\sc syn++} modelling of this feature suggests that this is the high velocity component of H\,{\sc i} and not a metal line such as Si\,{\sc ii}. However, such high velocity feature in inconspicuous in the spectra of SNe\,2014cy and 2015cz.  

\subsection{Status of SNe 2014cx, 2014cy and 2015cz in the parameter space of SNe II}

\begin{figure}
               \includegraphics[scale=0.48,clip, trim={0.6cm 0.8cm 2.0cm 2.40cm}]{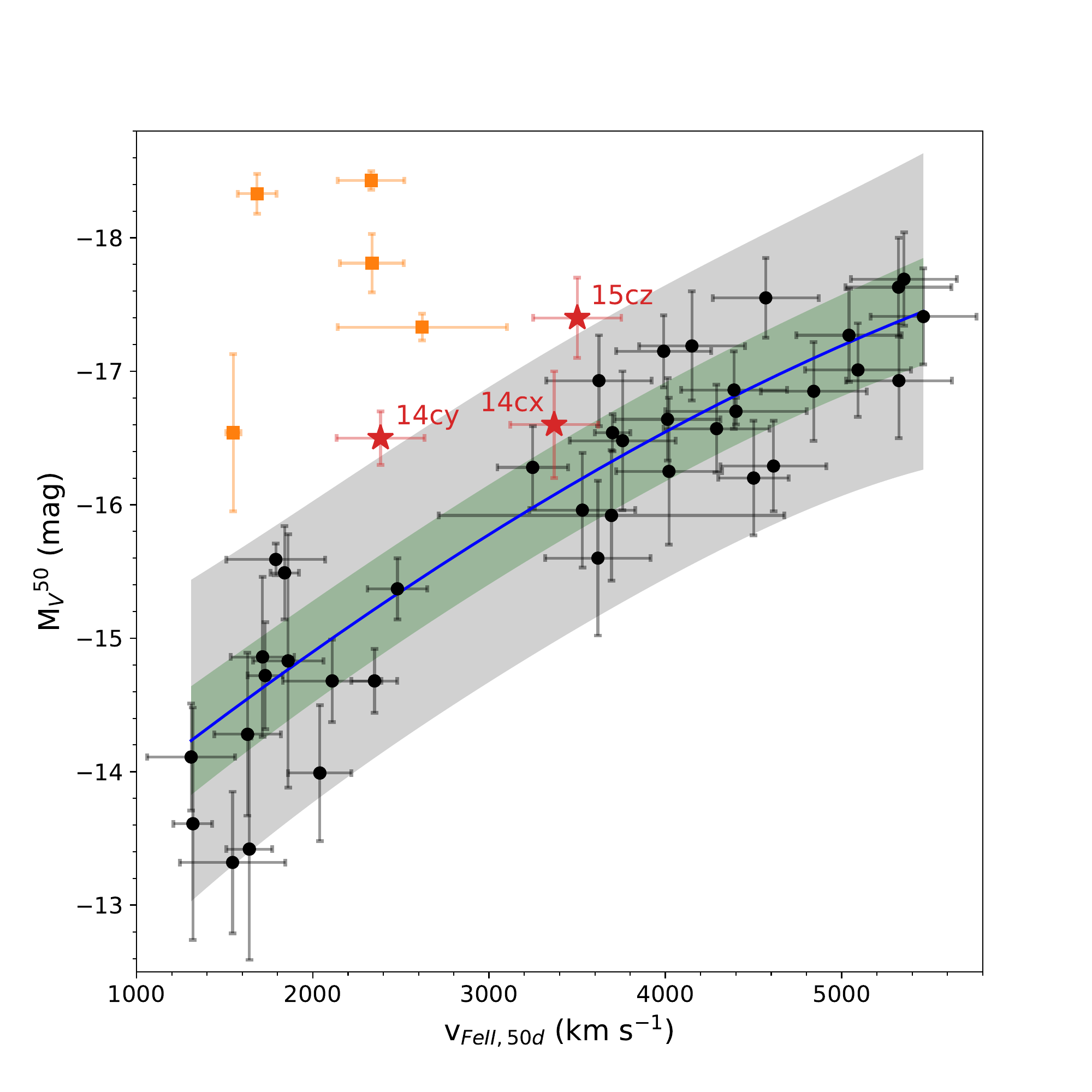}
  \caption{Position of SNe\,2014cx, 2014cy and 2015cz (labelled with red stars) in the plot of $V$-band absolute magnitude at 50$\,\rm{d}$ (M$_V^{50}$) versus the photospheric velocity at 50$\,\rm{d}$. The yellow squares represent luminous SNe\,II with low expansion velocity. The gaussian process fit is shown as the solid blue line and the 1$\sigma$ and 3$\sigma$ fits are shown with shaded green and grey regions respectively.\label{ch6:disc}}
\end{figure}

In Figure\,\ref{ch6:disc3}, we have used the sample of SNe\,II from \cite{2014ApJ...786...67A} and \cite{2014ApJ...786L..15G,2017ApJ...850...90G} to discern the location of the three SNe in the parameter space (plateau length (t$_{PT}$), plateau slope (s$_2$) and pEW of H$\beta$ and Fe\,{\sc ii}\,$\lambda$5169 velocity) of SNe II, i.e. whether the three SNe falls within the 3$\sigma$ limit of the correlation suggested by the other authors \citep{2014ApJ...786...67A,2014ApJ...786L..15G,2017ApJ...850...90G}. Generally one would expect that the more typical SNe II to fall within the 3$\sigma$ limit while peculiar events might show some deviation.

On the left panel of the diagram, we find that a negative correlation exists between t$_{PT}$ and pEW of H$\beta$. Although moderate, such a correlation implies that SNe\,II with longer plateau length show weaker H$\beta$ absorption component. This can be physically interpreted as follows. The plateau length depends on the recombination time, which in turn depends on the envelope mass as well as the density of the gas. High density leads to rapid recombination and hence shorter plateau. Moreover, because of the rapid recombination, the density and hence the opacity of the recombined hydrogen will be higher which will result in stronger H-feature viz. higher pEW. SNe\,2014cx, 2014cy and 2015cz follows this trend and falls within the 3$\sigma$ limit of the fit.

\begin{figure}
\centering
               \includegraphics[scale=0.55, clip, trim={5.4cm 2.4cm 0.0cm 3.00cm}]{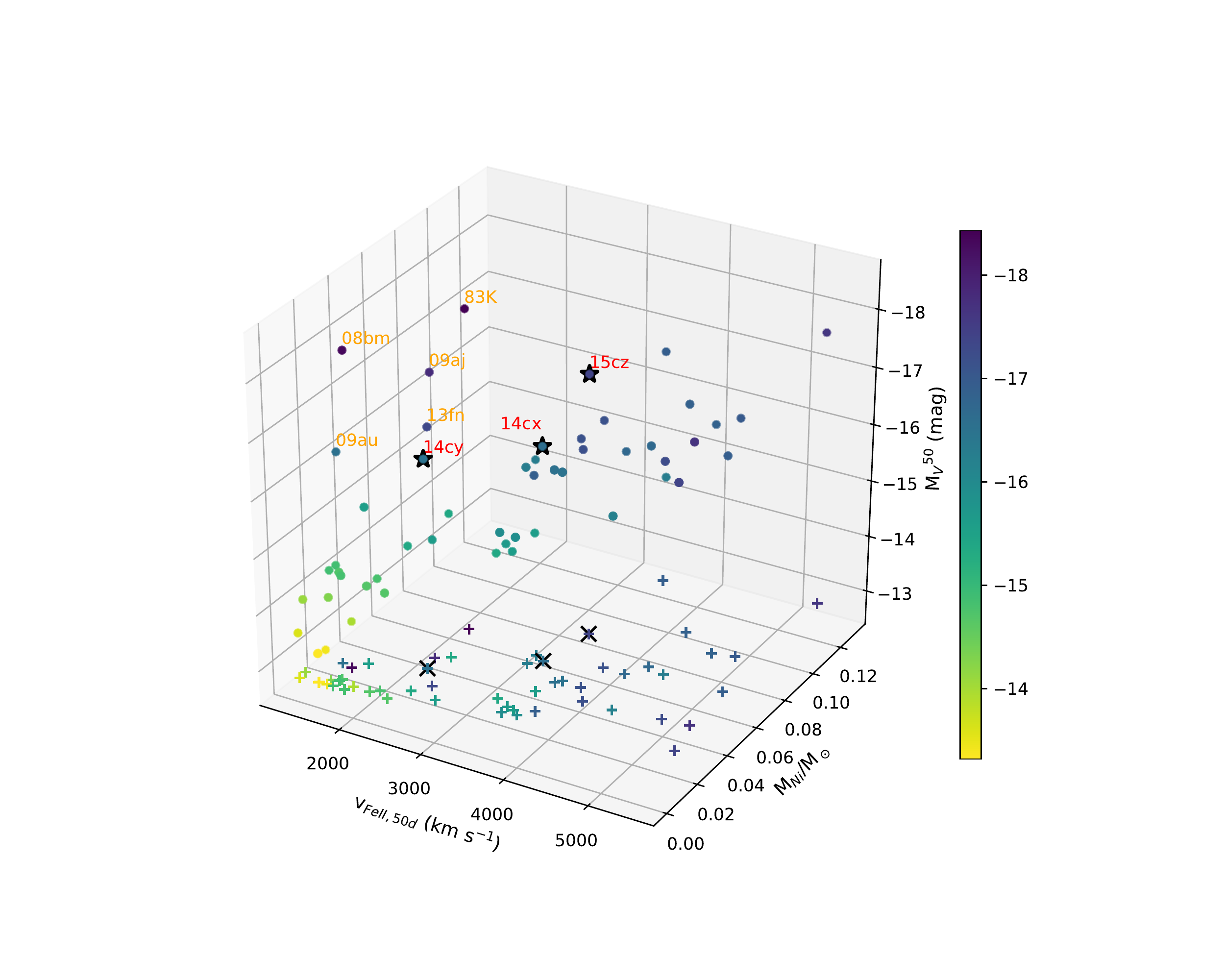}
  \caption{The location of SNe\,2014cx, 2014cy and 2015cz in the 3D phase space of Fe\,{\sc ii} $\lambda$5169 velocity at 50\,d, $^{56}$Ni mass and $V$-band absolute magnitude at 50\,d. The LLEV events are SNe 1983K, 2008bm, 2009aj, 2009au and LSQ13fn. The sample is colour-coded with the $V$-band absolute magnitudes. The projection of the SNe in the velocity-$^{56}$Ni mass plane is shown with \textquoteleft +'. SNe 2014cx, 2014cy and 2015cz are highlighted with \textquoteleft x'.} 
  \label{ch6:disc1}
\end{figure}

The right panel of Figure\,\ref{ch6:disc3} shows a positive correlation between the plateau slope and Fe\,{\sc ii}\,$\lambda$5169 velocity which has also been discussed in \cite{2014ApJ...786...67A} and \cite{2017ApJ...850...90G}. This correlation implies that the Fe\,{\sc ii}\,$\lambda$5169 velocity is higher in SNe\,II with steeper plateau slopes. The slope of the plateau depends on the density gradient of the envelope, viz. larger slopes could be arising from steeper density profiles. SNe\,II with smaller envelope masses generally have steeper density profiles. For a given explosion energy, a smaller envelope mass in turn implies that the expansion velocities will be higher. SNe\,2014cx and 2015cz lie within the 3$\sigma$ limit of the correlation plot, while SN\,2014cy does not. SN\,2014cy exhibits lower velocity which could be arising from the massive hydrogen envelope of this event as manifested by its long plateau.

In Figure \ref{ch6:disc}, the absolute magnitude in $V$-band at 50$\,\rm{d}$ is plotted against Fe\,{\sc ii}\,$\lambda$5169 velocity, a proxy for the photospheric velocity, for a sample of SNe\,II. The distribution of the events in this space implies that higher luminosity events have higher energies and hence higher expansion velocities in agreement with the findings of \cite{2002ApJ...566L..63H, 2014ApJ...786...67A}. SN\,2014cx lies within 3$\sigma$ limit of the fit while SNe\,2014cy and 2015cz lie at the edge of the 3$\sigma$ limit. Although SNe\,2014cy and 2015cz have lower expansion velocities as compared to their luminosities, it appears to be well separated from the LLEV events.

To put things in perspective, we compare the positions of SNe\,2014cx, 2014cy and 2015cz in the 3D phase space of the expansion velocity, $^{56}$Ni mass and luminosity in Figure\,\ref{ch6:disc1}. SNe\,2014cx follow conventional SNe\,II while SNe\,2014cy and 2015cz exhibit slightly higher absolute magnitudes. SN\,2014cy falls just below LLEV events, well separated from the typical SN\,II. We also note that SN\,2014cy has bluer colours similar to LLEV events, however metal features are stronger in this event in comparison to LLEV events (see Figure\,\ref{ch6:pEW_comp}).

\section{Summary}\label{sec:8}
In this paper, we present the photometric and spectroscopic analysis of three Type II SNe\,2014cx, 2014cy and 2015cz. SN\,2014cx is analogous to SNe\,IIP with a slowly declining plateau phase (0.2\,mag/50d) and an unusual short plateau ($\sim$86\,d). SNe\,2014cy and 2015cz show relatively large decline rates (0.88 and 1.64 mag/50d, respectively) at early times as seen in Type IIL SN. The absolute magnitudes lie in the range $-$16 to $-$18\,mag placing SNe\,2014cx, 2014cy and 2015cz among the normal luminosity Type II SN. SNe\,2014cy and 2015cz exhibit bluer (V-I)$_0$ colours and higher temperatures than SN 2014cx which has similar colours as that of the more typical Type IIP SNe\,1999em and 2017eaw. 

The spectra of SNe\,2014cx, 2014cy and 2015cz display strong hydrogen features typical of SNe\,II. High velocity features have been identified in the spectra of SN\,2014cx at 11600\,km~s$^{-1}$, which is possibly a signature of circumstellar interaction. However, the contribution of this interaction to luminosity is indiscernible. The spectra of SN\,2014cx is analogous to Type IIP SN spectra (e.g. SNe\,1999em and 2017eaw). Despite the early steep declining light curve in SNe\,2014cy and 2015cz similar to SNe\,IIL, they show strong H$\alpha$ absorption component unlike SNe\,IIL. At early times the spectra of SN\,2014cy show well-developed H\,{\sc i} lines similar to SN\,2005cs. However, sub-luminous SNe develops stronger metal features with respect to H Balmer lines in the plateau phase which is not the case in SN\,2014cy. 

The $^{56}$Ni mass yield from the events spans a range of 0.027\,-\,0.070\,M$_\odot$. The explosion parameters derived from the best fit semi-analytic model generated using the prescription of \cite{2016A&A...589A..53N} to the bolometric light curve yielded a total ejecta mass of $\sim$15.0, 22.2 and 18.7\,M$_\odot$ for SNe\,2014cx, 2014cy and 2015cz, a total explosion energy of 3.3, 4.8 and 6.0\,foe and an initial radius of 478, 507 and 608\,R$_\odot$ respectively. Adding a remnant mass of 1.5\,M$_\odot$ to the ejecta masses yield 16.5, 23.7 and 20.2\,M$_\odot$ as a lower estimate of the progenitor masses of SNe\,2014cx, 2014cy and 2015cz respectively. Only the progenitor mass of SN\,2014cx is well within the limits of progenitor masses computed from pre-explosion images of SNe\,II (8 - 18\,M$_\odot$, \citealt{2009ARA&A..47...63S,2015PASA...32...16S}), while for SNe 2014cy and 2015cz the values overshoots the upper limit. 

\section*{Acknowledgements}
We thank the referee for providing valuable comments on the manuscript. We thank Llu\'is Galbany for kindly providing us the spectra of NGC\,7742. This paper is based on observations collected at Copernico and Schmidt telescopes (Asiago, Italy) of the INAF - Osservatorio Astronomico di Padova. We acknowledge the scientific and technical staff of ARIES (DFOT and ST), IAO (Hanle) and CREST (Hosakote), for making possible a part of the observations reported here. The facilities at IAO and CREST are operated by the Indian Institute of Astrophysics, Bangalore. This research has made use of the NASA/IPAC Extragalactic Database (NED) which is operated by the Jet Propulsion Laboratory, California Institute of Technology, under contract with the National Aeronautics and Space Administration. We acknowledge the usage of the HyperLeda data base (http://leda.univ-lyon1.fr). Guoshoujing Telescope (the Large Sky Area Multi-Object Fiber Spectroscopic Telescope LAMOST) is a National Major Scientific Project built by the Chinese Academy of Sciences. Funding for the project has been provided by the National Development and Reform Commission. LAMOST is operated and managed by the National Astronomical Observatories, Chinese Academy of Sciences. X. Wang is supported by the National Natural Science Foundation of China (NSFC grants 11325313, 11633002, and 11761141001), and the National Program on Key Research and Development Project (grant no. 2016YFA0400803. SBP and KM acknowledges BRICS grant DST/IMRCD/BRICS/Pilotcall/ProFCheap/2017(G) for the present work. SBP and KM also acknowledge the DST/JSPS grant, DST/INT/JSPS/P/281/2018.

\section*{Data Availability}
The photometric data underlying this article is available in the article and the spectroscopic data will be made available on WISeREP.




\bibliographystyle{mnras}
\bibliography{ref_mn2} 




\appendix

\section{Bolometric light curve modelling: Trace plot and Corner plots}
\begin{figure}
\begin{center}
\includegraphics[scale=0.5, trim={1.5cm 0.8cm 2.5cm 1.0cm}]{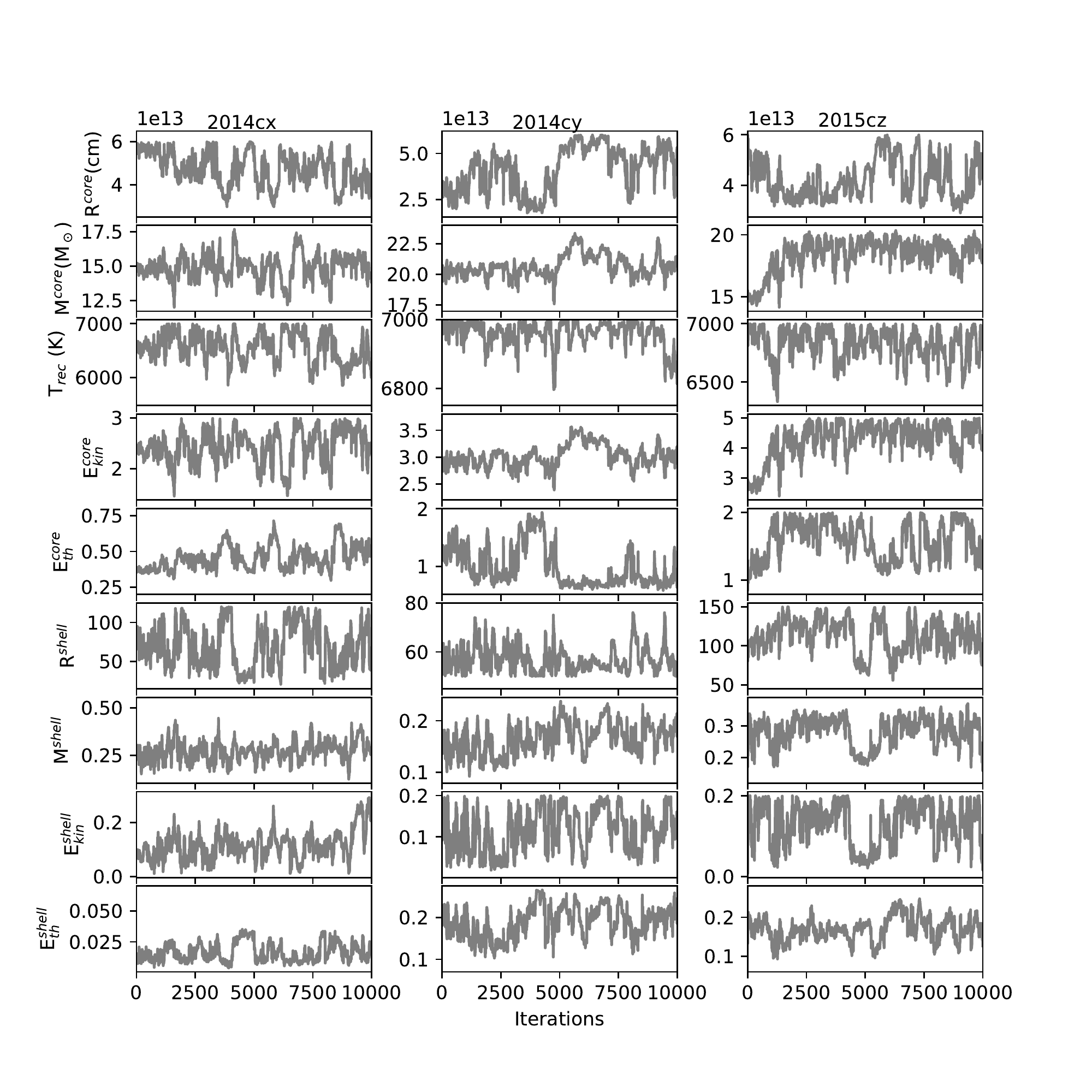}
\end{center}
\caption{The trace plots of the nine free paramaters for SNe 2014cx, 2014cy and 2015cz.}
\label{ch6:trace}
\end{figure}

\onecolumn

\begin{figure}
\begin{center}
\includegraphics[scale=0.55, height=0.6\textheight, width=1.1\textwidth, clip, trim={3.0cm 1.5cm 0.0cm 1.0cm}]{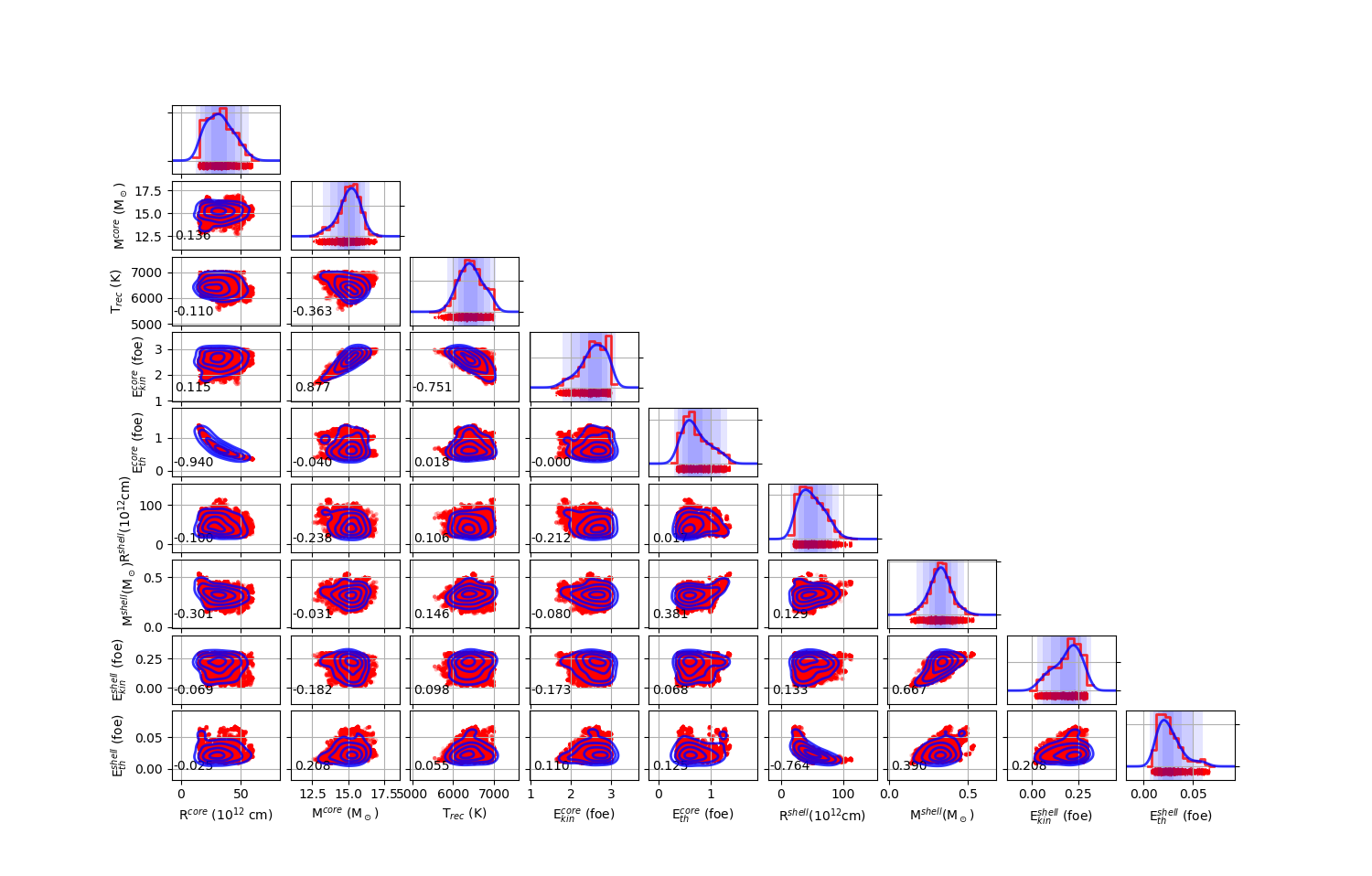}
\end{center}
\caption{The corner plot generated for SN 2014cx to show the covariance between the parameters. The diagonal sub-plots shows the kernel density estimates of the parameters. The correlation coefficient between the parameters can be found in each of the subplots.}
\label{ch6:corner_cx}
\end{figure}

\begin{figure*}
\begin{center}
\includegraphics[scale=0.55, height=0.6\textheight, width=1.1\textwidth, clip, trim={3.0cm 1.5cm 0.0cm 1.0cm}]{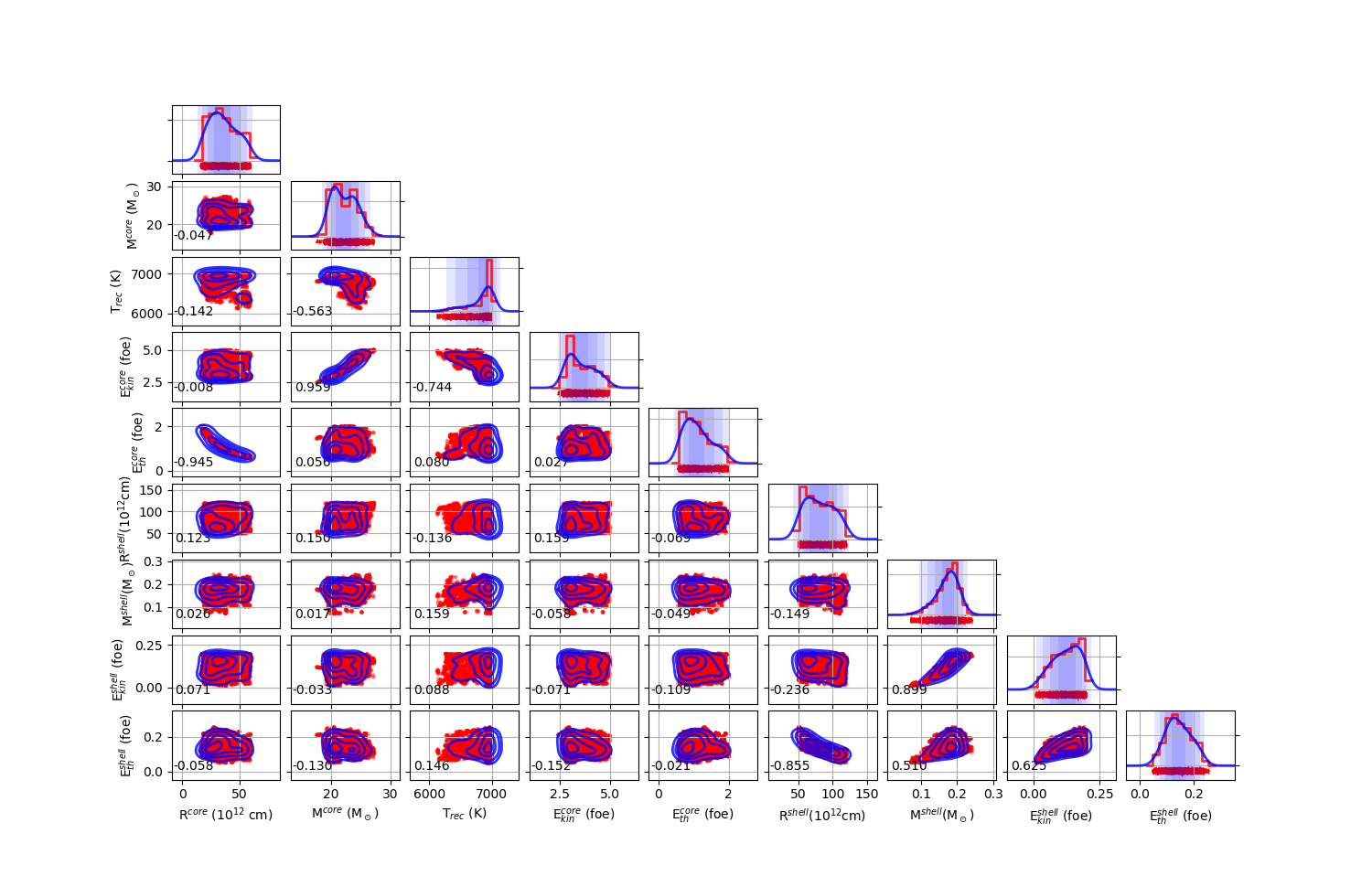}
\end{center}
\caption{The corner plot generated for SN 2014cy to show the covariance between the parameters. The diagonal sub-plots shows the kernel density estimates of the parameters. The correlation coefficient between the parameters can be found in each of the subplots.}
\label{ch6:corner_cy}
\end{figure*}

\begin{figure*}
\begin{center}
\includegraphics[scale=0.55, height=0.6\textheight, width=1.1\textwidth, clip, trim={3.0cm 1.5cm 0.0cm 1.0cm}]{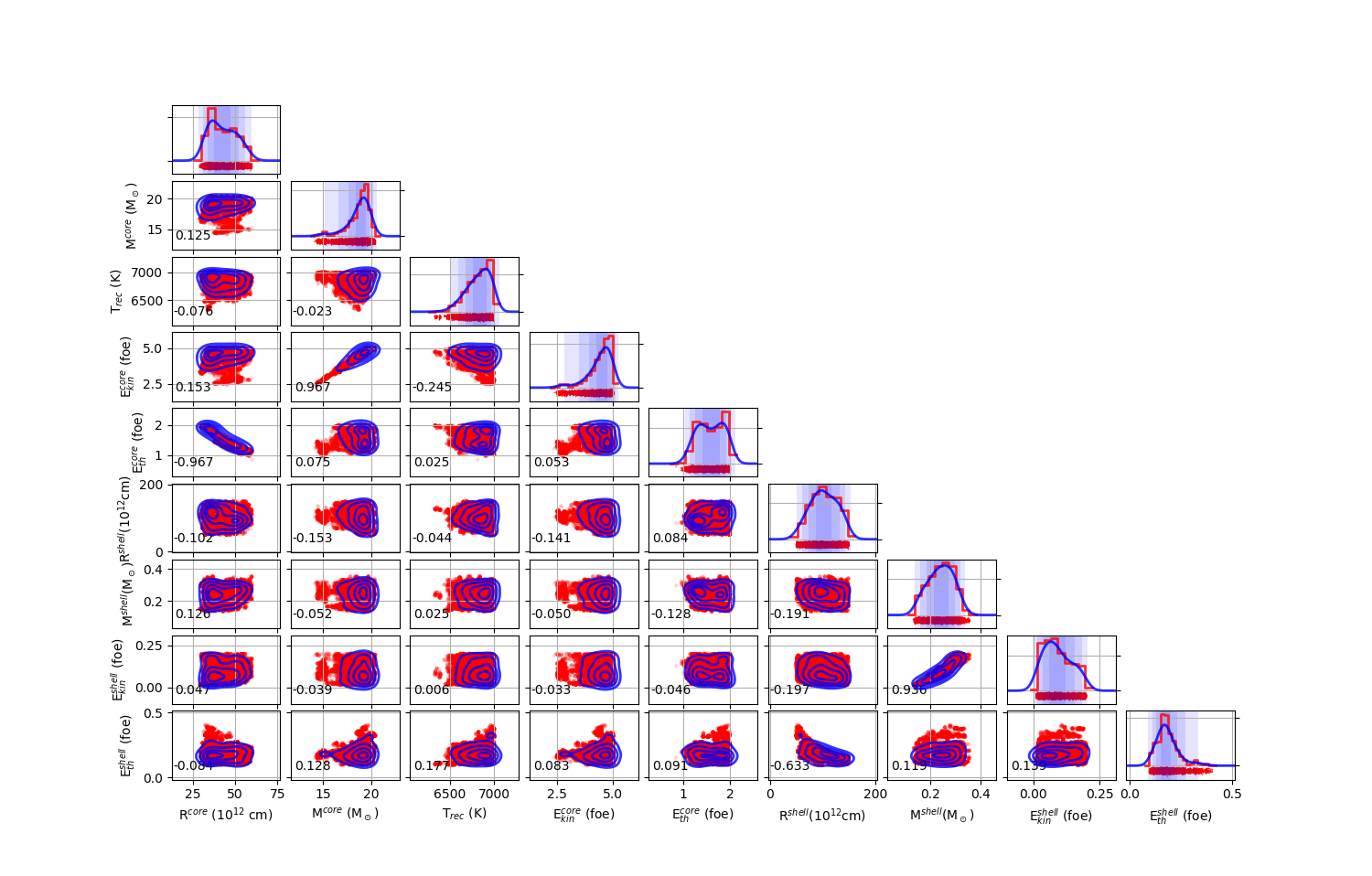}
\end{center}
\caption{The corner plot generated for SN 2015cz to show the covariance between the parameters. The diagonal sub-plots shows the kernel density estimates of the parameters. The correlation coefficient between the parameters can be found in each of the subplots.}
\label{ch6:corner_cz}
\end{figure*}



\bsp	
\label{lastpage}
\end{document}